\documentclass[aps,pra,twocolumn,superscriptaddress,floatfix,longbibliography]{revtex4-2}

\usepackage{graphicx}
\usepackage{amsmath}
\usepackage{amssymb}
\usepackage{bm}
\usepackage{url}

\newcommand{\dmu}{d\mu}

\newcommand{\amu}{a_{\dmu}}
\newcommand{\GC}{\Gamma_{\mathrm{C}}}
\newcommand{\eV}{\,\mathrm{eV}}
\newcommand{\meV}{\,\mathrm{meV}}

\begin{document}
\title{Response of the near-threshold resonances of the $\bm{dt\mu}$ molecular ion to an effective atomic $2s$--$2p$ splitting}
\author{Yi Zhang}
\email{yzhang@hainanu.edu.cn}
\affiliation{Center for Theoretical Physics, Hainan University, Haikou 570228, China}
\affiliation{School of Physics and Optoelectronic Engineering, Hainan University, Haikou 570228, China}
\date{September 13, 2026}
\begin{abstract}
In the nonrelativistic three-body Coulomb problem, the
near-threshold resonances of the deuterium--tritium
muonic molecular ion ($dt\mu$) are supported by the long-range
dipole coupling between the degenerate atomic $2s$ and
$2p$ states. We study how an atomic $2s$--$2p$ splitting $\Delta$ changes
their positions and decay widths.
Complex-absorbing-potential calculations resolve additional members
of the $L^\pi=0^+$ and $1^-$ series, whose geometric spacing agrees
with the $1/R^2$ dipole attraction ($R$ is the atom--nucleus separation).
The splitting is introduced through an effective short-range potential
in a spin-independent, nonrelativistic Hamiltonian, without an explicit
Lamb-shift calculation.
As $\Delta$ grows, the resonances approach the $d\mu(2s)+t$
threshold and their Coulomb decay widths decrease, the distance to
threshold shrinking faster than the width. At $\Delta=0.2\eV$,
close to the physical splitting in $d\mu$,
the seventh $1^-$ member lies within $2\meV$ of the threshold
with a full width of about $4\meV$, compared with $48$ and
$15.6\meV$ at $\Delta=0$, and in neither symmetry do the three-body
calculations resolve an eighth member below the threshold. A
quantum-defect model calibrated only at $\Delta=0$ reproduces this
response and predicts that the eighth members reach the threshold
already at $\Delta=0.12\eV$ ($1^-$) and $0.15\eV$ ($0^+$), leaving
seven members of each series at $\Delta=0.2\eV$. The splitting thus
replaces the infinite accumulation expected for degenerate $2s$ and
$2p$ levels by a finite series.
\end{abstract}
\maketitle
\raggedbottom

\section{Introduction}
\label{sec:introduction}

A negative muon can bind to a nucleus to form a muonic atom, or to two
nuclei to form a muonic molecular ion. Because the muon is about 207
times heavier than the electron, these systems are small, and their
levels are sensitive to electron vacuum polarization, muon
self-energy, and finite nuclear size~\cite{pachucki2024}. The small
internuclear separation of a muonic molecule also enhances nuclear
tunneling, so that the muon can catalyze fusion of hydrogen
isotopes~\cite{breunlich1989,ponomarev1990}. In the muon-catalyzed
fusion ($\mu$CF) cycle, the rate-limiting step is the formation of a
new $dt\mu$ ion after each D--T fusion, where $d$ and $t$ denote the
deuteron and triton. It proceeds resonantly through the Vesman
mechanism~\cite{vesman1967,ponomarev1990}: the binding energy of a
weakly bound $dt\mu$ state below the $t\mu(1s)+d$
threshold~\cite{froelich1992} is taken up by a rovibrational level of
the $[(dt\mu)dee]$ complex. That resonance belongs to the complex; the
$dt\mu$ state itself is bound.

Molecules can also form before the muonic atom reaches its ground
state. Collisions of $n=2$ atoms in the atomic
cascade~\cite{jensen2002cascade} can form excited molecular ions, for
example through $t\mu(2s)+\mathrm{D}_2\to[(dt\mu^\ast)dee]$
\cite{froelich1995,wallenius1996}, where the asterisk denotes a
resonance of the three-body $dt\mu$ ion. Such a state lies above lower
dissociation thresholds and has a finite lifetime even without fusion.
Its nonradiative decay $dt\mu^\ast\to d\mu(1s)+t$ transfers the muon
from the triton to the deuteron and can reduce the cycling
rate~\cite{froelich1995}. Radiative decay can instead leave a bound
molecule that feeds subsequent fusion, with appreciable fractions for
some $dd\mu^\ast$ and $dt\mu^\ast$ resonances~\cite{yamashita2025}, or
accompany dissociation. The emitted X rays carry spectral profiles
characteristic of the initial resonance~\cite{lindroth2003,yamashita2025},
and Toyama \textit{et al.}~\cite{toyama2026} recently resolved
molecular features in X rays from $dd\mu^\ast$ radiative dissociation
with transition-edge-sensor microcalorimeters at J-PARC, so that
calculated molecular states can now be compared directly with
experiment.

These processes require the positions and widths of the resonances:
positions enter the energy balance for formation, and Coulomb decay
widths describe loss into open dissociation channels in competition
with radiative decay. We consider $dt\mu$, the molecule of the D--T
$\mu$CF cycle, for which earlier three-body calculations of this
energy region provide benchmarks. Because the atomic reduced masses
are unequal, its $t\mu(n=2)+d$ and $d\mu(n=2)+t$ thresholds are
separated by $12.0\eV$. The resonances between them approach
$d\mu(n=2)+t$ from below, and the near-threshold members decay mainly
into the open $t\mu(n=2)+d$ channel (Sec.~\ref{sec:method}). In the
homonuclear $dd\mu$ and $tt\mu$ ions the two $n=2$ thresholds
coincide and this channel is absent. We study these nonrelativistic
three-body Coulomb resonances and examine how an atomic $2s$--$2p$
splitting changes their positions and widths.

Previous three-body studies provide benchmarks for this spectrum.
Shimamura~\cite{shimamura1989} and Hara and Ishihara~\cite{hara1989}
calculated states of muonic molecular ions below the $n=2$ atomic
levels, Korobov~\cite{korobov1996window} treated $t\mu(2s,2p)+d$
scattering between the two $n=2$ thresholds of $dt\mu$ in a
four-channel adiabatic approximation, and K{\i}l{\i}\c{c}
\textit{et al.}~\cite{kilic2004,kilic2005} calculated resonance
energies and Coulomb and radiative decay properties. Wen and
Zhu~\cite{wenzhu2025} combined an extended stochastic variational
method with complex scaling and resolved additional shallow
resonances in several muonic systems; the same approach has since
identified Gailitis--Damburg sequences in purely leptonic three-body
systems~\cite{wen2026lepton}. These calculations, like the Coulomb
baseline of the present work, assume exactly degenerate atomic levels.
Near the $d\mu(n=2)+t$ threshold the atom--nucleus motion extends to
large distances while the atomic motion remains compact, and we
construct the basis to resolve both scales.

Without atomic splitting, dipole coupling between the degenerate
$2s$ and $2p$ states generates a long-range $1/R^2$ attraction, with
$R$ the atom--nucleus separation, that supports an infinite geometric
sequence of resonances accumulating at the $d\mu(n=2)+t$
threshold~\cite{gailitis1963,gailitis1982,shimamura1989}. Such a
sequence is a Coulomb analog of the discrete scale invariance of
Efimov states~\cite{braaten2006,nishida2022,wen2026lepton}. In real
muonic atoms the Lamb shift, dominated by electron vacuum
polarization, removes this degeneracy by lowering $2s$ relative to
$2p$. The measured
$2S$--$2P_{1/2}$ Lamb shifts are $202.37\meV$ in muonic hydrogen and
$202.88\meV$ in $d\mu$~\cite{pachucki2024,pohl2016}, and theory gives
about $0.22\eV$ for $t\mu$~\cite{muT2017}. This splitting is small
compared with the atomic binding energy but can exceed the distance
of a shallow resonance from the $d\mu(2s)+t$ threshold. Where the
dipole coupling becomes weak compared with the splitting, the
attraction falls as $1/R^4$ and the condition for infinite
accumulation no longer holds~\cite{karr2012,karr2013}.

This truncation mechanism has precedents. Wallenius and
Kamimura~\cite{walleniusKamimura1996} included the Uehling potential
in the $dt\mu$ Hamiltonian and found a finite series of metastable
states on the adiabatic $3\sigma$ potential, relevant to formation in
$t\mu(2s)$--$\mathrm{D}_2$ collisions. Karr and Hilico~\cite{karr2012}
described weakly bound $p^+\mu^-x^-$ resonances with an effective
Yukawa potential calibrated to the atomic $2S$ and $2P$
vacuum-polarization shifts, checked it against an Uehling calculation
of a deeper $dt\mu$ resonance, and later treated low vibrational
resonances with the Uehling potential directly~\cite{karr2013}. These
studies establish the truncation mechanism and the use of an
effective atomic potential. The present study addresses instead the
shallow $0^+$ and $1^-$ members approaching the upper $d\mu(n=2)+t$
threshold.

We calculate the $L^\pi=0^+$ and $1^-$ resonances with the three-body
complex-absorbing-potential (CAP) method~\cite{riss1993}, expanding the
wave function in explicitly correlated Gaussians and atomic-channel
functions that describe the compact atomic and the extended
atom--nucleus motion together~\cite{mitroy2013,hiyama2003}. After
establishing the pure-Coulomb spectrum, we introduce the splitting
$\Delta=E_{2p}-E_{2s}$ through a short-range Gaussian potential
calibrated separately for $d\mu$ and $t\mu$, impose the same $\Delta$
in both atoms, and follow each resonance as $\Delta$ is scanned in the
spin-independent Hamiltonian. To
identify the long-range part of the response, we compare the
three-body results with a multichannel quantum-defect
model~\cite{fano1970,greene1979,seaton1983} that treats the
$d\mu(2s)+t$ and $d\mu(2p)+t$ channels explicitly and represents
short-range correlation and decay into lower channels by a complex
boundary condition fitted at $\Delta=0$ and then held fixed.

Section~\ref{sec:method} presents the methods, Sec.~\ref{sec:results}
the Coulomb dipole series, and Sec.~\ref{sec:longrange} its response
to the splitting and the quantum-defect analysis;
Sec.~\ref{sec:conclusions} concludes. Numerical controls and the model
continuation and scattering analysis are given in the Supplemental
Material~\cite{supplemental}, which also includes
Refs.~\cite{korobov2000,newton1961}.

\section{Hamiltonian and theoretical description}
\label{sec:method}

\subsection{Hamiltonian and effective atomic splitting}
We study the $L^\pi=0^+$ and $1^-$ resonances below the
$d\mu(n=2)+t$ dissociation threshold. The system is described by
three distinguishable particles with a spin-independent,
nonrelativistic Hamiltonian. In electronic atomic units,
\begin{equation}
 H=H_{\mathrm C}+V_\Delta,\qquad
 H_{\mathrm C}=T_{\mathrm{rel}}-\frac{1}{r_{\mu d}}
              -\frac{1}{r_{\mu t}}+\frac{1}{r_{dt}},
 \label{eq:H}
\end{equation}
where $H_{\mathrm C}$ contains the Coulomb interactions and internal
kinetic energy, and $V_\Delta$ is the effective splitting potential
defined in Eq.~\eqref{eq:split}. With $\bm x_1=\bm r_d-\bm r_\mu$ and
$\bm x_2=\bm r_t-\bm r_\mu$, the latter is
\begin{equation}
 T_{\mathrm{rel}}=-\frac{\nabla_1^2}{2\mu_d}
                 -\frac{\nabla_2^2}{2\mu_t}
                 -\frac{\nabla_1\!\cdot\!\nabla_2}{m_\mu},
 \qquad \mu_a=\frac{m_\mu m_a}{m_\mu+m_a}.
\end{equation}
The masses are $m_d=3670.48296788$, $m_t=5496.92153573$, and
$m_\mu=206.7682830$, in electron-mass units~\cite{codata2018}.
Energies are measured from $d+t+\mu$ breakup, and lengths are
expressed in $\amu=a_0/\mu_d$ where indicated. For $H=H_{\mathrm C}$ ($\Delta=0$),
the $t\mu(n=2)+d$ and $d\mu(n=2)+t$ thresholds lie at
$-677.810418\eV$ and $-665.800121\eV$, respectively. The difference
arises from the unequal atomic reduced masses. Resonances between
these thresholds can decay through the lower $t\mu(n=2)+d$ channel
and the open $n=1$ channels. For an arrangement $a\mu+b$, we use
$\bm r=\bm r_a-\bm r_\mu$ for the atomic coordinate and
$\bm R=\bm r_b-\bm r_\mu-\kappa_a\bm r$ for the displacement of
nucleus $b$ from the atomic center of mass, where
$\kappa_a=m_a/(m_a+m_\mu)$.

To examine the effect of atomic $2s$--$2p$ splitting, we add a
short-range Gaussian potential between the muon and each nucleus,
\begin{equation}
 V_\Delta=\sum_{a=d,t}v_a(\Delta)
             \exp[-r_{\mu a}^2/\rho^2],\qquad \rho=0.05\amu,
 \label{eq:split}
\end{equation}
where $r_{\mu a}$ is the muon--nucleus distance, $\rho$ the range,
and $v_a(\Delta)$ the strength for nucleus $a$. For each atom, we
solve the isolated radial problem and choose $v_a(\Delta)$ so that
$E^{(a)}_{2p}-E^{(a)}_{2s}=\Delta$.
The prescribed splitting is the same for $d\mu$ and $t\mu$, but their
absolute thresholds are calculated separately. At $\Delta=0$ the
added potential vanishes. Scanning $\Delta$ therefore connects the
Coulomb spectrum to a spectrum with split atomic levels. This is
an effective description of atomic splitting without an explicit
Lamb-shift calculation, similar in spirit to the effective Yukawa
potential that Karr and Hilico calibrated to atomic
vacuum-polarization shifts~\cite{karr2012}; the Uehling potential
itself was used in Refs.~\cite{walleniusKamimura1996,karr2013}.
The potential also changes the atomic
radial functions, and the small displacement of the $2p$ level
is retained in the atomic calibration. Because $\rho$ is small
compared with the atomic radius, $V_\Delta$ acts as a contact
potential: to first order it shifts each level in proportion to
$|\phi_{nl}(0)|^2$, so the $1s$ levels move by about eight times
the $2s$ shift, roughly $-1.6\eV$ at $\Delta=0.2\eV$, while the
$2p$ levels move by less than $0.1\meV$. The $1s$ shift is
immaterial here: the $n=1$ channels are open with about
$2\,\mathrm{keV}$ of kinetic energy.

The Hamiltonian is spin independent, so $\Delta$ corresponds to the
interval between the spin-averaged $2s$ and $2p$ levels. In $d\mu$ the
$2p$ fine-structure interval is $8.86\meV$ and the $2s$ hyperfine
splitting $6.28\meV$~\cite{krauth2016}, and the $2p$ level has further
hyperfine structure. The spin-averaged $2p$ level lies two thirds of
the fine-structure interval above $2P_{1/2}$, so the spin-averaged
splitting of $d\mu$ is $0.209\eV$, compared with its Lamb shift of
$0.203\eV$. The largest splitting used for the main results,
$\Delta=0.2\eV$, is therefore close to the physical value; the $1^-$
three-body calculations include one further point, $\Delta=0.24\eV$.
The spin-dependent splittings are small compared with $\Delta$, but
they exceed the distance of the shallowest members from threshold at
$\Delta=0.2\eV$. Because they lift
the remaining degeneracies of the $n=2$ level, they cannot restore a
$1/R^2$ tail; they do split and shift the thresholds on the scale of
those distances. The finite-splitting results below therefore
characterize the spin-independent model, and spectroscopic predictions
for the shallowest states of $dt\mu$ require the spin-dependent
threshold structure.

For each resonance we write
\begin{equation}
 E=E_r-i\Gamma/2,\qquad
 B(\Delta)=E^{(d)}_{2s}(\Delta)-E_r(\Delta).
 \label{eq:binding}
\end{equation}
Thus $B$ is the distance below the $d\mu(2s)+t$ threshold.
A positive $B$ does not preclude decay into lower channels.
All widths are full widths. We denote the width from the three-body
CAP calculation by $\GC$ and that from the reduced model by
$\Gamma$. Radiative decay and nuclear fusion are not included
in this Hamiltonian.

\subsection{Three-body calculation of near-threshold resonances}
As the Coulomb sequence approaches the $d\mu(n=2)+t$ threshold,
the relative atom--nucleus motion extends to larger distances,
while the atomic motion remains compact. We represent both scales
by expanding the wave function as
\begin{equation}
 \Psi_{LM}=\sum_i c_i\Phi^{\mathrm{in}}_{i,LM}
          +\sum_i d_i\Phi^{\mathrm{ext}}_{i,LM}
          +\sum_{\alpha\nu}f_{\alpha\nu}\Xi_{\alpha\nu,LM}.
 \label{eq:basis-expansion}
\end{equation}
The first two sums contain compact and extended correlation functions.
The last contains contracted atomic-channel functions, with $\alpha$
specifying the arrangement and channel type and $\nu$ the radial
and oscillatory parameters. All three sets enter the same generalized
eigenproblem. Their linear coefficients are determined together.

The correlation functions use explicitly correlated Gaussians
\cite{suzukivarga1998,mitroy2013}, with global-vector angular factors
\cite{suzuki1998gvr}:
\begin{equation}
 \begin{split}
 G_{KLM}(A,\bm u;\bm x)
   &=|\bm v|^{2K}\mathcal Y_{LM}(\bm v)
      e^{-\bm x^T A\bm x},\\
 \bm v&=u_1\bm x_1+u_2\bm x_2.
 \end{split}
 \label{eq:ecg-form}
\end{equation}
Here $A$ is positive definite, $\mathcal Y_{LM}(\bm v)
=v^L Y_{LM}(\hat{\bm v})$, and the exponent has no factor of $1/2$.
For the correlation blocks we write
$\bm x^T A\bm x=z_{dt}r_{dt}^2+z_{\mu d}r_{\mu d}^2
+z_{\mu t}r_{\mu t}^2$. The scalar $L=0$ blocks use $K=0$;
the $L=1$ compact block includes $K=0$ and $1$. We calculate with
$M=0$, for which the angular factors are real.

The compact blocks are optimized for the lowest Coulomb state of
the corresponding $L$ by stochastic variational selection~\cite{varga1995}. Trial
Gaussian parameters are compared through the lowest generalized
eigenvalue, and functions are added or replaced to lower that energy.
The extended correlation blocks instead sample pair lengths
logarithmically over the $n=2$ spatial scales, with
$z_{ab}=1/(2s_{ab}^2)$. These functions are not optimized separately
for each resonance. For $L=1$, their global vectors are chosen from
$\bm x_1$, $\bm x_2$, and $\bm x_1-\bm x_2$. The parameter ranges,
block sizes, and overlap selection are specified in the Supplemental
Material.

For each atom we first represent the radial orbitals by
\begin{equation}
 \phi^{(a)}_{nl}(r)=r^l\sum_p D^{(a,nl)}_p e^{-\beta^{(a,nl)}_p r^2}.
 \label{eq:atomic-contraction}
\end{equation}
The coefficients follow from the isolated-atom radial eigenproblem.
For the original orbitals, $\beta_p=\beta_0 q_a^{p-1}$ with 16 terms; we optimize
$\beta_0$ and $q_a$ by minimizing the radial eigenvalue associated
with the chosen $(n,l)$ state. The relative motion is expanded on
a geometric grid,
\begin{equation}
 R_\nu=R_{\min}q_R^{\nu-1},\qquad
 a_\nu=\frac{1}{2R_\nu^2},
 \label{eq:channel-grid}
\end{equation}
with real functions
\begin{equation}
 F^{c,s}_{\nu\omega}(R)=e^{-a_\nu R^2}
 \begin{cases}
 \cos(\omega a_\nu R^2),\\
 \sin(\omega a_\nu R^2).
 \end{cases}
 \label{eq:channel-range}
\end{equation}
These are the real combinations of complex-range Gaussians
\cite{hiyama2003,ohtsubo2013,takenaka2014}; the sine function is omitted at $\omega=0$.
The principal grid uses $R_{\min}=0.3\amu$, $q_R=1.06$, and a
nominal upper range of $400\amu$, with $\omega=0,3,10$.
The overlap, kinetic, and potential matrices are therefore real symmetric.

Both $d\mu+t$ and $t\mu+d$ arrangements include $1s$, $2s$, and
$2p$ atomic functions. For $L=0$, the $s$ channels are products
$\phi^{(a)}_{ns}(r)F^{c,s}_{\nu\omega}(R)$. The $p$ channel is
implemented by an antisymmetric pair of cross-correlated Gaussians:
\begin{equation}
 \begin{split}
 \Xi^{(a)}_{2p,\nu\omega}
 &=F^{c,s}_{\nu\omega}(R)\sum_p D^{(a,2p)}_p e^{-\beta_p r^2}
   \frac{\sinh(2\gamma_{p\nu}\bm r\cdot\bm R)}{2\gamma_{p\nu}},\\
 \gamma_{p\nu}&=\xi\sqrt{a_\nu\beta_p},\qquad \xi=0.03.
 \end{split}
 \label{eq:p-channel-pair}
\end{equation}
As $\xi\to0$, the last factor tends to $\bm r\cdot\bm R$,
which supplies the scalar coupling of two $p$ waves. At finite
$\xi$ the function contains higher odd angular components. Their
effect is checked against an exact scalar angular construction in
the Supplemental Material.

For $L=1$, the $K=0$ channel functions are
\begin{equation}
 \Xi^{(a)}_{nl,\nu\omega,1M}
 =F^{c,s}_{\nu\omega}(R)\sum_p D^{(a,nl)}_p
       e^{-\beta_p r^2}\mathcal Y_{1M}(\bm v),
 \label{eq:l1-channel}
\end{equation}
with $\bm v=\bm R$ for $ns$ and $\bm v=\bm r$ for $2p$.
We supplement these with real $K=1$ contractions obtained by
replacing $\mathcal Y_{1M}(\bm v)$ by
$|\bm v|^2\mathcal Y_{1M}(\bm v)$ and taking
$\bm v=\cos\vartheta\,\bm r+\sin\vartheta\,\bm R$.
Both $2s$ and $2p$ radial coefficients are used, with
$\vartheta=0.3,0.6,0.9,1.2$ radians and $\omega=0$.
These functions include the $(l_{\mathrm{atom}},l_R)=(1,2)$
component absent from the separable $K=0$ channels.

The splitting scan also requires an accurate representation of
the changing atomic orbitals. A contraction fixed to the Coulomb
$2s$ orbital can introduce an error comparable to the small energy
separations under study. We therefore solve the atomic problem in
an enlarged space of 24 Gaussian terms with fixed geometric
exponents, or 32 terms as a control, at $\Delta=0$ and at
$\Delta_*=0.3\eV$. From the normalized $2s$ radial functions we form
\begin{equation}
 \begin{split}
 s_a&=\langle\phi^{(a,0)}_{2s}|\phi^{(a,\Delta_*)}_{2s}\rangle,\\
 \phi^{(a)}_{\mathrm{resp}}
 &=\frac{\phi^{(a,\Delta_*)}_{2s}-s_a\phi^{(a,0)}_{2s}}
         {\sqrt{1-|s_a|^2}}.
 \end{split}
 \label{eq:radial-response}
\end{equation}
The inner product uses the radial measure $r^2dr$. The phases
are chosen so that $s_a$ is real and positive. The $K=0$ $2s$
channels contain both $\phi^{(a,0)}_{2s}$ and
$\phi^{(a)}_{\mathrm{resp}}$; the $K=1$ $2s$ channels use the
improved zero-splitting contraction. We call the molecular bases with
these contractions enhanced and label them by the number of $2s$
terms (24-term or 32-term); the original bases use the 16-term $2s$
orbital and no response function. The $1s$ and $2p$ orbitals are
the 16-term functions in both cases. All nonlinear basis parameters
and atomic contraction coefficients remain fixed during the
$\Delta$ and $\eta$ scans. The molecular coefficients in
Eq.~\eqref{eq:basis-expansion} remain free. Isolated-atom tests
check its accuracy over the scanned range of $\Delta$. These tests constrain
the atomic representation; the sensitivity of molecular positions
and widths is assessed separately.

To extract the complex energies, we add a CAP in the large-$R$
region~\cite{jolicard1985,riss1993,riss1998} and solve
\begin{equation}
 \begin{gathered}
 (H-i\eta W)c=ESc,\\
 W(R)=(R-R_{\mathrm{CAP}})^2\theta(R-R_{\mathrm{CAP}}),
 \end{gathered}
 \label{eq:cap}
\end{equation}
where $S$ is the overlap matrix and $\eta>0$ is the absorber strength.
In the principal calculations, the absorber acts in the
$t\mu(n=2)$ coefficient block. Outgoing flux in the open $n=1$
channels is not absorbed; those channels are represented on the
finite grid. To test whether this omission affects the reported
widths, we added an all-contraction atomic $1s$ projector absorber
with an independent strength in both arrangements; for both $L$ it
changes the seventh member by less than $0.1\meV$ in position and
width (Supplemental Material~\cite{supplemental}). The
widths $\GC$ reported below are thus obtained with absorption in
the $t\mu(n=2)$ channels. For a fixed basis and absorber onset,
we follow the eigenvalues as $\eta$ varies. We select representatives
at interior local minima of $|dE/d\ln\eta|$ and obtain
$\GC=-2\operatorname{Im}E$; no first-order correction in $\eta$ is
applied, and the selected speeds $v=|dE/d\ln\eta|$ listed with the
controls bound its size. Trajectories are extracted before
assigning sequence labels or comparing with earlier spectra.

We test whether the changes with $\Delta$ can be resolved against
the dependence on the numerical representation. The controls vary
the atomic radial expansion, the range of the atom--nucleus grid,
the basis construction, the overlap cutoff, and the absorber onset.
Because the contracted basis is strongly overcomplete, near-dependent
directions are removed by a cutoff $\epsilon_S$ on the eigenvalues
of the normalized overlap matrix. A retained space is used in the
splitting scan only if its zero-splitting spectrum contains the
members being followed with the dipole spacing of
Sec.~\ref{sec:geometric} and with mutually consistent stationary
points; this criterion involves no literature values and is applied
in Sec.~\ref{sec:lamb}.
Fixed-strength snapshots without an interior stationary minimum
are kept as diagnostics only.
Because the target states are embedded in open dissociation
continua, the complex CAP poles carry no variational upper bound.
We therefore judge the response to $\Delta$ against the spread
over these controls, treating positions and widths separately.
Quoted ranges are the observed spreads over the tested settings;
they are not total error bounds, and the displayed digits do not
imply a uniform accuracy. The atomic tests, scan parameters,
trajectory-selection criteria, eigenproblem residuals, and full
control tables are given in the Supplemental Material.

\subsection{Long-range dipole coupling and quantum-defect model}
\label{sec:qdefect}
At large separation in the $d\mu(n=2)+t$ arrangement, the electric
field of the triton couples the degenerate atomic $2s$ and $2p$ states.
The leading dipole interaction varies as $1/R^2$. Including the
centrifugal terms, the coupled radial equation at degeneracy is
\begin{equation}
 \left[-\frac{1}{2\mu_R}\frac{d^2}{dR^2}
       +\frac{\mathsf M}{2\mu_RR^2}-\mathcal E\right]\bm u=0,
 \label{eq:dipole}
\end{equation}
where $\mu_R=m_t(m_d+m_\mu)/(m_d+m_t+m_\mu)$ and
$\mathcal E=E-E^{(d)}_{2p}(0)$. The matrix $\mathsf M$ contains
both dipole and centrifugal contributions. The angular channels
$(l_{\mathrm{atom}},l_R)$ are $(0,0),(1,1)$ for $L=0$ and
$(0,1),(1,0),(1,2)$ for $L=1$. Thus the two atomic thresholds
correspond to two and three radial channels, respectively.
The lowest eigenvalue $\lambda_{\min}$ of $\mathsf M$ defines
$\alpha_L=\sqrt{-\lambda_{\min}-1/4}$, with values $8.2734$ and
$8.1551$ for $L=0$ and $1$. These values agree with the closed-form
expressions of Ref.~\cite{wen2026lepton} evaluated with
$\eta=\mu_R/\mu_d$. The attractive channel supports the
Gailitis--Damburg sequence~\cite{gailitis1963,gailitis1982,shimamura1989},
\begin{equation}
 B_j-i\Gamma_j/2=E_{\mathrm{ref}}
       \exp[-2\pi(j-\mu)/\alpha_L],
 \label{eq:GD}
\end{equation}
where $\mu$ is a complex dipole quantum defect. We choose
$E_{\mathrm{ref}}=1\eV$ as the reference energy defining this
quantum-defect convention. A different positive reference energy
shifts $\operatorname{Re}\mu$ by a constant and leaves the spectrum
unchanged. Here $j$ labels a molecular sequence
member, distinct from the atomic principal quantum number $n=2$.
The left side is the conjugated binding energy
$E_{\mathrm{thr}}-E_j^*$. An approximately constant defect gives
geometric spacing, which can be compared with the three-body
Coulomb spectrum.

To determine the finite-splitting spectrum, we use a multichannel
quantum-defect model~\cite{fano1970,greene1979,seaton1983}.
The atomic splitting enters the coupled equations directly:
\begin{equation}
 \left[-\frac{1}{2\mu_R}\frac{d^2}{dR^2}
       +\frac{\mathsf M}{2\mu_RR^2}
       -\Delta P_{2s}-\mathcal E\right]\bm u=0.
 \label{eq:mqdt}
\end{equation}
Here $P_{2s}$ projects onto the atomic $2s$ channel, and $\mathcal E$
uses the Coulomb $2p$ reference defined above. The $2s$ diagonal
threshold is lowered by $\Delta$ while the $2p$ reference is fixed.
Both the dipole coupling and the centrifugal terms remain in
$\mathsf M$. We integrate these equations in the logarithmic
coordinate $x=\ln(R/\amu)$ with $\bm u=\sqrt R\,\bm v$;
the transformed equations are given in the Supplemental Material.

Short-range correlation and loss into eliminated channels are
represented by a complex logarithmic-derivative boundary at
$R_m=20\amu$. The attractive-channel boundary is
linear in $\varepsilon=-\mathcal E=E^{(d)}_{2p}(0)-E$, in eV:
$L_a(\varepsilon)=c_0+c_1\varepsilon$, where
$L_a=(1/v_a)\,dv_a/dx$ at $R_m$. Thus $c_0$ is dimensionless,
$c_1$ is in $\mathrm{eV}^{-1}$, and the energy reference remains
the Coulomb $2p$ threshold at finite splitting; it is not reset to
the moving $2s$ threshold. The other channels take regular
solutions. We fit the coefficients to the $\Delta=0$ resonances
from one three-body basis and keep them fixed throughout the
splitting scan. The boundary retains its energy dependence, but no
additional dependence on $\Delta$ is fitted. Each three-body basis
has its own calibration, named after that basis: a 24-term or
32-term (enhanced) calibration, or an original-input calibration. Comparing the finite-splitting predictions
with the corresponding three-body results then shows how much of
the response follows from the changing long-range coupling alone.

The model keeps the $2p$ reference at its Coulomb value and places
the $2s$ threshold at $E^{(d)}_{2p}(0)-\Delta$. The three-body atomic
calibration includes a small $2p$ shift, so the two calculations use
slightly different dissociation thresholds. We retain each calculation's
own threshold when comparing $B$. The Supplemental Material gives
the threshold offset and the conversion to a common breakup energy
reference. Near the $2s$ threshold, we also continue the model
solutions along complex radial contours and retain the momentum
branch of each channel. This distinguishes the continuation of a
complex energy from its identification as an outgoing resonance
pole~\cite{eden1964}. The matching procedure and the corresponding elastic
scattering analysis are presented in the Supplemental Material.

\subsection{Software, AI assistance, and numerical verification}
\label{sec:software}
The compact $L=0$ block was optimized with the original
explicitly correlated Gaussian (ECG) stochastic variational program
jointly developed by the author and collaborators and previously
applied to positronium complexes~\cite{zhang2020ps2,zhang2023psh};
this program is not distributed. The $L=1$ compact block was
generated by a Python implementation of the same selection, which is
included in the reproduction archive. The frozen nonlinear parameters
of both blocks are inputs to the present calculations.

AI tools assisted, under the author's direction, with the Python
implementation, the execution and analysis of the calculations, and
the preparation of figures and tables.\footnote{Claude Code
(Anthropic; Claude Fable~5.1 and Claude Opus~5.5) and OpenAI Codex
(GPT-6 Astra), September 2026.} The author defined the
problem and methods, checked the reported values against the stored
calculations and their input checksums, and is responsible for the
content of the paper.

The overlap, kinetic, and potential matrices agree with a
collaborator's separate implementation to relative Frobenius
differences below $2\times10^{-12}$ for 160 sampled contracted
functions, and the calculations reproduce the two-body thresholds
and the resonance benchmarks of Table~\ref{tab:baseline}.
\section{Resonance spectrum: the dipole series near $d\mu(n=2)+t$}
\label{sec:results}
\label{sec:series}
\label{sec:geometric}

Table~\ref{tab:baseline} lists the resonances of the Coulomb
Hamiltonian between the two $n=2$ thresholds together with earlier
three-body results~\cite{kilic2004,kilic2005,wenzhu2025}. The deeper
members agree with these calculations. The first two $0^+$ positions
agree with K{\i}l{\i}\c{c} \textit{et al.} to the $0.01\eV$
precision of the published values, and their widths to within $1\%$.
The $1^-$ series reproduces the published sequence, with larger
numerical variations for its broad deep members. The sixth and higher
members of both series were not resolved in the earlier calculations.

\begin{table*}[t]
\caption{\label{tab:baseline}Resonance positions $E_r$
and full widths $\GC$ for the Coulomb Hamiltonian, in eV, with $d+t+\mu$ breakup as zero.
Literature columns give $E_r$ ($\Gamma$): Wen--Zhu,
Ref.~\cite{wenzhu2025}, Table 4; K{\i}l{\i}\c{c},
Ref.~\cite{kilic2005}, Tables B.12/B.13.
Parenthesized numbers give, in units of the last digit, the
comparison scales defined in the Supplemental Material, which also
specifies the basis used for each entry; they are sensitivities, not
statistical errors. The dagger marks the eighth $1^-$ member.}
\begin{ruledtabular}
\begin{tabular}{clllll}
$L^\pi$ & $j$ & $E_r$ & $\GC$ & Wen--Zhu & K{\i}l{\i}\c{c}\\\hline
$0^+$ & 1 & -670.810(13) & 1.273(71) & $-670.73$ (1.2) & $-670.81$ (1.2637) \\
$0^+$ & 2 & -668.130(5) & 0.635(37) & $-668.18$ (0.8) & $-668.13$ (0.6294) \\
$0^+$ & 3 & -666.885(2) & 0.304(19) & $-666.87$ (0.4) & --- \\
$0^+$ & 4 & -666.307(1) & 0.144(9) & $-666.30$ (0.2) & --- \\
$0^+$ & 5 & -666.037(1) & 0.068(4) & $-666.03$ & --- \\
$0^+$ & 6 & -665.911(1) & 0.032(2) & --- & --- \\
$0^+$ & 7 & -665.851(1) & 0.015(1) & --- & --- \\
\hline
$1^-$ & 1 & -670.65(16) & 1.43(33) & $-670.62$ (1.4) & $-670.549$ (1.472) \\
$1^-$ & 2 & -668.066(73) & 0.764(99) & $-668.03$ (0.8) & $-667.988$ (0.720) \\
$1^-$ & 3 & -666.837(23) & 0.394(64) & $-666.82$ (0.4) & --- \\
$1^-$ & 4 & -666.276(7) & 0.185(33) & $-666.25$ (0.2) & --- \\
$1^-$ & 5 & -666.019(2) & 0.086(16) & $-665.97$ & --- \\
$1^-$ & 6 & -665.9035(15) & 0.0314(33) & --- & --- \\
$1^-$ & 7 & -665.8481(9) & 0.0156(12) & --- & --- \\
$1^-$ & $8^\dagger$ & -665.8224(4) & 0.0081(9) & --- & --- \\
\end{tabular}
\end{ruledtabular}
\end{table*}

The present $0^+$ spectra show no stationary pole below the first
member. Between the $t\mu(n=2)+d$ threshold and the member at
$-670.81\eV$ lie 15 eigenvalues with $\GC<8\eV$. Each of them moves by
at least $0.12\eV$ when the absorber strength is changed by a factor
of about three, whereas the listed members move by less than $1\meV$
(Supplemental Material~\cite{supplemental}). A resonance about
$0.8\eV$ above this threshold, suggested by an extrapolation of the
series~\cite{shimamura1989}, is thus not found; the adiabatic cross
sections of Ref.~\cite{korobov1996window} likewise show no such
resonance. The $0^+$ series in this window therefore begins at
$-670.8\eV$.

At the other end of the series, the calculations that include the
response function of Eq.~\eqref{eq:radial-response} resolve an eighth
and a ninth $0^+$ member. Table~\ref{tab:sequence} lists the shallowest members of both
series with the ratios of successive bindings and widths. For $0^+$,
members five to nine give binding ratios of $2.138$ and width ratios
of $2.13$--$2.14$. For $1^-$, $B_6/B_7=2.156$ and $B_7/B_8=2.157$,
with width ratios of $2.01$ and $1.94$.

\begin{table}[tbp]
\caption{\label{tab:sequence}Near-threshold Coulomb resonances,
$B=E_{2s}-E_r$ and full widths $\GC$ in meV, with the successive
ratios $r_B=B_{j-1}/B_j$ and $r_\Gamma=\GC^{(j-1)}/\GC^{(j)}$; the
dipole values are $2.137$ ($0^+$) and $2.161$ ($1^-$). The bases
used and the sensitivity of the shallowest $1^-$ members to the
retained subspace are given in the Supplemental
Material~\cite{supplemental}. The dagger marks the eighth $1^-$
member, assigned by its spacing and stability.}
\begin{ruledtabular}
\begin{tabular}{ccrrcc}
$L^\pi$ & $j$ & $B$ & $\GC$ & $r_B$ & $r_\Gamma$\\\hline
$0^+$ & 5 & 237.602 & 72.122 & --- & --- \\
$0^+$ & 6 & 111.152 & 33.842 & 2.138 & 2.131 \\
$0^+$ & 7 & 52.000 & 15.855 & 2.138 & 2.134 \\
$0^+$ & 8 & 24.325 & 7.423 & 2.138 & 2.136 \\
$0^+$ & 9 & 11.375 & 3.475 & 2.138 & 2.136 \\
$1^-$ & 6 & 103.42 & 31.36 & --- & --- \\
$1^-$ & 7 & 47.96 & 15.62 & 2.156 & 2.008 \\
$1^-$ & $8^\dagger$ & 22.23 & 8.06 & 2.157 & 1.938 \\
\end{tabular}
\end{ruledtabular}
\end{table}

The shallowest assigned members are stable under the available basis
and range controls. The ninth $0^+$ member ($B=11.4\meV$) moves by
$0.10\meV$ when the channel grid is doubled, and the eighth $1^-$
member by $0.3\meV$ when the atomic radial representation is
improved. Two candidates closer to the threshold are not assigned: a
$0^+$ candidate near $5\meV$ moves by $0.9\meV$ under the same grid
change, and a stationary $1^-$ candidate near $4.5\meV$ does not fall
at the next place of the sequence, near $10\meV$. The widths depend
more strongly on the radial representation than the positions do.
Improving the atomic radial representation moves the sixth and seventh
$1^-$ positions by less than $0.4\meV$ but lowers their widths by
$20\%$ and $16\%$. The comparison scales of Table~\ref{tab:baseline}
are defined in the Supplemental Material; they refer to the listed
controls and do not cover every systematic change. The complete
radial, cutoff, and range
comparisons are given in the Supplemental Material.

The ratios in Table~\ref{tab:sequence} identify the members as a
Gailitis--Damburg sequence. For an approximately constant complex
quantum defect, Eq.~\eqref{eq:GD} gives the same ratio
$\exp(2\pi/\alpha_L)$ between successive bindings and between
successive widths, and the dipole matrices of Sec.~\ref{sec:qdefect}
fix it at $2.137$ for $0^+$ and $2.161$ for $1^-$, independently of
any short-range input. The binding ratios agree with these values to
within $0.3\%$, and this agreement assigns the sequence numbers,
including that of the eighth $1^-$ member. The $1^-$ width ratios
scatter more because these widths are more sensitive to the radial
basis. The resolved members extend the numerical test of the $1/R^2$
interpretation~\cite{shimamura1989} by three to four members beyond
earlier work, although a finite set of levels does not by itself
demonstrate an infinite accumulation.

The ratio depends on the mass configuration. In the leptonic systems
of Ref.~\cite{wen2026lepton}, an electron spectator makes the ratio so
large that only two or three members per channel are resolved, whereas
in $\mu^+\mu^+e^-$ the ratio approaches unity and the scaling
parameters fitted to about twenty members remain below the dipole
limit. The masses of $dt\mu$ give an intermediate ratio of about
$2.14$, at which eight or nine members per series are resolved and
their spacing can be compared directly with the dipole value.

\section{Response to atomic $2s$--$2p$ splitting}
\label{sec:longrange}

\subsection{Resonance positions and decay widths}
\label{sec:lamb}
\begin{figure*}[t]
\includegraphics[width=0.96\textwidth]{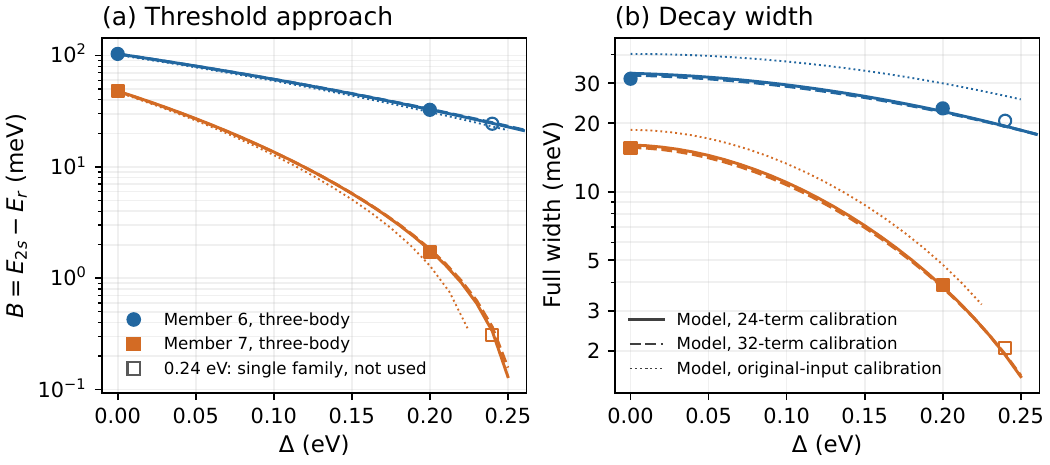}
\caption{\label{fig:v2-response}Response of the sixth and seventh
$1^-$ members to effective atomic splitting: (a) distance below
$d\mu(2s)+t$ and (b) full decay width. Filled symbols are the
three-body 24-term results at $\Delta=0$ and $0.2\eV$ (circles,
member six; squares, member seven); the open symbols at $0.24\eV$
come from a single family whose shallow candidate is not resolved
and are not used. Solid and dashed curves are quantum-defect
predictions with boundaries calibrated only to the zero-splitting
members 5--8 of the enhanced 24-term and 32-term bases; dotted
curves use the original-input calibration. No boundary
parameter is refitted at finite splitting. Bindings use each
calculation's own $2s$ threshold. Both vertical scales are
logarithmic. Plotted from stored numerical data using AI-assisted
plotting code; see Sec.~\ref{sec:software}.}
\end{figure*}

Atomic splitting shifts the studied near-threshold resonances toward
$d\mu(2s)+t$ and reduces their Coulomb decay widths. The fractional
decrease in the distance below threshold is larger than that in the
width, so the narrowing does not make the shallowest resonance better
isolated. Figure~\ref{fig:v2-response} shows this response for the
sixth and seventh $1^-$ members, with each binding measured from the
$2s$ threshold of the same calculation.

At $\Delta=0.2\eV$, the seventh $1^-$ member lies about $1.7\meV$
below threshold with a full width of about $4\meV$, compared with
$47.96$ and $15.62\meV$ at zero splitting. Its width then exceeds
its distance below threshold. The more deeply bound sixth member
has a smaller fractional response: its binding changes from
$103.4$ to $32.6\meV$ and its width from $31.4$ to $23.2\meV$.
The seventh $0^+$ member follows the same pattern, with binding
reduced from $52.0$ to $3.68\meV$ and width from $15.9$ to
$5.38\meV$. Over the retained spaces that pass the zero-splitting
criteria of Sec.~\ref{sec:method}, the seventh $1^-$ binding at
$0.2\eV$ ranges from $1.58$ to $1.83\meV$. This spread covers only
these retained spaces and does not establish convergence with respect
to the $L=1$ radial range. The complete controls, including the
cutoff dependence of the $0^+$ width, are given in the
Supplemental Material~\cite{supplemental}.

\subsection{Effect of splitting on the long-range attraction}
The approach to the threshold follows from the loss of the
$1/R^2$ attraction at large distance. At finite splitting the
dipole coupling becomes weak relative to $\Delta$ beyond some
separation. In a two-level reduction without centrifugal terms, a
$2s$ channel coupled to a $2p$ combination by $C/R^2$, with real
$C$ and diagonal energies $0$ and $\Delta>0$ relative to the $2s$
threshold, has the lower eigenvalue
\begin{equation}
 V_-(R)=\frac{\Delta}{2}
       -\sqrt{\frac{\Delta^2}{4}+\frac{C^2}{R^4}},
 \label{eq:split-tail}
\end{equation}
which equals $-|C|/R^2$ at degeneracy but, for $|C|/R^2\ll\Delta$,
\begin{equation}
 V_-(R)=-\frac{C^2}{\Delta R^4}+O(R^{-8}).
 \label{eq:polarization-tail}
\end{equation}
The attractive $1/R^2$ tail therefore no longer extends to
arbitrarily large separation, and the long-range condition for
infinite geometric accumulation at the lower threshold is removed.
In the language of few-body universality, the splitting introduces an
infrared length scale, of order $(|C|/\Delta)^{1/2}$, beyond which the
discrete scale invariance of the degenerate series is lost, much as a
finite scattering length limits the number of Efimov
states~\cite{braaten2006}.
The full coupled equations retain the centrifugal terms, so $V_-$
is not the complete effective potential, and the weaker attraction
by itself determines neither the complex energies nor the widths,
which also depend on short-range correlation and on loss into the
lower channels. The quantum-defect model includes both the modified long-range
coupling and short-range loss.

\subsection{Quantum-defect description of the spectral response}
\label{sec:model}
With the short-range boundary fixed by the zero-splitting spectrum,
the quantum-defect model accounts for both the approach to threshold
and the reduction in width. The solid and dashed curves in
Fig.~\ref{fig:v2-response} and the values in
Table~\ref{tab:v2-model} show its predictions without a finite-splitting
refit. At $\Delta=0.2\eV$, the principal model predicts the seventh
$1^-$ member at $B=1.79\meV$ and $\Gamma=3.83\meV$, compared with
$1.73$ and $3.89\meV$ in the three-body calculation. For the seventh
$0^+$ member, the corresponding values are $3.78$ and $5.44\meV$,
compared with $3.68$ and $5.38\meV$. Across the finite-splitting
$1^-$ entries of Table~\ref{tab:v2-model}, the model and three-body
values differ by at most about $0.4\meV$ in position and $0.9\meV$
in width. The full comparison, calibration dependence, and conversion between
energy references are given in the Supplemental Material~\cite{supplemental}.

\begin{table*}[t]
\caption{\label{tab:v2-model}Three-body $1^-$ results and
quantum-defect predictions for the sixth and seventh members.
$B$ and full widths are in meV. The three-body column uses the
enhanced 24-term representation. Model columns use boundaries
calibrated at $\Delta=0$ to the 24-term and 32-term spectra,
with the same parameters used at finite $\Delta$.
Bindings are relative to each calculation's own $2s$ threshold.
The complete comparison and threshold conversion are given in
the Supplemental Material~\cite{supplemental}.}
\begin{ruledtabular}
\begin{tabular}{crrrrrrr}
 & & \multicolumn{2}{c}{Three-body, 24 terms}
 & \multicolumn{2}{c}{Model, 24-term input}
 & \multicolumn{2}{c}{Model, 32-term input}\\
$j$ & $\Delta$ (eV) & $B$ & $\GC$ & $B$ & $\Gamma$ & $B$ & $\Gamma$\\\hline
6 & 0.00 & 103.42 & 31.36 & 103.47 & 33.13 & 103.21 & 32.51 \\
6 & 0.20 & 32.55 & 23.24 & 32.73 & 22.53 & 32.98 & 22.36 \\
7 & 0.00 & 47.96 & 15.62 & 47.95 & 16.05 & 47.69 & 15.66 \\
7 & 0.20 & 1.73 & 3.89 & 1.79 & 3.83 & 1.83 & 3.79 \\
\end{tabular}
\end{ruledtabular}
\end{table*}

Because one boundary describes both positions and widths, we
attribute the calculated response mainly to the modified long-range
dipole mixing. Short-range loss still sets the absolute widths, and
the agreement shows only that the zero-splitting boundary remains
adequate up to $\Delta=0.2\eV$.

The splitting removes the series from its shallow end. In the model,
the real part of each member reaches the $2s$ threshold at a splitting
five to six times its zero-splitting binding: $\Delta\approx0.12$ and
$0.26\eV$ for the eighth and seventh $1^-$ members with the enhanced
calibrations, and $0.15$ and $0.33\eV$ for the eighth and seventh
$0^+$ members with the original-input calibration (Supplemental
Material~\cite{supplemental}). The three-body spectra follow the same
order. At degeneracy both sectors have an eighth member below
threshold, and a ninth $0^+$ member is resolved; at $\Delta=0.2\eV$ no
stationary pole lies between the seventh member and the $2s$ threshold
in any admitted $1^-$ retained space or in either $0^+$ channel grid.
At a splitting of this size, close to the physical value for
$d\mu$, the model thus predicts seven members of each series below the
$2s$ threshold instead of an infinite sequence; the three-body
calculations resolve these seven and no eighth, consistent with this
prediction. The seventh $1^-$ member is
itself close to its limit: the three-body candidate at
$\Delta=0.24\eV$ is not resolved across the tested grids, so no
critical splitting is assigned to it.
\section{Conclusions}
\label{sec:conclusions}

The resolved $0^+$ and $1^-$ resonances extend the Coulomb spectrum
of $dt\mu$ toward the $d\mu(n=2)+t$ threshold, and their positions
and Coulomb decay widths respond strongly to an effective atomic
$2s$--$2p$ splitting. For the seventh $1^-$ member, a splitting
$\Delta=0.2\eV$, close to the physical value for $d\mu$, reduces the
distance below $d\mu(2s)+t$ from about
$48$ to about $1.7\meV$ and the full width from $15.6$ to about
$4\meV$, and the three-body calculations resolve no eighth member
below the threshold in either symmetry. An atomic splitting small compared with the atomic binding
energy thus changes the near-threshold molecular spectrum
substantially. These results refer to the spin-independent
Hamiltonian; the fine and hyperfine structure of $d\mu(n=2)$, of
several meV, must be included before the shallowest states are
compared with experiment.

The geometric spacing of the Coulomb members agrees with the
$1/R^2$ attraction generated by dipole coupling between degenerate
$2s$ and $2p$ states. Atomic splitting weakens this mixing at large
separations, where the dipole attraction instead falls as $1/R^4$.
The long-range condition supporting infinite geometric accumulation
is consequently removed, which explains why the near-threshold
spectrum is so sensitive to the atomic degeneracy on which the
Coulomb series relies.

The quantum-defect results support this interpretation
quantitatively. With its short-range boundary fixed once at zero
splitting, the two-threshold model reproduces the near-threshold
positions and widths of the enhanced bases at $\Delta=0.2\eV$ in both
angular-momentum sectors to within $1\meV$, below the $1.8\meV$
residuals of the calibration itself,
so the response to the splitting is carried mainly by the
long-range dipole mixing. The model brings the eighth members to the
threshold at $\Delta=0.12$--$0.15\eV$, below the physical splitting,
consistent with the three-body spectra, which resolve no eighth member
at $\Delta=0.2\eV$: the splitting turns the infinite accumulation
expected for degenerate $2s$ and $2p$ levels into a finite series.
Because a single zero-splitting calibration then yields the full
dependence on the splitting at negligible cost, this combination of
three-body calculations with a multichannel quantum-defect model
offers a practical route to near-threshold spectra in other systems
where a small level splitting lifts the degeneracy that supports a
dipole series, such as other muonic molecular ions and positronium
complexes.

The positions and widths obtained here provide spectroscopic
benchmarks for the formation, radiative decay, and X-ray spectra of
excited muonic molecules. Now that molecular features in the X rays
from radiative dissociation of muonic molecules can be resolved
experimentally~\cite{toyama2026}, calculated $dt\mu^\ast$ spectra near
the $d\mu(n=2)+t$ threshold will require these resonance parameters,
including the effect of the atomic splitting, which reshapes the
shallowest part of the spectrum.

\begin{acknowledgments}
	This work was supported by the National Natural Science Foundation of China
	under Grant Nos.~12393821, 12364059, and 12547103.  Yi Zhang was also
	supported by the Innovational Fund for Scientific and Technological Personnel
	of Hainan Province under Grant No.~KJRC2023B11.

The AI tools identified in Sec.~\ref{sec:software} also assisted with
drafting and editing the text from the author's outline.
\end{acknowledgments}

\paragraph*{Data availability.}
The data that support the findings of this article are openly
available~\cite{zenodo2026tdmu}, embargo periods may apply. The deposit
holds the code and data package: the Python implementation, frozen
basis inputs, archived CAP spectra, and the table- and figure-generation
materials with a reproduction guide. Its files are accessible to the
editors and referees during the review and will become public upon
publication. The numerical comparisons are included in the Supplemental
Material~\cite{supplemental}. The ECG stochastic optimizer used for the
compact $L=0$ basis block is not distributed; reproduction starts from
the frozen basis inputs.
\bibliography{refs}

\begin{thebibliography}{50}%
\makeatletter
\providecommand \@ifxundefined [1]{%
 \@ifx{#1\undefined}
}%
\providecommand \@ifnum [1]{%
 \ifnum #1\expandafter \@firstoftwo
 \else \expandafter \@secondoftwo
 \fi
}%
\providecommand \@ifx [1]{%
 \ifx #1\expandafter \@firstoftwo
 \else \expandafter \@secondoftwo
 \fi
}%
\providecommand \natexlab [1]{#1}%
\providecommand \enquote  [1]{``#1''}%
\providecommand \bibnamefont  [1]{#1}%
\providecommand \bibfnamefont [1]{#1}%
\providecommand \citenamefont [1]{#1}%
\providecommand \href@noop [0]{\@secondoftwo}%
\providecommand \href [0]{\begingroup \@sanitize@url \@href}%
\providecommand \@href[1]{\@@startlink{#1}\@@href}%
\providecommand \@@href[1]{\endgroup#1\@@endlink}%
\providecommand \@sanitize@url [0]{\catcode `\\12\catcode `\$12\catcode
  `\&12\catcode `\#12\catcode `\^12\catcode `\_12\catcode `\%12\relax}%
\providecommand \@@startlink[1]{}%
\providecommand \@@endlink[0]{}%
\providecommand \url  [0]{\begingroup\@sanitize@url \@url }%
\providecommand \@url [1]{\endgroup\@href {#1}{\urlprefix }}%
\providecommand \urlprefix  [0]{URL }%
\providecommand \Eprint [0]{\href }%
\providecommand \doibase [0]{https://doi.org/}%
\providecommand \selectlanguage [0]{\@gobble}%
\providecommand \bibinfo  [0]{\@secondoftwo}%
\providecommand \bibfield  [0]{\@secondoftwo}%
\providecommand \translation [1]{[#1]}%
\providecommand \BibitemOpen [0]{}%
\providecommand \bibitemStop [0]{}%
\providecommand \bibitemNoStop [0]{.\EOS\space}%
\providecommand \EOS [0]{\spacefactor3000\relax}%
\providecommand \BibitemShut  [1]{\csname bibitem#1\endcsname}%
\let\auto@bib@innerbib\@empty
%</preamble>
\bibitem [{\citenamefont {Pachucki}\ \emph {et~al.}(2024)\citenamefont
  {Pachucki}, \citenamefont {Lensky}, \citenamefont {Hagelstein}, \citenamefont
  {Li~Muli}, \citenamefont {Bacca},\ and\ \citenamefont {Pohl}}]{pachucki2024}%
  \BibitemOpen
  \bibfield  {author} {\bibinfo {author} {\bibfnamefont {K.}~\bibnamefont
  {Pachucki}}, \bibinfo {author} {\bibfnamefont {V.}~\bibnamefont {Lensky}},
  \bibinfo {author} {\bibfnamefont {F.}~\bibnamefont {Hagelstein}}, \bibinfo
  {author} {\bibfnamefont {S.~S.}\ \bibnamefont {Li~Muli}}, \bibinfo {author}
  {\bibfnamefont {S.}~\bibnamefont {Bacca}},\ and\ \bibinfo {author}
  {\bibfnamefont {R.}~\bibnamefont {Pohl}},\ }\bibfield  {title} {\bibinfo
  {title} {Comprehensive theory of the {L}amb shift in light muonic atoms},\
  }\href {https://doi.org/10.1103/revmodphys.96.015001} {\bibfield  {journal}
  {\bibinfo  {journal} {Rev. Mod. Phys.}\ }\textbf {\bibinfo {volume} {96}},\
  \bibinfo {pages} {015001} (\bibinfo {year} {2024})}\BibitemShut {NoStop}%
\bibitem [{\citenamefont {Breunlich}\ \emph {et~al.}(1989)\citenamefont
  {Breunlich}, \citenamefont {Kammel}, \citenamefont {Cohen},\ and\
  \citenamefont {Leon}}]{breunlich1989}%
  \BibitemOpen
  \bibfield  {author} {\bibinfo {author} {\bibfnamefont {W.~H.}\ \bibnamefont
  {Breunlich}}, \bibinfo {author} {\bibfnamefont {P.}~\bibnamefont {Kammel}},
  \bibinfo {author} {\bibfnamefont {J.~S.}\ \bibnamefont {Cohen}},\ and\
  \bibinfo {author} {\bibfnamefont {M.}~\bibnamefont {Leon}},\ }\bibfield
  {title} {\bibinfo {title} {Muon-catalyzed fusion},\ }\href
  {https://doi.org/10.1146/annurev.ns.39.120189.001523} {\bibfield  {journal}
  {\bibinfo  {journal} {Annu. Rev. Nucl. Part. Sci.}\ }\textbf {\bibinfo
  {volume} {39}},\ \bibinfo {pages} {311} (\bibinfo {year} {1989})}\BibitemShut
  {NoStop}%
\bibitem [{\citenamefont {Ponomarev}(1990)}]{ponomarev1990}%
  \BibitemOpen
  \bibfield  {author} {\bibinfo {author} {\bibfnamefont {L.~I.}\ \bibnamefont
  {Ponomarev}},\ }\bibfield  {title} {\bibinfo {title} {Muon catalysed
  fusion},\ }\href {https://doi.org/10.1080/00107519008222019} {\bibfield
  {journal} {\bibinfo  {journal} {Contemp. Phys.}\ }\textbf {\bibinfo {volume}
  {31}},\ \bibinfo {pages} {219} (\bibinfo {year} {1990})}\BibitemShut
  {NoStop}%
\bibitem [{\citenamefont {Vesman}(1967)}]{vesman1967}%
  \BibitemOpen
  \bibfield  {author} {\bibinfo {author} {\bibfnamefont {E.~A.}\ \bibnamefont
  {Vesman}},\ }\bibfield  {title} {\bibinfo {title} {Concerning one possible
  mechanism of production of the mesomolecular ion $(dd\mu)^+$},\ }\href@noop
  {} {\bibfield  {journal} {\bibinfo  {journal} {JETP Lett.}\ }\textbf
  {\bibinfo {volume} {5}},\ \bibinfo {pages} {91} (\bibinfo {year}
  {1967})}\BibitemShut {NoStop}%
\bibitem [{\citenamefont {Froelich}(1992)}]{froelich1992}%
  \BibitemOpen
  \bibfield  {author} {\bibinfo {author} {\bibfnamefont {P.}~\bibnamefont
  {Froelich}},\ }\bibfield  {title} {\bibinfo {title} {Muon catalysed fusion:
  chemical confinement of nuclei within the muonic molecule $dt\mu$},\ }\href
  {https://doi.org/10.1080/00018739200101533} {\bibfield  {journal} {\bibinfo
  {journal} {Adv. Phys.}\ }\textbf {\bibinfo {volume} {41}},\ \bibinfo {pages}
  {405} (\bibinfo {year} {1992})}\BibitemShut {NoStop}%
\bibitem [{\citenamefont {Jensen}\ and\ \citenamefont
  {Markushin}(2002)}]{jensen2002cascade}%
  \BibitemOpen
  \bibfield  {author} {\bibinfo {author} {\bibfnamefont {T.~S.}\ \bibnamefont
  {Jensen}}\ and\ \bibinfo {author} {\bibfnamefont {V.~E.}\ \bibnamefont
  {Markushin}},\ }\bibfield  {title} {\bibinfo {title} {Collisional
  deexcitation of exotic hydrogen atoms in highly excited states. {II}. cascade
  calculations},\ }\href {https://doi.org/10.1140/epjd/e2002-00208-x}
  {\bibfield  {journal} {\bibinfo  {journal} {Eur. Phys. J. D}\ }\textbf
  {\bibinfo {volume} {21}},\ \bibinfo {pages} {271} (\bibinfo {year}
  {2002})}\BibitemShut {NoStop}%
\bibitem [{\citenamefont {Froelich}\ and\ \citenamefont
  {Wallenius}(1995)}]{froelich1995}%
  \BibitemOpen
  \bibfield  {author} {\bibinfo {author} {\bibfnamefont {P.}~\bibnamefont
  {Froelich}}\ and\ \bibinfo {author} {\bibfnamefont {J.}~\bibnamefont
  {Wallenius}},\ }\bibfield  {title} {\bibinfo {title} {Resonance sidepath in
  muon catalyzed fusion},\ }\href {https://doi.org/10.1103/PhysRevLett.75.2108}
  {\bibfield  {journal} {\bibinfo  {journal} {Phys. Rev. Lett.}\ }\textbf
  {\bibinfo {volume} {75}},\ \bibinfo {pages} {2108} (\bibinfo {year}
  {1995})}\BibitemShut {NoStop}%
\bibitem [{\citenamefont {Wallenius}\ and\ \citenamefont
  {Froelich}(1996)}]{wallenius1996}%
  \BibitemOpen
  \bibfield  {author} {\bibinfo {author} {\bibfnamefont {J.}~\bibnamefont
  {Wallenius}}\ and\ \bibinfo {author} {\bibfnamefont {P.}~\bibnamefont
  {Froelich}},\ }\bibfield  {title} {\bibinfo {title} {Formation of metastable
  $dt\mu$ molecules in $t\mu(2s)$--$\mathrm{D}_2$ collisions},\ }\href
  {https://doi.org/10.1103/PhysRevA.54.1171} {\bibfield  {journal} {\bibinfo
  {journal} {Phys. Rev. A}\ }\textbf {\bibinfo {volume} {54}},\ \bibinfo
  {pages} {1171} (\bibinfo {year} {1996})}\BibitemShut {NoStop}%
\bibitem [{\citenamefont {Yamashita}\ \emph {et~al.}(2025)\citenamefont
  {Yamashita}, \citenamefont {Yasuda},\ and\ \citenamefont
  {Kino}}]{yamashita2025}%
  \BibitemOpen
  \bibfield  {author} {\bibinfo {author} {\bibfnamefont {T.}~\bibnamefont
  {Yamashita}}, \bibinfo {author} {\bibfnamefont {K.}~\bibnamefont {Yasuda}},\
  and\ \bibinfo {author} {\bibfnamefont {Y.}~\bibnamefont {Kino}},\ }\bibfield
  {title} {\bibinfo {title} {Radiative decay of muonic molecules in resonance
  states},\ }\href {https://doi.org/10.1103/physreva.111.012811} {\bibfield
  {journal} {\bibinfo  {journal} {Phys. Rev. A}\ }\textbf {\bibinfo {volume}
  {111}},\ \bibinfo {pages} {012811} (\bibinfo {year} {2025})}\BibitemShut
  {NoStop}%
\bibitem [{\citenamefont {Lindroth}\ \emph {et~al.}(2003)\citenamefont
  {Lindroth}, \citenamefont {Wallenius},\ and\ \citenamefont
  {Jonsell}}]{lindroth2003}%
  \BibitemOpen
  \bibfield  {author} {\bibinfo {author} {\bibfnamefont {E.}~\bibnamefont
  {Lindroth}}, \bibinfo {author} {\bibfnamefont {J.}~\bibnamefont
  {Wallenius}},\ and\ \bibinfo {author} {\bibfnamefont {S.}~\bibnamefont
  {Jonsell}},\ }\bibfield  {title} {\bibinfo {title} {Decay rates of excited
  muonic molecular ions},\ }\href {https://doi.org/10.1103/PhysRevA.68.032502}
  {\bibfield  {journal} {\bibinfo  {journal} {Phys. Rev. A}\ }\textbf {\bibinfo
  {volume} {68}},\ \bibinfo {pages} {032502} (\bibinfo {year} {2003})},\
  \bibinfo {note} {erratum: Phys. Rev. A \textbf{69}, 059903(E)
  (2004)}\BibitemShut {NoStop}%
\bibitem [{\citenamefont {Toyama}\ \emph {et~al.}(2026)\citenamefont {Toyama}
  \emph {et~al.}}]{toyama2026}%
  \BibitemOpen
  \bibfield  {author} {\bibinfo {author} {\bibfnamefont {Y.}~\bibnamefont
  {Toyama}} \emph {et~al.},\ }\bibfield  {title} {\bibinfo {title} {Direct
  observation of muonic molecules in resonance states critical to muon
  catalyzed fusion},\ }\href {https://doi.org/10.1126/sciadv.aed3321}
  {\bibfield  {journal} {\bibinfo  {journal} {Science Advances}\ }\textbf
  {\bibinfo {volume} {12}},\ \bibinfo {pages} {eaed3321} (\bibinfo {year}
  {2026})}\BibitemShut {NoStop}%
\bibitem [{\citenamefont {Shimamura}(1989)}]{shimamura1989}%
  \BibitemOpen
  \bibfield  {author} {\bibinfo {author} {\bibfnamefont {I.}~\bibnamefont
  {Shimamura}},\ }\bibfield  {title} {\bibinfo {title} {Series of resonance
  states of muonic molecules},\ }\href
  {https://doi.org/10.1103/physreva.40.4863} {\bibfield  {journal} {\bibinfo
  {journal} {Phys. Rev. A}\ }\textbf {\bibinfo {volume} {40}},\ \bibinfo
  {pages} {4863} (\bibinfo {year} {1989})}\BibitemShut {NoStop}%
\bibitem [{\citenamefont {Hara}\ and\ \citenamefont
  {Ishihara}(1989)}]{hara1989}%
  \BibitemOpen
  \bibfield  {author} {\bibinfo {author} {\bibfnamefont {S.}~\bibnamefont
  {Hara}}\ and\ \bibinfo {author} {\bibfnamefont {T.}~\bibnamefont
  {Ishihara}},\ }\bibfield  {title} {\bibinfo {title} {Bound and resonant
  states of muonic molecules below the $n=2$ level of muonic atoms},\ }\href
  {https://doi.org/10.1103/PhysRevA.40.4232} {\bibfield  {journal} {\bibinfo
  {journal} {Phys. Rev. A}\ }\textbf {\bibinfo {volume} {40}},\ \bibinfo
  {pages} {4232} (\bibinfo {year} {1989})}\BibitemShut {NoStop}%
\bibitem [{\citenamefont {Korobov}(1996)}]{korobov1996window}%
  \BibitemOpen
  \bibfield  {author} {\bibinfo {author} {\bibfnamefont {V.~I.}\ \bibnamefont
  {Korobov}},\ }\bibfield  {title} {\bibinfo {title} {Resonances in the $dt\mu$
  molecular ion between $t\mu(n=2)$ and $d\mu(n=2)$ thresholds},\ }\href
  {https://doi.org/10.1007/BF02227641} {\bibfield  {journal} {\bibinfo
  {journal} {Hyperfine Interact.}\ }\textbf {\bibinfo {volume} {101}},\
  \bibinfo {pages} {333} (\bibinfo {year} {1996})}\BibitemShut {NoStop}%
\bibitem [{\citenamefont {K{\i}l{\i}\c{c}}\ \emph {et~al.}(2004)\citenamefont
  {K{\i}l{\i}\c{c}}, \citenamefont {Karr},\ and\ \citenamefont
  {Hilico}}]{kilic2004}%
  \BibitemOpen
  \bibfield  {author} {\bibinfo {author} {\bibfnamefont {S.}~\bibnamefont
  {K{\i}l{\i}\c{c}}}, \bibinfo {author} {\bibfnamefont {J.-P.}\ \bibnamefont
  {Karr}},\ and\ \bibinfo {author} {\bibfnamefont {L.}~\bibnamefont {Hilico}},\
  }\bibfield  {title} {\bibinfo {title} {Coulombic and radiative decay rates of
  the resonances of the exotic molecular ions $pp\mu$, $pp\pi$, $dd\mu$,
  $dd\pi$, and $dt\mu$},\ }\href {https://doi.org/10.1103/physreva.70.042506}
  {\bibfield  {journal} {\bibinfo  {journal} {Phys. Rev. A}\ }\textbf {\bibinfo
  {volume} {70}},\ \bibinfo {pages} {042506} (\bibinfo {year}
  {2004})}\BibitemShut {NoStop}%
\bibitem [{\citenamefont {K{\i}l{\i}\c{c}}(2005)}]{kilic2005}%
  \BibitemOpen
  \bibfield  {author} {\bibinfo {author} {\bibfnamefont {S.}~\bibnamefont
  {K{\i}l{\i}\c{c}}},\ }\emph {\bibinfo {title} {Le probl\`eme Coulombien \`a
  trois corps : \'etats li\'es, r\'esonances et interaction avec la lumi\`ere,
  vers la spectroscopie vibrationnelle dans {$\mathrm{H}_2^+$}}},\ \href
  {https://www.lkb.fr/iontrap/wp-content/uploads/sites/15/2025/01/Senem-Kilic.pdf}
  {Ph.D. thesis},\ \bibinfo  {school} {Universit\'e Pierre et Marie Curie}
  (\bibinfo {year} {2005}),\ \bibinfo {note} {defended 7 October 2005; Appendix
  B, Tables B.12 and B.13}\BibitemShut {NoStop}%
\bibitem [{\citenamefont {Wen}\ and\ \citenamefont {Zhu}(2026)}]{wenzhu2025}%
  \BibitemOpen
  \bibfield  {author} {\bibinfo {author} {\bibfnamefont {L.-Z.}\ \bibnamefont
  {Wen}}\ and\ \bibinfo {author} {\bibfnamefont {S.-L.}\ \bibnamefont {Zhu}},\
  }\bibfield  {title} {\bibinfo {title} {Bound and resonant states of muonic
  few-body {C}oulomb systems: extended stochastic variational approach},\
  }\href {https://doi.org/10.1103/zld1-bzl5} {\bibfield  {journal} {\bibinfo
  {journal} {Phys. Rev. A}\ }\textbf {\bibinfo {volume} {113}},\ \bibinfo
  {pages} {032815} (\bibinfo {year} {2026})}\BibitemShut {NoStop}%
\bibitem [{\citenamefont {Wen}\ \emph {et~al.}(2026)\citenamefont {Wen},
  \citenamefont {Ma}, \citenamefont {Shen},\ and\ \citenamefont
  {Zhu}}]{wen2026lepton}%
  \BibitemOpen
  \bibfield  {author} {\bibinfo {author} {\bibfnamefont {L.-Z.}\ \bibnamefont
  {Wen}}, \bibinfo {author} {\bibfnamefont {Y.}~\bibnamefont {Ma}}, \bibinfo
  {author} {\bibfnamefont {Z.-H.}\ \bibnamefont {Shen}},\ and\ \bibinfo
  {author} {\bibfnamefont {S.-L.}\ \bibnamefont {Zhu}},\ }\href@noop {}
  {\bibinfo {title} {Bound and resonant spectra of few-lepton {C}oulomb
  systems}} (\bibinfo {year} {2026}),\ \Eprint
  {https://arxiv.org/abs/2608.21901} {arXiv:2608.21901 [physics.atom-ph]}
  \BibitemShut {NoStop}%
\bibitem [{\citenamefont {Gailitis}\ and\ \citenamefont
  {Damburg}(1963)}]{gailitis1963}%
  \BibitemOpen
  \bibfield  {author} {\bibinfo {author} {\bibfnamefont {M.}~\bibnamefont
  {Gailitis}}\ and\ \bibinfo {author} {\bibfnamefont {R.}~\bibnamefont
  {Damburg}},\ }\bibfield  {title} {\bibinfo {title} {The influence of close
  coupling on the threshold behaviour of cross sections of electron-hydrogen
  scattering},\ }\href {https://doi.org/10.1088/0370-1328/82/2/305} {\bibfield
  {journal} {\bibinfo  {journal} {Proc. Phys. Soc.}\ }\textbf {\bibinfo
  {volume} {82}},\ \bibinfo {pages} {192} (\bibinfo {year} {1963})}\BibitemShut
  {NoStop}%
\bibitem [{\citenamefont {Gailitis}(1982)}]{gailitis1982}%
  \BibitemOpen
  \bibfield  {author} {\bibinfo {author} {\bibfnamefont {M.}~\bibnamefont
  {Gailitis}},\ }\bibfield  {title} {\bibinfo {title} {The influence of finite
  masses on the threshold behaviour of scattering in a three charged particle
  system},\ }\href {https://doi.org/10.1088/0022-3700/15/19/012} {\bibfield
  {journal} {\bibinfo  {journal} {J. Phys. B}\ }\textbf {\bibinfo {volume}
  {15}},\ \bibinfo {pages} {3423} (\bibinfo {year} {1982})}\BibitemShut
  {NoStop}%
\bibitem [{\citenamefont {Braaten}\ and\ \citenamefont
  {Hammer}(2006)}]{braaten2006}%
  \BibitemOpen
  \bibfield  {author} {\bibinfo {author} {\bibfnamefont {E.}~\bibnamefont
  {Braaten}}\ and\ \bibinfo {author} {\bibfnamefont {H.-W.}\ \bibnamefont
  {Hammer}},\ }\bibfield  {title} {\bibinfo {title} {Universality in few-body
  systems with large scattering length},\ }\href
  {https://doi.org/10.1016/j.physrep.2006.03.001} {\bibfield  {journal}
  {\bibinfo  {journal} {Phys. Rep.}\ }\textbf {\bibinfo {volume} {428}},\
  \bibinfo {pages} {259} (\bibinfo {year} {2006})}\BibitemShut {NoStop}%
\bibitem [{\citenamefont {Nishida}(2022)}]{nishida2022}%
  \BibitemOpen
  \bibfield  {author} {\bibinfo {author} {\bibfnamefont {Y.}~\bibnamefont
  {Nishida}},\ }\bibfield  {title} {\bibinfo {title} {Efimovian states of three
  charged particles},\ }\href {https://doi.org/10.1103/PhysRevA.105.L010802}
  {\bibfield  {journal} {\bibinfo  {journal} {Phys. Rev. A}\ }\textbf {\bibinfo
  {volume} {105}},\ \bibinfo {pages} {L010802} (\bibinfo {year}
  {2022})}\BibitemShut {NoStop}%
\bibitem [{\citenamefont {Pohl}\ \emph {et~al.}(2016)\citenamefont {Pohl},
  \citenamefont {Nez}, \citenamefont {Fernandes} \emph {et~al.}}]{pohl2016}%
  \BibitemOpen
  \bibfield  {author} {\bibinfo {author} {\bibfnamefont {R.}~\bibnamefont
  {Pohl}}, \bibinfo {author} {\bibfnamefont {F.}~\bibnamefont {Nez}}, \bibinfo
  {author} {\bibfnamefont {L.~M.~P.}\ \bibnamefont {Fernandes}}, \emph
  {et~al.},\ }\bibfield  {title} {\bibinfo {title} {Laser spectroscopy of
  muonic deuterium},\ }\href {https://doi.org/10.1126/science.aaf2468}
  {\bibfield  {journal} {\bibinfo  {journal} {Science}\ }\textbf {\bibinfo
  {volume} {353}},\ \bibinfo {pages} {669} (\bibinfo {year}
  {2016})}\BibitemShut {NoStop}%
\bibitem [{\citenamefont {Karshenboim}\ \emph {et~al.}(2017)\citenamefont
  {Karshenboim}, \citenamefont {Korzinin}, \citenamefont {Shelyuto},\ and\
  \citenamefont {Ivanov}}]{muT2017}%
  \BibitemOpen
  \bibfield  {author} {\bibinfo {author} {\bibfnamefont {S.~G.}\ \bibnamefont
  {Karshenboim}}, \bibinfo {author} {\bibfnamefont {E.~Y.}\ \bibnamefont
  {Korzinin}}, \bibinfo {author} {\bibfnamefont {V.~A.}\ \bibnamefont
  {Shelyuto}},\ and\ \bibinfo {author} {\bibfnamefont {V.~G.}\ \bibnamefont
  {Ivanov}},\ }\bibfield  {title} {\bibinfo {title} {Theory of the {L}amb shift
  in muonic tritium and the muonic $^{3}$he ion},\ }\href
  {https://doi.org/10.1103/physreva.96.022505} {\bibfield  {journal} {\bibinfo
  {journal} {Phys. Rev. A}\ }\textbf {\bibinfo {volume} {96}},\ \bibinfo
  {pages} {022505} (\bibinfo {year} {2017})}\BibitemShut {NoStop}%
\bibitem [{\citenamefont {Karr}\ and\ \citenamefont {Hilico}(2012)}]{karr2012}%
  \BibitemOpen
  \bibfield  {author} {\bibinfo {author} {\bibfnamefont {J.-P.}\ \bibnamefont
  {Karr}}\ and\ \bibinfo {author} {\bibfnamefont {L.}~\bibnamefont {Hilico}},\
  }\bibfield  {title} {\bibinfo {title} {Why three-body physics does not solve
  the proton-radius puzzle},\ }\href
  {https://doi.org/10.1103/PhysRevLett.109.103401} {\bibfield  {journal}
  {\bibinfo  {journal} {Phys. Rev. Lett.}\ }\textbf {\bibinfo {volume} {109}},\
  \bibinfo {pages} {103401} (\bibinfo {year} {2012})}\BibitemShut {NoStop}%
\bibitem [{\citenamefont {Karr}\ and\ \citenamefont {Hilico}(2013)}]{karr2013}%
  \BibitemOpen
  \bibfield  {author} {\bibinfo {author} {\bibfnamefont {J.-P.}\ \bibnamefont
  {Karr}}\ and\ \bibinfo {author} {\bibfnamefont {L.}~\bibnamefont {Hilico}},\
  }\bibfield  {title} {\bibinfo {title} {Analytical matrix elements of the
  {Uehling} potential in three-body systems and applications to exotic
  molecules},\ }\href {https://doi.org/10.1103/physreva.87.012506} {\bibfield
  {journal} {\bibinfo  {journal} {Phys. Rev. A}\ }\textbf {\bibinfo {volume}
  {87}},\ \bibinfo {pages} {012506} (\bibinfo {year} {2013})}\BibitemShut
  {NoStop}%
\bibitem [{\citenamefont {Wallenius}\ and\ \citenamefont
  {Kamimura}(1996)}]{walleniusKamimura1996}%
  \BibitemOpen
  \bibfield  {author} {\bibinfo {author} {\bibfnamefont {J.}~\bibnamefont
  {Wallenius}}\ and\ \bibinfo {author} {\bibfnamefont {M.}~\bibnamefont
  {Kamimura}},\ }\bibfield  {title} {\bibinfo {title} {Resonance energies of
  meta-stable $dt\mu^*$},\ }\href {https://doi.org/10.1007/BF02227638}
  {\bibfield  {journal} {\bibinfo  {journal} {Hyperfine Interact.}\ }\textbf
  {\bibinfo {volume} {101}},\ \bibinfo {pages} {319} (\bibinfo {year}
  {1996})}\BibitemShut {NoStop}%
\bibitem [{\citenamefont {Riss}\ and\ \citenamefont {Meyer}(1993)}]{riss1993}%
  \BibitemOpen
  \bibfield  {author} {\bibinfo {author} {\bibfnamefont {U.~V.}\ \bibnamefont
  {Riss}}\ and\ \bibinfo {author} {\bibfnamefont {H.-D.}\ \bibnamefont
  {Meyer}},\ }\bibfield  {title} {\bibinfo {title} {Calculation of resonance
  energies and widths using the complex absorbing potential method},\ }\href
  {https://doi.org/10.1088/0953-4075/26/23/021} {\bibfield  {journal} {\bibinfo
   {journal} {J. Phys. B}\ }\textbf {\bibinfo {volume} {26}},\ \bibinfo {pages}
  {4503} (\bibinfo {year} {1993})}\BibitemShut {NoStop}%
\bibitem [{\citenamefont {Mitroy}\ \emph {et~al.}(2013)\citenamefont {Mitroy},
  \citenamefont {Bubin}, \citenamefont {Horiuchi}, \citenamefont {Suzuki},
  \citenamefont {Adamowicz}, \citenamefont {Cencek}, \citenamefont {Szalewicz},
  \citenamefont {Komasa}, \citenamefont {Blume},\ and\ \citenamefont
  {Varga}}]{mitroy2013}%
  \BibitemOpen
  \bibfield  {author} {\bibinfo {author} {\bibfnamefont {J.}~\bibnamefont
  {Mitroy}}, \bibinfo {author} {\bibfnamefont {S.}~\bibnamefont {Bubin}},
  \bibinfo {author} {\bibfnamefont {W.}~\bibnamefont {Horiuchi}}, \bibinfo
  {author} {\bibfnamefont {Y.}~\bibnamefont {Suzuki}}, \bibinfo {author}
  {\bibfnamefont {L.}~\bibnamefont {Adamowicz}}, \bibinfo {author}
  {\bibfnamefont {W.}~\bibnamefont {Cencek}}, \bibinfo {author} {\bibfnamefont
  {K.}~\bibnamefont {Szalewicz}}, \bibinfo {author} {\bibfnamefont
  {J.}~\bibnamefont {Komasa}}, \bibinfo {author} {\bibfnamefont
  {D.}~\bibnamefont {Blume}},\ and\ \bibinfo {author} {\bibfnamefont
  {K.}~\bibnamefont {Varga}},\ }\bibfield  {title} {\bibinfo {title} {Theory
  and application of explicitly correlated {G}aussians},\ }\href
  {https://doi.org/10.1103/revmodphys.85.693} {\bibfield  {journal} {\bibinfo
  {journal} {Rev. Mod. Phys.}\ }\textbf {\bibinfo {volume} {85}},\ \bibinfo
  {pages} {693} (\bibinfo {year} {2013})}\BibitemShut {NoStop}%
\bibitem [{\citenamefont {Hiyama}\ \emph {et~al.}(2003)\citenamefont {Hiyama},
  \citenamefont {Kino},\ and\ \citenamefont {Kamimura}}]{hiyama2003}%
  \BibitemOpen
  \bibfield  {author} {\bibinfo {author} {\bibfnamefont {E.}~\bibnamefont
  {Hiyama}}, \bibinfo {author} {\bibfnamefont {Y.}~\bibnamefont {Kino}},\ and\
  \bibinfo {author} {\bibfnamefont {M.}~\bibnamefont {Kamimura}},\ }\bibfield
  {title} {\bibinfo {title} {{G}aussian expansion method for few-body
  systems},\ }\href {https://doi.org/10.1016/s0146-6410(03)90015-9} {\bibfield
  {journal} {\bibinfo  {journal} {Prog. Part. Nucl. Phys.}\ }\textbf {\bibinfo
  {volume} {51}},\ \bibinfo {pages} {223} (\bibinfo {year} {2003})}\BibitemShut
  {NoStop}%
\bibitem [{\citenamefont {Fano}(1970)}]{fano1970}%
  \BibitemOpen
  \bibfield  {author} {\bibinfo {author} {\bibfnamefont {U.}~\bibnamefont
  {Fano}},\ }\bibfield  {title} {\bibinfo {title} {Quantum defect theory of $l$
  uncoupling in $\mathrm{H}_2$ as an example of channel-interaction
  treatment},\ }\href {https://doi.org/10.1103/physreva.2.353} {\bibfield
  {journal} {\bibinfo  {journal} {Phys. Rev. A}\ }\textbf {\bibinfo {volume}
  {2}},\ \bibinfo {pages} {353} (\bibinfo {year} {1970})}\BibitemShut {NoStop}%
\bibitem [{\citenamefont {Greene}\ \emph {et~al.}(1979)\citenamefont {Greene},
  \citenamefont {Fano},\ and\ \citenamefont {Strinati}}]{greene1979}%
  \BibitemOpen
  \bibfield  {author} {\bibinfo {author} {\bibfnamefont {C.~H.}\ \bibnamefont
  {Greene}}, \bibinfo {author} {\bibfnamefont {U.}~\bibnamefont {Fano}},\ and\
  \bibinfo {author} {\bibfnamefont {G.}~\bibnamefont {Strinati}},\ }\bibfield
  {title} {\bibinfo {title} {General form of the quantum-defect theory},\
  }\href {https://doi.org/10.1103/physreva.19.1485} {\bibfield  {journal}
  {\bibinfo  {journal} {Phys. Rev. A}\ }\textbf {\bibinfo {volume} {19}},\
  \bibinfo {pages} {1485} (\bibinfo {year} {1979})}\BibitemShut {NoStop}%
\bibitem [{\citenamefont {Seaton}(1983)}]{seaton1983}%
  \BibitemOpen
  \bibfield  {author} {\bibinfo {author} {\bibfnamefont {M.~J.}\ \bibnamefont
  {Seaton}},\ }\bibfield  {title} {\bibinfo {title} {Quantum defect theory},\
  }\href {https://doi.org/10.1088/0034-4885/46/2/002} {\bibfield  {journal}
  {\bibinfo  {journal} {Rep. Prog. Phys.}\ }\textbf {\bibinfo {volume} {46}},\
  \bibinfo {pages} {167} (\bibinfo {year} {1983})}\BibitemShut {NoStop}%
\bibitem [{sup()}]{supplemental}%
  \BibitemOpen
  \href@noop {} {}\bibinfo {note} {See Supplemental Material at [URL will be
  inserted by publisher] for atomic calibration, basis and absorber controls,
  trajectory selection, precision diagnostics, and reduced-model
  comparisons}\BibitemShut {NoStop}%
\bibitem [{\citenamefont {Korobov}(2000)}]{korobov2000}%
  \BibitemOpen
  \bibfield  {author} {\bibinfo {author} {\bibfnamefont {V.~I.}\ \bibnamefont
  {Korobov}},\ }\bibfield  {title} {\bibinfo {title} {Coulomb three-body
  bound-state problem: Variational calculations of nonrelativistic energies},\
  }\href {https://doi.org/10.1103/physreva.61.064503} {\bibfield  {journal}
  {\bibinfo  {journal} {Phys. Rev. A}\ }\textbf {\bibinfo {volume} {61}},\
  \bibinfo {pages} {064503} (\bibinfo {year} {2000})}\BibitemShut {NoStop}%
\bibitem [{\citenamefont {Newton}(1961)}]{newton1961}%
  \BibitemOpen
  \bibfield  {author} {\bibinfo {author} {\bibfnamefont {R.~G.}\ \bibnamefont
  {Newton}},\ }\bibfield  {title} {\bibinfo {title} {Structure of the
  many-channel {S} matrix},\ }\href {https://doi.org/10.1063/1.1703698}
  {\bibfield  {journal} {\bibinfo  {journal} {J. Math. Phys.}\ }\textbf
  {\bibinfo {volume} {2}},\ \bibinfo {pages} {188} (\bibinfo {year}
  {1961})}\BibitemShut {NoStop}%
\bibitem [{\citenamefont {Tiesinga}\ \emph {et~al.}(2021)\citenamefont
  {Tiesinga}, \citenamefont {Mohr}, \citenamefont {Newell},\ and\ \citenamefont
  {Taylor}}]{codata2018}%
  \BibitemOpen
  \bibfield  {author} {\bibinfo {author} {\bibfnamefont {E.}~\bibnamefont
  {Tiesinga}}, \bibinfo {author} {\bibfnamefont {P.~J.}\ \bibnamefont {Mohr}},
  \bibinfo {author} {\bibfnamefont {D.~B.}\ \bibnamefont {Newell}},\ and\
  \bibinfo {author} {\bibfnamefont {B.~N.}\ \bibnamefont {Taylor}},\ }\bibfield
   {title} {\bibinfo {title} {{CODATA} recommended values of the fundamental
  physical constants: 2018},\ }\href
  {https://doi.org/10.1103/revmodphys.93.025010} {\bibfield  {journal}
  {\bibinfo  {journal} {Rev. Mod. Phys.}\ }\textbf {\bibinfo {volume} {93}},\
  \bibinfo {pages} {025010} (\bibinfo {year} {2021})}\BibitemShut {NoStop}%
\bibitem [{\citenamefont {Krauth}\ \emph {et~al.}(2016)\citenamefont {Krauth},
  \citenamefont {Diepold}, \citenamefont {Franke}, \citenamefont {Antognini},
  \citenamefont {Kottmann},\ and\ \citenamefont {Pohl}}]{krauth2016}%
  \BibitemOpen
  \bibfield  {author} {\bibinfo {author} {\bibfnamefont {J.~J.}\ \bibnamefont
  {Krauth}}, \bibinfo {author} {\bibfnamefont {M.}~\bibnamefont {Diepold}},
  \bibinfo {author} {\bibfnamefont {B.}~\bibnamefont {Franke}}, \bibinfo
  {author} {\bibfnamefont {A.}~\bibnamefont {Antognini}}, \bibinfo {author}
  {\bibfnamefont {F.}~\bibnamefont {Kottmann}},\ and\ \bibinfo {author}
  {\bibfnamefont {R.}~\bibnamefont {Pohl}},\ }\bibfield  {title} {\bibinfo
  {title} {Theory of the $n=2$ levels in muonic deuterium},\ }\href
  {https://doi.org/10.1016/j.aop.2015.12.006} {\bibfield  {journal} {\bibinfo
  {journal} {Ann. Phys. (N.Y.)}\ }\textbf {\bibinfo {volume} {366}},\ \bibinfo
  {pages} {168} (\bibinfo {year} {2016})}\BibitemShut {NoStop}%
\bibitem [{\citenamefont {Suzuki}\ and\ \citenamefont
  {Varga}(1998)}]{suzukivarga1998}%
  \BibitemOpen
  \bibfield  {author} {\bibinfo {author} {\bibfnamefont {Y.}~\bibnamefont
  {Suzuki}}\ and\ \bibinfo {author} {\bibfnamefont {K.}~\bibnamefont {Varga}},\
  }\href {https://doi.org/10.1007/3-540-49541-x} {\emph {\bibinfo {title}
  {Stochastic Variational Approach to Quantum-Mechanical Few-Body Problems}}},\
  \bibinfo {series} {Lecture Notes in Physics Monographs}, Vol.~\bibinfo
  {volume} {54}\ (\bibinfo  {publisher} {Springer},\ \bibinfo {address}
  {Berlin},\ \bibinfo {year} {1998})\BibitemShut {NoStop}%
\bibitem [{\citenamefont {Suzuki}\ \emph {et~al.}(1998)\citenamefont {Suzuki},
  \citenamefont {Usukura},\ and\ \citenamefont {Varga}}]{suzuki1998gvr}%
  \BibitemOpen
  \bibfield  {author} {\bibinfo {author} {\bibfnamefont {Y.}~\bibnamefont
  {Suzuki}}, \bibinfo {author} {\bibfnamefont {J.}~\bibnamefont {Usukura}},\
  and\ \bibinfo {author} {\bibfnamefont {K.}~\bibnamefont {Varga}},\ }\bibfield
   {title} {\bibinfo {title} {New description of orbital motion with arbitrary
  angular momenta},\ }\href {https://doi.org/10.1088/0953-4075/31/1/007}
  {\bibfield  {journal} {\bibinfo  {journal} {J. Phys. B}\ }\textbf {\bibinfo
  {volume} {31}},\ \bibinfo {pages} {31} (\bibinfo {year} {1998})}\BibitemShut
  {NoStop}%
\bibitem [{\citenamefont {Varga}\ and\ \citenamefont
  {Suzuki}(1995)}]{varga1995}%
  \BibitemOpen
  \bibfield  {author} {\bibinfo {author} {\bibfnamefont {K.}~\bibnamefont
  {Varga}}\ and\ \bibinfo {author} {\bibfnamefont {Y.}~\bibnamefont {Suzuki}},\
  }\bibfield  {title} {\bibinfo {title} {Precise solution of few-body problems
  with the stochastic variational method on a correlated {Gaussian} basis},\
  }\href {https://doi.org/10.1103/PhysRevC.52.2885} {\bibfield  {journal}
  {\bibinfo  {journal} {Phys. Rev. C}\ }\textbf {\bibinfo {volume} {52}},\
  \bibinfo {pages} {2885} (\bibinfo {year} {1995})}\BibitemShut {NoStop}%
\bibitem [{\citenamefont {Ohtsubo}\ \emph {et~al.}(2013)\citenamefont
  {Ohtsubo}, \citenamefont {Fukushima}, \citenamefont {Kamimura},\ and\
  \citenamefont {Hiyama}}]{ohtsubo2013}%
  \BibitemOpen
  \bibfield  {author} {\bibinfo {author} {\bibfnamefont {S.-I.}\ \bibnamefont
  {Ohtsubo}}, \bibinfo {author} {\bibfnamefont {Y.}~\bibnamefont {Fukushima}},
  \bibinfo {author} {\bibfnamefont {M.}~\bibnamefont {Kamimura}},\ and\
  \bibinfo {author} {\bibfnamefont {E.}~\bibnamefont {Hiyama}},\ }\bibfield
  {title} {\bibinfo {title} {Complex-scaling calculation of three-body
  resonances using complex-range {G}aussian basis functions: Application to
  $3\alpha$ resonances in $^{12}${C}},\ }\href
  {https://doi.org/10.1093/ptep/ptt048} {\bibfield  {journal} {\bibinfo
  {journal} {Prog. Theor. Exp. Phys.}\ }\textbf {\bibinfo {volume} {2013}},\
  \bibinfo {pages} {073D02} (\bibinfo {year} {2013})}\BibitemShut {NoStop}%
\bibitem [{\citenamefont {Takenaka}\ \emph {et~al.}(2014)\citenamefont
  {Takenaka}, \citenamefont {Otani}, \citenamefont {Iwasaki}, \citenamefont
  {Mimura},\ and\ \citenamefont {Ito}}]{takenaka2014}%
  \BibitemOpen
  \bibfield  {author} {\bibinfo {author} {\bibfnamefont {Y.}~\bibnamefont
  {Takenaka}}, \bibinfo {author} {\bibfnamefont {R.}~\bibnamefont {Otani}},
  \bibinfo {author} {\bibfnamefont {M.}~\bibnamefont {Iwasaki}}, \bibinfo
  {author} {\bibfnamefont {K.}~\bibnamefont {Mimura}},\ and\ \bibinfo {author}
  {\bibfnamefont {M.}~\bibnamefont {Ito}},\ }\bibfield  {title} {\bibinfo
  {title} {{Gaussian} expansion methods under the absorbing boundary
  condition},\ }\href {https://doi.org/10.1093/ptep/ptu135} {\bibfield
  {journal} {\bibinfo  {journal} {Prog. Theor. Exp. Phys.}\ }\textbf {\bibinfo
  {volume} {2014}},\ \bibinfo {pages} {113D04} (\bibinfo {year}
  {2014})}\BibitemShut {NoStop}%
\bibitem [{\citenamefont {Jolicard}\ and\ \citenamefont
  {Austin}(1985)}]{jolicard1985}%
  \BibitemOpen
  \bibfield  {author} {\bibinfo {author} {\bibfnamefont {G.}~\bibnamefont
  {Jolicard}}\ and\ \bibinfo {author} {\bibfnamefont {E.~J.}\ \bibnamefont
  {Austin}},\ }\bibfield  {title} {\bibinfo {title} {Optical potential
  stabilisation method for predicting resonance levels},\ }\href
  {https://doi.org/10.1016/0009-2614(85)87164-5} {\bibfield  {journal}
  {\bibinfo  {journal} {Chem. Phys. Lett.}\ }\textbf {\bibinfo {volume}
  {121}},\ \bibinfo {pages} {106} (\bibinfo {year} {1985})}\BibitemShut
  {NoStop}%
\bibitem [{\citenamefont {Riss}\ and\ \citenamefont {Meyer}(1998)}]{riss1998}%
  \BibitemOpen
  \bibfield  {author} {\bibinfo {author} {\bibfnamefont {U.~V.}\ \bibnamefont
  {Riss}}\ and\ \bibinfo {author} {\bibfnamefont {H.-D.}\ \bibnamefont
  {Meyer}},\ }\bibfield  {title} {\bibinfo {title} {The transformative complex
  absorbing potential method: a bridge between complex absorbing potentials and
  smooth exterior scaling},\ }\href
  {https://doi.org/10.1088/0953-4075/31/10/016} {\bibfield  {journal} {\bibinfo
   {journal} {J. Phys. B}\ }\textbf {\bibinfo {volume} {31}},\ \bibinfo {pages}
  {2279} (\bibinfo {year} {1998})}\BibitemShut {NoStop}%
\bibitem [{\citenamefont {Eden}\ and\ \citenamefont {Taylor}(1964)}]{eden1964}%
  \BibitemOpen
  \bibfield  {author} {\bibinfo {author} {\bibfnamefont {R.~J.}\ \bibnamefont
  {Eden}}\ and\ \bibinfo {author} {\bibfnamefont {J.~R.}\ \bibnamefont
  {Taylor}},\ }\bibfield  {title} {\bibinfo {title} {Poles and shadow poles in
  the many-channel {S} matrix},\ }\href
  {https://doi.org/10.1103/PhysRev.133.B1575} {\bibfield  {journal} {\bibinfo
  {journal} {Phys. Rev.}\ }\textbf {\bibinfo {volume} {133}},\ \bibinfo {pages}
  {B1575} (\bibinfo {year} {1964})}\BibitemShut {NoStop}%
\bibitem [{\citenamefont {Zhang}\ \emph {et~al.}(2020)\citenamefont {Zhang},
  \citenamefont {Wu}, \citenamefont {Qian}, \citenamefont {Varga},
  \citenamefont {Han},\ and\ \citenamefont {Zhang}}]{zhang2020ps2}%
  \BibitemOpen
  \bibfield  {author} {\bibinfo {author} {\bibfnamefont {Y.}~\bibnamefont
  {Zhang}}, \bibinfo {author} {\bibfnamefont {M.-S.}\ \bibnamefont {Wu}},
  \bibinfo {author} {\bibfnamefont {Y.}~\bibnamefont {Qian}}, \bibinfo {author}
  {\bibfnamefont {K.}~\bibnamefont {Varga}}, \bibinfo {author} {\bibfnamefont
  {H.-L.}\ \bibnamefont {Han}},\ and\ \bibinfo {author} {\bibfnamefont {J.-Y.}\
  \bibnamefont {Zhang}},\ }\bibfield  {title} {\bibinfo {title} {Doubly excited
  states of the positronium molecule},\ }\href
  {https://doi.org/10.1103/PhysRevA.102.012825} {\bibfield  {journal} {\bibinfo
   {journal} {Phys. Rev. A}\ }\textbf {\bibinfo {volume} {102}},\ \bibinfo
  {pages} {012825} (\bibinfo {year} {2020})}\BibitemShut {NoStop}%
\bibitem [{\citenamefont {Zhang}\ \emph {et~al.}(2023)\citenamefont {Zhang},
  \citenamefont {Wu}, \citenamefont {Yan}, \citenamefont {Varga}, \citenamefont
  {Yan},\ and\ \citenamefont {Zhang}}]{zhang2023psh}%
  \BibitemOpen
  \bibfield  {author} {\bibinfo {author} {\bibfnamefont {Y.}~\bibnamefont
  {Zhang}}, \bibinfo {author} {\bibfnamefont {M.-S.}\ \bibnamefont {Wu}},
  \bibinfo {author} {\bibfnamefont {G.-A.}\ \bibnamefont {Yan}}, \bibinfo
  {author} {\bibfnamefont {K.}~\bibnamefont {Varga}}, \bibinfo {author}
  {\bibfnamefont {Z.-C.}\ \bibnamefont {Yan}},\ and\ \bibinfo {author}
  {\bibfnamefont {J.-Y.}\ \bibnamefont {Zhang}},\ }\bibfield  {title} {\bibinfo
  {title} {Doubly excited autodissociating resonant states of positronium
  hydride},\ }\href {https://doi.org/10.1103/PhysRevA.108.052813} {\bibfield
  {journal} {\bibinfo  {journal} {Phys. Rev. A}\ }\textbf {\bibinfo {volume}
  {108}},\ \bibinfo {pages} {052813} (\bibinfo {year} {2023})}\BibitemShut
  {NoStop}%
\bibitem [{Note1()}]{Note1}%
  \BibitemOpen
  \bibinfo {note} {Claude Code (Anthropic; Claude Fable~5.1 and Claude
  Opus~5.5) and OpenAI Codex (GPT-6 Astra), September 2026.}\BibitemShut
  {Stop}%
\bibitem [{\citenamefont {Zhang}(2026)}]{zenodo2026tdmu}%
  \BibitemOpen
  \bibfield  {author} {\bibinfo {author} {\bibfnamefont {Y.}~\bibnamefont
  {Zhang}},\ }\href {https://doi.org/10.5281/zenodo.23018724} {\bibinfo {title}
  {Code and data for ``{R}esponse of the near-threshold resonances of the
  $dt\mu$ molecular ion to an effective atomic $2s$--$2p$ splitting''}},\
  \bibinfo {howpublished} {Zenodo} (\bibinfo {year} {2026}),\ \bibinfo {note}
  {doi:10.5281/zenodo.23018724}\BibitemShut {NoStop}%
\end{thebibliography}%


\begin{thebibliography}{8}%
\makeatletter
\providecommand \@ifxundefined [1]{%
 \@ifx{#1\undefined}
}%
\providecommand \@ifnum [1]{%
 \ifnum #1\expandafter \@firstoftwo
 \else \expandafter \@secondoftwo
 \fi
}%
\providecommand \@ifx [1]{%
 \ifx #1\expandafter \@firstoftwo
 \else \expandafter \@secondoftwo
 \fi
}%
\providecommand \natexlab [1]{#1}%
\providecommand \enquote  [1]{``#1''}%
\providecommand \bibnamefont  [1]{#1}%
\providecommand \bibfnamefont [1]{#1}%
\providecommand \citenamefont [1]{#1}%
\providecommand \href@noop [0]{\@secondoftwo}%
\providecommand \href [0]{\begingroup \@sanitize@url \@href}%
\providecommand \@href[1]{\@@startlink{#1}\@@href}%
\providecommand \@@href[1]{\endgroup#1\@@endlink}%
\providecommand \@sanitize@url [0]{\catcode `\\12\catcode `\$12\catcode
  `\&12\catcode `\#12\catcode `\^12\catcode `\_12\catcode `\%12\relax}%
\providecommand \@@startlink[1]{}%
\providecommand \@@endlink[0]{}%
\providecommand \url  [0]{\begingroup\@sanitize@url \@url }%
\providecommand \@url [1]{\endgroup\@href {#1}{\urlprefix }}%
\providecommand \urlprefix  [0]{URL }%
\providecommand \Eprint [0]{\href }%
\providecommand \doibase [0]{https://doi.org/}%
\providecommand \selectlanguage [0]{\@gobble}%
\providecommand \bibinfo  [0]{\@secondoftwo}%
\providecommand \bibfield  [0]{\@secondoftwo}%
\providecommand \translation [1]{[#1]}%
\providecommand \BibitemOpen [0]{}%
\providecommand \bibitemStop [0]{}%
\providecommand \bibitemNoStop [0]{.\EOS\space}%
\providecommand \EOS [0]{\spacefactor3000\relax}%
\providecommand \BibitemShut  [1]{\csname bibitem#1\endcsname}%
\let\auto@bib@innerbib\@empty
%</preamble>
\bibitem [{\citenamefont {Wen}\ and\ \citenamefont {Zhu}(2026)}]{wenzhu2025}%
  \BibitemOpen
  \bibfield  {author} {\bibinfo {author} {\bibfnamefont {L.-Z.}\ \bibnamefont
  {Wen}}\ and\ \bibinfo {author} {\bibfnamefont {S.-L.}\ \bibnamefont {Zhu}},\
  }\bibfield  {title} {\bibinfo {title} {Bound and resonant states of muonic
  few-body {C}oulomb systems: extended stochastic variational approach},\
  }\href {https://doi.org/10.1103/zld1-bzl5} {\bibfield  {journal} {\bibinfo
  {journal} {Phys. Rev. A}\ }\textbf {\bibinfo {volume} {113}},\ \bibinfo
  {pages} {032815} (\bibinfo {year} {2026})}\BibitemShut {NoStop}%
\bibitem [{\citenamefont {K{\i}l{\i}\c{c}}(2005)}]{kilic2005}%
  \BibitemOpen
  \bibfield  {author} {\bibinfo {author} {\bibfnamefont {S.}~\bibnamefont
  {K{\i}l{\i}\c{c}}},\ }\emph {\bibinfo {title} {Le probl\`eme Coulombien \`a
  trois corps : \'etats li\'es, r\'esonances et interaction avec la lumi\`ere,
  vers la spectroscopie vibrationnelle dans {$\mathrm{H}_2^+$}}},\ \href
  {https://www.lkb.fr/iontrap/wp-content/uploads/sites/15/2025/01/Senem-Kilic.pdf}
  {Ph.D. thesis},\ \bibinfo  {school} {Universit\'e Pierre et Marie Curie}
  (\bibinfo {year} {2005}),\ \bibinfo {note} {defended 7 October 2005; Appendix
  B, Tables B.12 and B.13}\BibitemShut {NoStop}%
\bibitem [{\citenamefont {K{\i}l{\i}\c{c}}\ \emph {et~al.}(2004)\citenamefont
  {K{\i}l{\i}\c{c}}, \citenamefont {Karr},\ and\ \citenamefont
  {Hilico}}]{kilic2004}%
  \BibitemOpen
  \bibfield  {author} {\bibinfo {author} {\bibfnamefont {S.}~\bibnamefont
  {K{\i}l{\i}\c{c}}}, \bibinfo {author} {\bibfnamefont {J.-P.}\ \bibnamefont
  {Karr}},\ and\ \bibinfo {author} {\bibfnamefont {L.}~\bibnamefont {Hilico}},\
  }\bibfield  {title} {\bibinfo {title} {Coulombic and radiative decay rates of
  the resonances of the exotic molecular ions $pp\mu$, $pp\pi$, $dd\mu$,
  $dd\pi$, and $dt\mu$},\ }\href {https://doi.org/10.1103/physreva.70.042506}
  {\bibfield  {journal} {\bibinfo  {journal} {Phys. Rev. A}\ }\textbf {\bibinfo
  {volume} {70}},\ \bibinfo {pages} {042506} (\bibinfo {year}
  {2004})}\BibitemShut {NoStop}%
\bibitem [{\citenamefont {Korobov}(1996)}]{korobov1996window}%
  \BibitemOpen
  \bibfield  {author} {\bibinfo {author} {\bibfnamefont {V.~I.}\ \bibnamefont
  {Korobov}},\ }\bibfield  {title} {\bibinfo {title} {Resonances in the $dt\mu$
  molecular ion between $t\mu(n=2)$ and $d\mu(n=2)$ thresholds},\ }\href
  {https://doi.org/10.1007/BF02227641} {\bibfield  {journal} {\bibinfo
  {journal} {Hyperfine Interact.}\ }\textbf {\bibinfo {volume} {101}},\
  \bibinfo {pages} {333} (\bibinfo {year} {1996})}\BibitemShut {NoStop}%
\bibitem [{\citenamefont {Newton}(1961)}]{newton1961}%
  \BibitemOpen
  \bibfield  {author} {\bibinfo {author} {\bibfnamefont {R.~G.}\ \bibnamefont
  {Newton}},\ }\bibfield  {title} {\bibinfo {title} {Structure of the
  many-channel {S} matrix},\ }\href {https://doi.org/10.1063/1.1703698}
  {\bibfield  {journal} {\bibinfo  {journal} {J. Math. Phys.}\ }\textbf
  {\bibinfo {volume} {2}},\ \bibinfo {pages} {188} (\bibinfo {year}
  {1961})}\BibitemShut {NoStop}%
\bibitem [{\citenamefont {Eden}\ and\ \citenamefont {Taylor}(1964)}]{eden1964}%
  \BibitemOpen
  \bibfield  {author} {\bibinfo {author} {\bibfnamefont {R.~J.}\ \bibnamefont
  {Eden}}\ and\ \bibinfo {author} {\bibfnamefont {J.~R.}\ \bibnamefont
  {Taylor}},\ }\bibfield  {title} {\bibinfo {title} {Poles and shadow poles in
  the many-channel {S} matrix},\ }\href
  {https://doi.org/10.1103/PhysRev.133.B1575} {\bibfield  {journal} {\bibinfo
  {journal} {Phys. Rev.}\ }\textbf {\bibinfo {volume} {133}},\ \bibinfo {pages}
  {B1575} (\bibinfo {year} {1964})}\BibitemShut {NoStop}%
\bibitem [{\citenamefont {Korobov}(2000)}]{korobov2000}%
  \BibitemOpen
  \bibfield  {author} {\bibinfo {author} {\bibfnamefont {V.~I.}\ \bibnamefont
  {Korobov}},\ }\bibfield  {title} {\bibinfo {title} {Coulomb three-body
  bound-state problem: Variational calculations of nonrelativistic energies},\
  }\href {https://doi.org/10.1103/physreva.61.064503} {\bibfield  {journal}
  {\bibinfo  {journal} {Phys. Rev. A}\ }\textbf {\bibinfo {volume} {61}},\
  \bibinfo {pages} {064503} (\bibinfo {year} {2000})}\BibitemShut {NoStop}%
\bibitem [{\citenamefont {Zhang}(2026)}]{zenodo2026tdmu}%
  \BibitemOpen
  \bibfield  {author} {\bibinfo {author} {\bibfnamefont {Y.}~\bibnamefont
  {Zhang}},\ }\href {https://doi.org/10.5281/zenodo.23018724} {\bibinfo {title}
  {Code and data for ``{R}esponse of the near-threshold resonances of the
  $dt\mu$ molecular ion to an effective atomic $2s$--$2p$ splitting''}},\
  \bibinfo {howpublished} {Zenodo} (\bibinfo {year} {2026}),\ \bibinfo {note}
  {doi:10.5281/zenodo.23018724}\BibitemShut {NoStop}%
\end{thebibliography}%
\end{document}

\title{Supplemental Material: Response of the near-threshold resonances of the $dt\mu$ molecular ion to an effective atomic $2s$--$2p$ splitting}
\author{Yi Zhang}
\date{13 September 2026}
\maketitle

This Supplemental Material gives the numerical controls for the
nonrelativistic Coulomb baseline and the effective splitting Hamiltonian of the
main text. The enhanced $L=1$ results comprise 304 spectral points computed in
two sets: 160 points, and 144 points for the overlap-cut sequence and the
radial-range extension.
Parameter ranges describe the stated configurations; displayed digits support
comparison and are not a uniform physical error estimate. Energies use
$d+t+\mu$ breakup as zero, and every width is a full width.

\section{Atomic calibration and comparison of energy origins}

All molecular positions $E_r$ use $d+t+\mu$ breakup as zero. In each atomic problem,
an attractive Gaussian of range $0.05\amu$ is calibrated nonlinearly to
$E_{2p}-E_{2s}=\Delta$, separately for $d\mu$ and $t\mu$. The $2p$ level also shifts;
the calibration does not impose $E_{2s}=E_{2p}(0)-\Delta$.
\begin{table}[tbp]
\caption{\label{tab:s-atomic}Nonlinear atomic calibration for Gaussian range $0.05a_{d\mu}$. Energies are in eV from breakup; $r_\Delta=E_{2p}-E_{2s}-\Delta$. Extra digits record the numerical convention and do not imply physical accuracy.}
\begin{ruledtabular}
\begin{tabular}{ccrrr}
Atom & $\Delta$ & $E_{2s}$ & $E_{2p}$ & $r_\Delta$ \\
\hline
$d\mu$ & 0.0 & $-665.800120873$ & $-665.800120873$ & $1.14\times10^{-13}$ \\
$d\mu$ & 0.2 & $-666.000185608$ & $-665.800185608$ & $-3.14\times10^{-12}$ \\
$t\mu$ & 0.0 & $-677.810417892$ & $-677.810417892$ & $1.14\times10^{-13}$ \\
$t\mu$ & 0.2 & $-678.010485036$ & $-677.810485036$ & $-3.14\times10^{-12}$ \\
\end{tabular}
\end{ruledtabular}
\end{table}

The original frozen $2s$ contractions have an atomic energy error of
about $1.8\meV$ at $\Delta=0.2\eV$. The enhanced representation
combines an improved contraction with a radial-response contraction.
The isolated-atom errors are below $0.019\meV$ for 24 Gaussian terms
and $0.012\meV$ for 32 terms over $0\leq\Delta\leq0.35\eV$.
Both spaces remain fixed throughout the molecular scans. These atomic
tests are separate from the molecular position and width controls.

The reduced model keeps its $2p$ reference at the Coulomb value and places its lower
threshold at $E_{2p}(0)-\Delta$. Let $B_m=E_{2s,m}-E_m$ and
$B_c=E_{2s,c}-E_c$ denote the model and molecular bindings. Their difference obeys
\begin{equation}
 B_m-B_c=(E_{2s,m}-E_{2s,c})-(E_m-E_c).
\end{equation}
For $d\mu(2s)+t$ at $\Delta=0.2\eV$, the two lower thresholds differ by
$0.064735191\meV$.
A comparison of native binding energies therefore differs from a comparison of
breakup-referenced positions. Both diagnostics are retained; neither represents
a new physical calibration of the reduced model.

\section{Basis construction and fixed radial response}
\label{sec:s-basis-construction}
The principal enhanced spaces contain 6400 contracted functions for
$L=0$ and 9000 for $L=1$. They are assembled as follows:
\begin{center}
\begin{tabular}{lrr}
Component & $L=0$ & $L=1$\\\hline
Compact correlation block & 400 & 1000\\
Extended correlation block & 1000 & 1000\\
Original atomic-channel contractions & 3750 & 3750\\
Additional $K=1$ channel contractions & 0 & 2000\\
Added $2s$ radial-response contractions & 1250 & 1250\\\hline
Total before overlap truncation & 6400 & 9000
\end{tabular}
\end{center}
Replacing the original $2s$ contraction by its improved counterpart
does not change the number of functions; adding the response contraction
does. The $L=0$ range extension is described separately below.

The compact blocks are optimized for the lowest Coulomb eigenvalue
in their respective angular-momentum sectors. The $L=0$ block is
the 400-function real ECG set obtained with three stochastic
optimization sweeps. The $L=1$ block is a 1000-function stochastic
variational expansion containing 601 $K=0$ and 399 $K=1$ functions.
For its growth, trial pair lengths are sampled logarithmically in
$s_{dt}=0.3$--$8\amu$ and $s_{\mu d},s_{\mu t}=0.2$--$6\amu$.
The global-vector direction is sampled in the $(\bm x_1,\bm x_2)$,
$(\bm r_d,\bm R_d)$, or $(\bm r_t,\bm R_t)$ coordinate frame.
Candidates are ranked by the lowest eigenvalue of the enlarged
generalized eigenproblem. A candidate is rejected if its maximum
normalized overlap with an existing function exceeds the working
overlap limit. This limit starts at $0.995$ and is relaxed after
unsuccessful rounds, up to $0.99999$.

The extended blocks are sampled independently of the resonance
positions. The pair-length intervals are $s_{dt}=2$--$60\amu$ and
$s_{\mu d},s_{\mu t}=1$--$30\amu$, with pair exponents
$z_{ab}=1/(2s_{ab}^2)$. For $L=1$ these functions have $K=0$ and
global vectors chosen from $\bm x_1$, $\bm x_2$, and
$\bm x_1-\bm x_2$. Changes of the random realization test this
part of the basis. They do not constitute a new optimization of
each resonance.

The relative-motion grid has 125 points,
$R_\nu=0.3(1.06)^{\nu-1}\amu$. The grid is extended through the
first point above the nominal $400\amu$ limit; its last point is
$412.1436\amu$. Each $1s$, $2s$, or $2p$ channel has one real
Gaussian and cosine and sine partners at $\omega=3,10$, giving
five functions per grid point. The six atomic channels in the two
arrangements therefore give 3750 contractions. In $L=1$, the
additional $K=1$ channels use the same grid, both $2s$ and $2p$
radial coefficient sets, and the four angles
$\vartheta=0.3,0.6,0.9,1.2$ radians. They are real and have no
oscillatory partners, giving 2000 further contractions.

The original radial orbitals have 16 Gaussian terms. Their coefficients
are obtained from the radial eigenproblem with measure $r^2dr$.
For a geometric exponent sequence, the initial exponent and ratio
are optimized in logarithmic variables by a Nelder--Mead search
from multiple starting points, minimizing the chosen atomic eigenvalue.
The $1s$ and $2p$ functions use the lowest radial root in their
angular sectors, and $2s$ uses the second $s$ root.
The enhanced $2s$ orbitals instead use a fixed geometric exponent
grid from $0.008\mu_a^2$ to $10^4\mu_a^2$ in electronic atomic
units, with 24 or 32 terms. The zero-splitting orbital and the
orthogonal response are constructed in that same space from
$\Delta=0$ and $0.3\eV$. The original $1s$ contraction is retained
when testing the enhanced radial span. The response is added to
the $K=0$ $2s$ channels in both arrangements; the $K=1$ $2s$
channels receive the improved base contraction only.

All formulas in the main text may be read as unnormalized basis
functions. In the matrix construction each primitive is normalized
individually. Its contraction coefficient is multiplied by the
corresponding primitive norm so that the intended contraction is
unchanged. In particular, the $L=0$ $p$-channel difference carries
weights $\pm D_p/(4\gamma_{p\nu})$ before this normalization.
Omitting the factor $1/\gamma_{p\nu}$ would change the atomic
radial combination because $\gamma_{p\nu}$ depends on $\beta_p$.

\section{Long-range equations and reference-energy convention}
\label{sec:s-long-range-equations}
In the channel orders specified in the main text, the dimensionless
dipole and centrifugal matrices are
\begin{equation}
 \mathsf M_{0}=\begin{pmatrix}0&-g/\sqrt3\\-g/\sqrt3&2\end{pmatrix},
 \qquad
 \mathsf M_{1}=\begin{pmatrix}
 2&-g/3&-\sqrt2 g/3\\
 -g/3&0&0\\
 -\sqrt2 g/3&0&6
 \end{pmatrix},\qquad
 g=6\sqrt3\,\frac{\mu_R}{\mu_d}.
\end{equation}
Here $P_{2s}=\operatorname{diag}(1,0)$ for $L=0$ and
$\operatorname{diag}(1,0,0)$ for $L=1$.
Writing $x=\ln(R/\amu)$ and $\bm u=\sqrt R\,\bm v$ gives
\begin{align}
 -\bm v''+\mathsf K_0\bm v
  &=\frac{2\bar\mu_R\mathcal E}{E_\mu}e^{2x}\bm v,\\
 \mathsf K_0&=\mathsf M+\tfrac14\mathsf I
  -\frac{2\bar\mu_R\Delta}{E_\mu}e^{2x}P_{2s},
\end{align}
where primes denote $x$ derivatives, $\bar\mu_R=\mu_R/\mu_d$,
and $E_\mu=\mu_d E_h$. These are the finite-splitting equations
used in the reduced calculation, including all listed angular channels.

The reference energy in the geometric law fixes a quantum-defect
convention. Under a change from $E_{\mathrm{ref}}$ to another positive
reference $E_{\mathrm{ref}}'$, the same complex energies are obtained with
\begin{equation}
 \mu'=\mu-\frac{\alpha_L}{2\pi}
          \ln\frac{E_{\mathrm{ref}}'}{E_{\mathrm{ref}}}.
\end{equation}
Only the real constant part of the defect changes. The choice
$E_{\mathrm{ref}}=1\eV$ therefore introduces no additional physical
scale or fit constraint.

\section{Coulomb spectrum: sensitivity conventions and radial comparisons}
\label{sec:s-spectrum}
Table~I of the main text keeps three sources distinct: the
untruncated $L=0$ QZ spectrum, the original-radial-basis stationary
values for the deeper $L=1$ members, and the enhanced stationary
values for its near-threshold members. The near-threshold Table~II of
the main text takes its $0^+$ entries from the enhanced basis with the
channel grid extended to $800\amu$ (Table~\ref{tab:s-rmax}) and its
$1^-$ entries from the enhanced 24-term basis with the $400\amu$ grid
(Table~\ref{tab:s-v2-sequence}), all at $\epsilon_S=10^{-12}$. The eighth $L=1$ entry is
assigned by its numerical stability and agreement with the dipole
sequence. The separate unassigned shallow candidate is discussed below.

Parenthesized entries refer to the last digits and record the following
comparison scales. They are not total error bounds or statistical intervals.
$0^+$: value
$=$ untruncated QZ endpoint of the original 5150-function basis;
$\delta E_r=\max(|\mathrm{QZ}-\mathrm{enh}|,1\meV)$ and
$\delta\GC=|\mathrm{QZ}-\mathrm{enh}|$, where ``enh'' is the enhanced basis at
$\epsilon_S=10^{-12}$ (retained-subspace dependence, $6\%$ of every width;
random-block realisations agree to $10\meV$). $1^-$, members 1--5: value
$=$ stationary pole of the original radial basis (26-point $\eta$ scan,
reproduced by the 192-point server family to $\le11\meV$); comparison scale $=$
largest displacement produced by the enhanced radial spans in
fixed-$\eta$ comparisons (radial-basis systematic; the 16-point enhanced scans
do not resolve stationary points for these broad members). $1^-$, members
6--8: value $=$ enhanced 24-term stationary pole; comparison scale $=$ larger of
the 24/32-term difference and the difference between two random extended
blocks (seeds) in the original radial basis; the latter is the larger for all
three members. Literature: Ref.~\cite{wenzhu2025}
Table~4 (widths $\pm0.2\eV$) and Ref.~\cite{kilic2005} Tables B.12/B.13
(full widths); the $J=0$ paper~\cite{kilic2004} gives $-670.81$ and $-668.13\eV$.

Below the first $0^+$ member no pole is assigned. In both untruncated-QZ
scans (two random extended blocks), 15 eigenvalues with $0<\GC<8\eV$ lie
between the $t\mu(n=2)+d$ threshold and $-670.81\eV$. Relative to
$\eta=0.3$, each of them moves by at least $0.12\eV$ at $\eta=0.095$ or
$0.95$, whereas the listed members move by less than $1\meV$ (the
least-moving ones lie between $-676.7$ and $-675.2\eV$ with
$\GC=2$--$5\eV$). In the
enhanced bases a broad candidate at $-675.92\eV$ with $\GC=3.83\eV$
appears only at $\epsilon_S=10^{-11}$ and $\Delta=0$; it is absent at
$10^{-12}$ and $10^{-10}$. The resonance once extrapolated to lie
$0.8\eV$ above the $t\mu(n=2)+d$ threshold is therefore not resolved,
in agreement with the adiabatic cross sections of
Ref.~\cite{korobov1996window}.

\begin{table}[t]
\caption{\label{tab:s-v2-sequence}$L=1$ representatives at $\Delta=0$, with
$B=E_{2s}-E_r$ and full width $\GC$ in meV. Each column pair comes from
one complete CAP family at $\epsilon_S=10^{-12}$ and
$R_{\mathrm{CAP}}=60\amu$. The last row has no assigned member number.}
\begin{ruledtabular}
\begin{tabular}{lrrrrrr}
 & \multicolumn{2}{c}{Original radial basis} & \multicolumn{2}{c}{Enhanced, 24 terms}
 & \multicolumn{2}{c}{Enhanced, 32 terms}\\
Member & $B$ & $\GC$ & $B$ & $\GC$ & $B$ & $\GC$\\\hline
6 & 103.06 & 39.01 & 103.42 & 31.36 & 103.49 & 31.12 \\
7 & 47.63 & 18.68 & 47.96 & 15.62 & 47.76 & 15.23 \\
8 & 21.92 & 8.13 & 22.23 & 8.06 & 21.98 & 7.80 \\
Unassigned & 3.85 & 1.47 & 4.55 & 1.18 & 4.52 & 1.21 \\
\end{tabular}
\end{ruledtabular}
\end{table}

The pre-existing descriptive quantum-defect fits, using either the original
$L=1$ range centres or one coherent run, predict positions near
$-665.8217$ and $-665.8218\eV$ and widths $7.8$ and $8.7\meV$.
This agreement supports the sequence interpretation across these input choices.
The fits and their input sets are kept separate in the supplemental material.

Improving the radial representation changes the sixth and seventh
$L=1$ positions by about $0.36$ and $0.33\meV$, but reduces their
widths by about $20\%$ and $16\%$. The eighth width changes by less
than $1\%$ relative to the original principal configuration.
The unassigned candidate near $B=4.55\meV$ does not fill the next
position of the geometric sequence, expected near $10\meV$.
Its ratio $\GC/B=0.26$ is also below the $0.30$--$0.36$ range of
members six to eight in the same basis. It is not assigned a sequence number.
The seed, cutoff, and radial-range tests below provide the
additional diagnostics for such weak candidates.

\section{Numerical controls for the enhanced $L=1$ basis}
\label{sec:s-l1}
The enhanced calculations use 9000 contractions, with 142000 primitives
for the 24-term atomic representation and 170000 for the 32-term one.
The Gaussian range is $0.05\amu$ and the atomic interval ratio is
$\Delta_t/\Delta_d=1$. The first set of calculations comprises
five settings and ten CAP families; the second adds four
settings and nine families (Table~\ref{tab:s-inventory};
Secs.~\ref{sec:s-l1-cut} and \ref{sec:s-l1-rmax}). All families
have 16 strengths $\eta=10^{-2+3j/15}$, $j=0,\ldots,15$.

\begin{table}[tbp]
\caption{\label{tab:s-inventory}Enhanced $L=1$ scan coverage.
The absorber onset is in $a_{d\mu}$; $N$ is the retained dimension.
The last four settings form the second set; the extended-grid
basis has 9672 contractions (Sec.~\ref{sec:s-l1-rmax}).}
\begin{ruledtabular}
\begin{tabular}{lccrrr}
Atomic span / seed & $\epsilon_S$ & $R_{\mathrm{CAP}}$ & $\Delta$ (eV) & Points & $N$\\\hline
24 terms / principal & $10^{-12}$ & 60 & $0,0.2,0.24$ & 48 & 5442\\
24 terms / principal & $10^{-10}$ & 60 & $0,0.2$ & 32 & 4667\\
24 terms / principal & $10^{-12}$ & 90 & $0.2$ & 16 & 5442\\
24 terms / second & $10^{-12}$ & 60 & $0,0.2$ & 32 & 5436\\
32 terms / principal & $10^{-12}$ & 60 & $0,0.2$ & 32 & 5442\\
24 terms / principal & $10^{-11}$ & 60 & $0,0.2$ & 32 & 5077\\
24 terms / principal & $3\times10^{-13}$ & 60 & $0,0.2$ & 32 & 5607\\
24 terms / principal & $10^{-13}$ & 60 & $0,0.2$ & 32 & 5753\\
24 terms / grid to $800\amu$ & $10^{-12}$ & 60 & $0,0.2,0.24$ & 48 & 5765\\
\end{tabular}
\end{ruledtabular}
\end{table}

The original radial-basis comparison contains a further 192 points in
12 families. The reference MQDT boundary fitted to the original radial
basis uses the separate \texttt{l1k1chonly} zero-splitting run at
$10^{-12}$, with positions and widths from that one run rather than the
original range centres; the refitted boundaries use the stationary
members $5$--$8$ of the enhanced 24-term and 32-term families
(Table~\ref{tab:s-model-l1}).
The maximum projected action residual among the 160 enhanced
points is $2.18\times10^{-12}$, and the largest difference between
the eigenvalue width and the recorded CAP-flux identity is
$3.69\times10^{-7}\eV=0.000369\meV$. There are no negative widths.
Among the 144 points of the second set the maximum residual is
$3.9\times10^{-11}$, below the $10^{-8}$ selection gate.
These quantify the retained eigenproblem; basis sensitivity is assessed
with changes of representation and overlap cutoff.

\subsection{Retained-subspace dependence}
\label{sec:s-l1-cut}
The overlap cut $\epsilon_S$ selects the retained subspace of the
9000-function basis. Table~\ref{tab:s-l1-cut} lists the
zero-splitting members six to eight for five cuts and for the
extended channel grid, all with the same 16-point $\eta$ scan and
the same selection rules. A retained space is admitted for the
splitting scan when (i) the three members have interior stationary
representatives with
successive binding ratios within $3\%$ of the dipole value
$2.161$, and (ii) all stationary minima of one member, on one or on
several $\eta$ trajectories, agree within the systematic
comparison scale of that member in Table~I of the main text ($1.5$,
$0.9$, and $0.4\meV$). Both criteria were fixed before the second
set was computed and use neither literature values nor the
finite-splitting results.

\begin{table}[tbp]
\caption{\label{tab:s-l1-cut}Zero-splitting members six to eight
of the enhanced 24-term $L=1$ basis for five overlap cuts and for
the extended channel grid; $B$ ($\GC$) in meV, $N$ the retained
dimension. Parentheses contain the full width of the selected
representative. A dagger indicates additional stationary minima
on the same or another $\eta$ trajectory; only the representative
with the lowest speed is printed here. The
last columns give the successive binding ratios (dipole value
$2.161$) and the admissibility verdict of Sec.~\ref{sec:s-l1-cut}.}
\begin{ruledtabular}
\begin{tabular}{lrlllccc}
Retained space & $N$ & Member 6 & Member 7 & Member 8 & $B_6/B_7$ & $B_7/B_8$ & Admitted\\\hline
$\epsilon_S=10^{-10}$ & 4667 & --- & --- & --- & --- & --- & no \\
$\epsilon_S=10^{-11}$ & 5077 & 102.25 (38.49) & 44.60 (19.06) & --- & --- & --- & no \\
24 Gaussian terms (principal) & 5442 & 103.42 (31.36) & 47.96 (15.62) & 22.23 (8.06) & 2.156 & 2.157 & yes \\
$\epsilon_S=3\times10^{-13}$ & 5607 & 90.89 (28.50)$^{\dagger}$ & 42.19 (12.22)$^{\dagger}$ & 20.45 (5.38)$^{\dagger}$ & 2.154 & 2.063 & no \\
$\epsilon_S=10^{-13}$ & 5753 & 103.22 (32.17)$^{\dagger}$ & 47.59 (15.88)$^{\dagger}$ & 21.94 (7.54)$^{\dagger}$ & 2.169 & 2.169 & yes \\
Channel grid to $800\amu$ & 5765 & 100.59 (35.42) & 45.79 (16.21) & 19.07 (6.95) & 2.197 & 2.401 & no \\
\end{tabular}
\end{ruledtabular}
\end{table}

The admissible window is narrow. At $10^{-11}$ the eighth member
has no stationary trajectory and the seventh is displaced by
$3.4\meV$; at $10^{-10}$ no dipole sequence remains (the stationary
candidates lie at $307$, $188$, $33$, $15$, and $4.9\meV$). The cut
$3\times10^{-13}$ keeps overlap eigenvalues within a factor of $46$
of the most negative computed eigenvalue, $-3.1\times10^{-12}$; its
members six to eight each show two stationary minima on one
$\eta$ trajectory that differ by $6.2$, $2.5$, and $0.9\meV$, so
criterion (ii) fails (the table lists the lower-speed minimum of
each pair, at $\eta=0.04$; the other lies at $\eta=1.6$). The cut
$10^{-13}$ (factor $15$) reproduces
the $10^{-12}$ members to $0.2$, $0.4$, and $0.3\meV$ and their
widths to $0.8$, $0.3$, and $0.5\meV$, within the Table~I
comparison scales. The second random block reproduces the sequence in
$\eta=2.51$ snapshots ($103.0$, $47.7$, and $22.1\meV$) but has no
interior stationary minimum at degeneracy (its speed is smallest
at the $\eta=10$ end of the scan). These snapshots remain useful
diagnostics but do not meet criterion (i), so this setting is
excluded from the admitted range. Since the retained eigenvalues at $10^{-13}$,
$5\times10^{-11}$ in the normalized overlap, are already
comparable with the double-precision noise of the overlap
elements, the agreement between $10^{-12}$ and $10^{-13}$ brackets
the working cut but does not demonstrate convergence in
$\epsilon_S$ in the usual sense.

\begin{table*}[tbp]
\caption{\label{tab:s-l1-controls}Seventh $1^-$ member at
$\Delta=0.2\eV$ over all retained spaces, atomic expansions, random
blocks, and absorber onsets. This table collects the controls for
the main-text finite-splitting results. $N$ is the retained dimension of the 9000-function
basis (9672 for the extended grid); $B$ and $\GC$ in meV; $\eta$ and
$v=|dE/d\ln\eta|$ (meV) refer to the selected stationary point, a
second stationary minimum of the same member, when present, in
parentheses. The last column is the admissibility verdict of
Sec.~\ref{sec:s-l1-cut}; the second-seed zero-splitting members
have only $\eta=2.51$ snapshots and fail the stationarity requirement.}
\begin{ruledtabular}
\begin{tabular}{lrllrrc}
Setting & $N$ & $B$ & $\GC$ & $\eta$ & $v$ & $\Delta=0$ test\\\hline
24 Gaussian terms (principal) & 5442 & 1.730 & 3.894 & 2.51 & 0.099 & yes \\
32 Gaussian terms & 5442 & 1.758 & 3.839 & 2.51 & 0.097 & yes \\
Second random seed & 5436 & 1.650 & 3.990 & 2.51 & 0.123 & no \\
$R_{\mathrm{CAP}}=90\amu$ & 5442 & 1.828 & 3.934 & 6.31 & 0.126 & yes \\
$\epsilon_S=10^{-13}$ & 5753 & 1.706 (1.577) & 3.992 (4.285) & 1.00 & 0.026 & yes \\
$\epsilon_S=3\times10^{-13}$ & 5607 & 0.543 (1.084) & 2.456 (3.056) & 0.06 & 0.051 & no \\
$\epsilon_S=10^{-11}$ & 5077 & 1.469 & 5.327 & 1.58 & 0.081 & no \\
$\epsilon_S=10^{-10}$ & 4667 & --- & --- & --- & --- & no \\
Channel grid to $800\amu$ & 5765 & 1.319 & 4.042 & 6.31 & 0.123 & no \\
Grid to $800\amu$, $\epsilon_S=10^{-11}$ & 5381 & 1.777 (1.670) & 4.254 (3.874) & 3.98 & 0.023 & no \\
Grid to $800\amu$, $\epsilon_S=5\times10^{-12}$ & 5505 & 1.160 & 3.148 & 3.98 & 0.020 & no \\
Grid to $800\amu$, $\epsilon_S=3\times10^{-12}$ & 5588 & 1.291 & 4.629 & 3.98 & 0.097 & no \\
Grid to $800\amu$, $\epsilon_S=2\times10^{-12}$ & 5655 & 1.502 (1.326) & 4.323 (3.968) & 1.00 & 0.045 & no \\
\end{tabular}
\end{ruledtabular}
\end{table*}

At $\Delta=0.2\eV$ the admissible retained spaces give the
seventh member at $B=1.58$--$1.83\meV$ and $\GC=3.84$--$4.29\meV$
and the sixth at $31.9$--$33.1$ and $23.1$--$26.4\meV$, the outer
values coming from the second stationary minimum of the
$10^{-13}$ trajectories (Table~\ref{tab:s-l1-controls}). The inadmissible spaces give $1.47/5.33$
($10^{-11}$), $0.54$--$1.08/2.46$--$3.06$ ($3\times10^{-13}$, two
minima of one trajectory), $1.32/4.04$ (extended grid), and
$0.11/3.15\meV$ ($10^{-10}$). The second-seed result at finite
splitting is $1.650/3.990\meV$ but is excluded because its
zero-splitting sequence has no interior stationary representatives.
These excluded settings do not support a consistent assignment by
the stated criterion. Their deviations also caution against
interpreting the admitted range as a bound on all numerical errors.
All ranges are observations over the listed settings, not
statistical confidence intervals.

The 24- and 32-term atomic expansions, the displaced absorber onset,
and the cut $\epsilon_S=10^{-13}$ meet the zero-splitting criteria
and define the admitted range quoted above. The absorber-onset
control inherits the principal-space test. Relative to the principal
result, the atomic expansion and absorber onset each change the
seventh-member position by less than $0.1\meV$. The second-seed
finite-splitting representative is retained as a diagnostic but
does not enter this range, as its zero-splitting sequence lacks
interior stationary minima.

\subsection{Radial range of the channel grid}
\label{sec:s-l1-rmax}
The channel grid was continued from $400$ to $800\amu$ with the
same ratio (12 points; 552 source contractions, 672 enhanced
contractions), all other functions unchanged. The first 9000
contractions of the extended basis are identical to the principal
basis, and its CAP selection differs only by the appended
$t\mu(n=2)$ functions. Table~\ref{tab:s-l1-rmax} pairs the
below-threshold representatives of the two grids at fixed
$\epsilon_S=10^{-12}$. Unlike the $L=0$ case
(Table~\ref{tab:s-rmax}), the extension changes the $L=1$ retained
subspace appreciably: $5442\to5765$ retained directions and
$1249\to1364$ negative overlap eigenvalues. Members six to eight
move by $-2.8$, $-2.2$, and $-3.2\meV$ at degeneracy although
their zero-width binding-length estimates, $1/\kappa=47$--$102\amu$, lie inside the
original grid, and the ratio $B_7/B_8=2.40$ fails criterion (i).
This comparison changes both the radial representation and the
retained subspace. The sign of a shift does not distinguish their effects:
complex CAP poles do not obey the variational monotonicity of
bound-state energies. At $\Delta=0.2\eV$ the
extended grid gives the seventh member at $B=1.32\meV$,
$\GC=4.04\meV$, and at $0.24\eV$ the shallow candidate moves from
$0.31$ to $0.03\meV$. The range test is thus inconclusive for
$L=1$: the range effect cannot be separated from the subspace
dependence at this cut. The extended-grid $1.32\meV$ candidate
does not enter the admitted $1.58$--$1.83\meV$ range; these two
sets must not be combined into a convergence interval. The
$0.24\eV$ candidate is not resolved across the tested grids. The
extended grid also hosts shallow stationary candidates at $3.5$,
$1.5$, and $0.25\meV$ at degeneracy that the original grid does
not; none fills the next place of the sequence
($\approx10\meV$).

Since the extension changes the retained subspace, the extended
basis was also projected at the tighter cuts
$\epsilon_S=2\times10^{-12}$, $3\times10^{-12}$, $5\times10^{-12}$,
and $10^{-11}$, whose retained dimensions ($5655$, $5588$, $5505$,
$5381$) bracket the $5442$ of the principal basis
(Table~\ref{tab:s-l1-r800-cut}). If the shifts at $10^{-12}$ were
a consequence of the larger retained dimension alone, one of these
cuts might recover the principal sequence, although equal retained
dimensions do not imply equal subspaces. None does. At
$5\times10^{-12}$ and $10^{-11}$ the eighth member has no
stationary trajectory and the seventh is displaced by $-7.5$ and
$-4.1\meV$; at $3\times10^{-12}$ and $2\times10^{-12}$ all three
members are present but the ratios are $2.28/2.37$ and
$2.22/2.38$, and at $2\times10^{-12}$ the sixth and seventh members
each have two stationary minima on different trajectories that
differ by $1.3\meV$ (criterion (ii)). The seventh member at
$0.2\eV$ spans $1.16$--$1.78\meV$ with widths $3.1$--$4.6\meV$
over these cuts. The appended functions change the conditioning
of the retained space, and none of the tested cuts recovers the
principal sequence. This construction therefore does not isolate
a radial-range error. We use it as a combined range/subspace
diagnostic and do not infer a converged radial limit from it.

\begin{table}[tbp]
\caption{\label{tab:s-l1-r800-cut}Zero-splitting members six to
eight of the extended-grid basis ($9672$ contractions) for five
overlap cuts; columns as in Table~\ref{tab:s-l1-cut}. The
$10^{-12}$ row repeats the extended-grid entry of that table.}
\begin{ruledtabular}
\begin{tabular}{lrlllccc}
Retained space & $N$ & Member 6 & Member 7 & Member 8 & $B_6/B_7$ & $B_7/B_8$ & Admitted\\\hline
Grid to $800\amu$, $\epsilon_S=10^{-11}$ & 5381 & 103.99 (36.85)$^{\dagger}$ & 43.87 (15.71)$^{\dagger}$ & --- & --- & --- & no \\
Grid to $800\amu$, $\epsilon_S=5\times10^{-12}$ & 5505 & 99.22 (34.81) & 40.42 (15.07) & --- & --- & --- & no \\
Grid to $800\amu$, $\epsilon_S=3\times10^{-12}$ & 5588 & 102.93 (39.71) & 45.22 (21.91) & 19.07 (13.79) & 2.276 & 2.371 & no \\
Grid to $800\amu$, $\epsilon_S=2\times10^{-12}$ & 5655 & 104.14 (34.21)$^{\dagger}$ & 46.91 (16.44)$^{\dagger}$ & 19.70 (7.17) & 2.220 & 2.381 & no \\
Grid to $800\amu$, $\epsilon_S=10^{-12}$ & 5765 & 100.59 (35.42) & 45.79 (16.21) & 19.07 (6.95) & 2.197 & 2.401 & no \\
\end{tabular}
\end{ruledtabular}
\end{table}

\begin{table}[tbp]
\caption{\label{tab:s-l1-rmax}Radial-range control of the enhanced
$L=1$ basis at $\epsilon_S=10^{-12}$: below-threshold
representatives with the channel grid ending at $400$ and
$800\amu$; $B$ and $\GC$ in meV, $\delta B=B_{800}-B_{400}$,
$\delta\GC=\GC^{800}-\GC^{400}$. Rows marked (new) are stationary
candidates present only in the extended grid.}
\begin{ruledtabular}
\begin{tabular}{clrrrrrr}
$\Delta$ (eV) & Member & $B_{400}$ & $\GC^{400}$ & $B_{800}$ & $\GC^{800}$ & $\delta B$ & $\delta\GC$\\\hline
0 & 6 & 103.420 & 31.362 & 100.592 & 35.419 & -2.829 & +4.057 \\
0 & 7 & 47.960 & 15.615 & 45.793 & 16.207 & -2.167 & +0.591 \\
0 & 8 & 22.231 & 8.061 & 19.074 & 6.949 & -3.157 & -1.112 \\
0 & (new) & --- & --- & 0.252 & 0.776 & --- & --- \\
0 & (new) & --- & --- & 1.513 & 1.135 & --- & --- \\
0 & (new) & --- & --- & 3.524 & 3.713 & --- & --- \\
0.2 & 6 & 32.552 & 23.235 & 30.479 & 26.082 & -2.073 & +2.847 \\
0.2 & 7 & 1.730 & 3.894 & 1.319 & 4.042 & -0.412 & +0.148 \\
0.24 & 6 & 24.411 & 20.500 & 22.580 & 22.893 & -1.831 & +2.394 \\
0.24 & 7 & 0.311 & 2.059 & 0.026 & 1.979 & -0.285 & -0.080 \\
\end{tabular}
\end{ruledtabular}
\end{table}

At $\Delta=0$, the unnumbered shallow trajectory gives
$B=4.552\meV$, $\GC=1.184\meV$ with 24 terms and
$B=4.518\meV$, $\GC=1.206\meV$ with 32 terms. The second-seed
trajectory exists at all 16 strengths. At the common $\eta=2.512$
it gives $B=4.529\meV$, $\GC=1.291\meV$; at the endpoint
$\eta=10$ it gives $B=4.550\meV$, $\GC=1.336\meV$.
Its global minimum speed occurs at the endpoint, with no selected
interior minimum. The same distinction between trajectory presence
and selection applies to neighbouring second-seed candidates at degeneracy.
The $10^{-10}$ principal-basis shallow representative is
$B=4.877\meV$, $\GC=3.466\meV$.

At $\Delta=0.24\eV$, the principal enhanced representative is
$B=0.311\meV$, $\GC=2.059\meV$, $\eta=2.512$, and the extended-grid
basis gives $B=0.026\meV$, $\GC=1.979\meV$
(Table~\ref{tab:s-l1-rmax}). In the global original/enhanced
one-to-one assignment, the original point assigned to this
candidate has another enhanced neighbour that is closer in complex
energy. The grid-induced displacement is comparable to the candidate's
distance from the threshold, and no 32-term or second-seed family
exists at this splitting. For this broad complex pole, the
asymptotic closed-channel scale is
\begin{equation}
 \ell=\frac{1}{\operatorname{Re}\kappa},\qquad
 \kappa=\sqrt{2\mu_R(B+i\GC/2)},\quad\operatorname{Re}\kappa>0,
\end{equation}
with energies and lengths converted consistently to atomic units.
The principal values give $\ell\simeq575\amu$; ignoring the width
would instead give $859\amu$. Neither scale is a molecular radius,
and the largest Gaussian range is not a hard spatial boundary.
The candidate is
therefore not resolved; it defines neither a critical splitting nor
a seventh-member continuation and is not used in the main text.

\subsection{Open $n=1$ channels: projector absorber for $L=1$}
\label{sec:s-l1-n1}
The production absorber acts in the $t\mu(n=2)$ coefficient block;
outgoing flux in the open $n=1$ channels, with about $2\,\mathrm{keV}$
of kinetic energy, is represented on the finite grid but not
absorbed. As for $L=0$ (Sec.~\ref{sec:s-l0-n1}), we add to the
principal $L=1$ basis at $\epsilon_S=10^{-12}$ an all-contraction
atomic-$1s$ projector absorber $\eta_1P_{1s}W_{R_0}P_{1s}$ in both
arrangements, with the $\Delta$-calibrated $1s$ orbital of each atom
as target, exterior quadrature from $R_0=6$ or $10\amu$ (order 512
per interval), and an independent strength $\eta_1$. The whole
Hamiltonian is rebuilt in its own canonical coordinates from the
frozen matrices; at $\eta_1=0$ the projected operators coincide
with the production ones to machine precision. The largest
exterior-projector eigenvalue is $1.0000076$, consistent with
$P\le G$ for the finite metric $G=X^{T}SX$. The $n=2$ strength
$\eta_2$ takes the values $1$, $2.51$, and $6.31$, bracketing the
stationary strength of the seventh member, and $\eta_1$ takes
$0$, $100$, $300$, and $1000$.

\begin{table}[tbp]
\caption{\label{tab:s-l1-n1}Effect of the additional open-$1s$
projector absorber on the sixth and seventh $1^-$ members of the
principal basis. For each $\Delta$ and $n=2$ strength $\eta_2$, the
$\eta_1=0$ values ($B$, $\GC$ in meV) are followed by the largest
displacement $|\delta B|$, $|\delta\GC|$ over $\eta_1=100$, $300$,
$1000$ and both onsets; the two onsets differ by less than
$0.007\meV$ throughout.}
\begin{ruledtabular}
\begin{tabular}{crrrrrrrrr}
 & & \multicolumn{4}{c}{Member 6} & \multicolumn{4}{c}{Member 7}\\
$\Delta$ (eV) & $\eta_2$ & $B$ & $\GC$ & $|\delta B|$ & $|\delta\GC|$ & $B$ & $\GC$ & $|\delta B|$ & $|\delta\GC|$\\\hline
0 & 1.00 & 102.650 & 31.773 & 0.238 & 0.199 & 47.561 & 15.823 & 0.060 & 0.030 \\
0 & 2.51 & 103.420 & 31.362 & 0.083 & 0.417 & 47.960 & 15.615 & 0.007 & 0.080 \\
0 & 6.31 & 103.857 & 32.428 & 0.067 & 0.386 & 48.198 & 16.213 & 0.074 & 0.039 \\
0.2 & 1.00 & 31.988 & 23.524 & 0.172 & 0.187 & 1.625 & 3.865 & 0.021 & 0.031 \\
0.2 & 2.51 & 32.552 & 23.235 & 0.055 & 0.336 & 1.730 & 3.894 & 0.003 & 0.042 \\
0.2 & 6.31 & 32.866 & 24.017 & 0.050 & 0.304 & 1.757 & 4.077 & 0.015 & 0.026 \\
\end{tabular}
\end{ruledtabular}
\end{table}

Table~\ref{tab:s-l1-n1} lists the $\eta_1=0$ values and the
largest displacements. At the stationary $n=2$ strength the seventh
member moves by at most $0.007\meV$ in position and $0.08\meV$ in
width at degeneracy and by $0.003$ and $0.04\meV$ at $0.2\eV$; over
all three $\eta_2$ values the bounds are $0.07/0.08$ and
$0.02/0.04\meV$. The sixth member is more sensitive, up to
$0.24\meV$ in position off the stationary strength and $0.42\meV$
($1.3\%$) in width; the eighth member at degeneracy changes by at
most $0.05/0.07\meV$. All widths decrease slightly as $\eta_1$
grows. Compared with the $L=0$ test ($0.0013/0.011\meV$), the
$L=1$ response is an order of magnitude larger but remains an
order of magnitude below the retained-subspace spread of
Sec.~\ref{sec:s-l1-cut}. The label $\GC$, obtained with absorption
in the $t\mu(n=2)$ block, is retained for the reported widths;
the added $n=1$ absorber produces the small changes listed here.
As for $L=0$, this is
an operator-definition control; it does not add the newly open
$d\mu(2s)$ outgoing boundary above the split lower threshold, and
block absorption is not a branching fraction.

\section{Reference scale and deeper $L=1$ diagnostics}
The broadest trajectories are less stationary than the shallow ones.
To show their scale without applying a shallow absolute-speed requirement,
Table~\ref{tab:s-deep} compares all three enhanced $10^{-12}$
representations at the common $\eta=1$. Member labels are diagnostic
correspondences with the earlier sequence; these snapshots are not
newly selected stationary pole values. The original-basis column uses
the same $\eta$, so shifts are not hidden by selecting different
absorber strengths or different reference matches.

\begin{table}[tbp]
\caption{\label{tab:s-deep}Fixed-$\eta=1$ $L=1$ spectra at degeneracy,
in eV. Enhanced ranges cover 24 terms, 32 terms, and the second seed
at $10^{-12}$. Reference entries are $E_r/\Gamma$ with full widths.
The thesis source is Table B.13 of Ref.~\cite{kilic2005}.}
\begin{ruledtabular}
\begin{tabular}{crllll}
Member & Original $E_r/\GC$ & Enhanced $E_r$ & Enhanced $\GC$ & Wen--Zhu\cite{wenzhu2025} & K\i l\i \c{c}\cite{kilic2005}\\\hline
1 & $-670.561$ / $1.394$ & $-670.447$ to $-670.396$ & $1.618$--$1.719$ & $-670.62$ / $1.4$ & $-670.549$ / $1.472$ \\
2 & $-668.051$ / $0.725$ & $-667.995$ to $-667.978$ & $0.626$--$0.669$ & $-668.03$ / $0.8$ & $-667.988$ / $0.720$ \\
3 & $-666.843$ / $0.360$ & $-666.829$ to $-666.820$ & $0.295$--$0.313$ & $-666.82$ / $0.4$ & --- \\
4 & $-666.282$ / $0.172$ & $-666.279$ to $-666.275$ & $0.139$--$0.147$ & $-666.25$ / $0.2$ & --- \\
5 & $-666.023$ / $0.081$ & $-666.022$ to $-666.021$ & $0.065$--$0.069$ & $-665.97$ / unresolved & --- \\
6 & $-665.903$ / $0.038$ & $-665.903$ to $-665.902$ & $0.031$--$0.032$ & --- & --- \\
7 & $-665.848$ / $0.018$ & $-665.848$ to $-665.847$ & $0.015$--$0.016$ & --- & --- \\
\end{tabular}
\end{ruledtabular}
\end{table}

The first enhanced snapshot moves away from the thesis reference by
about $0.10$--$0.15\eV$, while the original snapshot is closer.
For the second and third members the positions retain an overall
correspondence with the published sequence. Width differences remain
at the level of tens of percent for some broad members. The full
width is always $-2\operatorname{Im}E$; a printed imaginary part
must be doubled before making these comparisons.

\section{Projection, absorption, and eigenvector diagnostics}

The complex absorbing potential (CAP) method uses real Hermitian matrices
$H,S,W$ in a finite retained space,
\begin{equation}
 (H-i\eta W)c=E S c,\qquad E=E_r-i\Gamma/2,
 \qquad \Gamma=2\eta\frac{c^\dagger Wc}{c^\dagger Sc}.
 \label{eq:s-flux}
\end{equation}
The last identity follows by multiplying the eigenproblem by $c^\dagger$ and
taking its imaginary part. It assumes a positive physical norm and Hermitian
$H,S,W$; numerical violations of those assumptions must be checked separately.
Decomposing $W$ into coefficient blocks yields absorption diagnostics for those
blocks. It does not establish asymptotic branching fractions into physical channels.

The canonical construction first normalizes the overlap in the original contracted basis,
\begin{equation}
 D_{ii}=\sqrt{S_{ii}},\quad D^{-1}SD^{-1}=U\,\mathrm{diag}(s_j)U^T,
 \quad X=D^{-1}U_{\rm keep}\,\mathrm{diag}(s_j^{-1/2}),
 \quad s_j>\epsilon_S s_{\max}.
\end{equation}
In exact arithmetic, $G=X^T S X=I$. In finite arithmetic, $G$ must be inspected
explicitly. Comparing an identity-metric calculation with one using the measured
$G$ probes projection arithmetic; it does not repair inaccurate input integrals.
Likewise, lowering $\epsilon_S$ admits additional directions whose matrix elements
may be dominated by cancellation. An observed negative eigenvalue of the computed
$S$ bounds from below the correction needed to restore positive semidefiniteness;
it is not an upper bound on the full matrix error.

The production projected standard problem is $Ar=Er$, with
$A=H_p-i\eta W_p$. A left eigenvector satisfies $l^\dagger A=E l^\dagger$.
For a simple eigenvalue, its first-order response to a perturbation is
\begin{equation}
 \delta E=\frac{l^\dagger\delta A\,r}{l^\dagger r}+O(\|\delta A\|^2),
 \qquad \kappa_2(E)=\frac{\|l\|_2\|r\|_2}{|l^\dagger r|},
 \qquad
 \frac{dE}{d\ln\eta}=\frac{l^\dagger(-i\eta W_p)r}{l^\dagger r}.
 \label{eq:s-condition}
\end{equation}
Here $\kappa_2$ refers to perturbations of this standard matrix in these coordinates;
it is not a basis-independent physical error estimate. Left and right action
residuals are computed separately. For a fixed nonidentity metric $G$, the
denominator in the derivative is instead $l^\dagger G r$; if the metric varies,
the numerator includes $-E\,\delta G$.

\begin{table}[tbp]
\caption{\label{tab:s-condition}Selected-point standard-eigenproblem diagnostics at $\epsilon_S=10^{-12}$. $v=|dE/d\ln\eta|$ is evaluated with left/right eigenvectors and $v_{\rm grid}$ is the stored finite-difference speed. ``Deep'' and ``shallow'' describe the selected representatives, not a certified ordering of physical states.}
\begin{ruledtabular}
\begin{tabular}{lclrrrr}
Basis & $\Delta$ (eV) & Candidate & $\eta$ & $\kappa_2$ & $v$ (meV) & $v/v_{\rm grid}$ \\
\hline
Response 24 & 0.0 & Shallow & 0.06310 & 1.3189 & 0.000317 & 1.0015 \\
Response 24 & 0.0 & Deep & 0.10000 & 12.1649 & 0.061015 & 0.7202 \\
Response 24 & 0.2 & Shallow & 0.03981 & 1.3760 & 0.000097 & 1.0647 \\
Response 24 & 0.2 & Deep & 0.10000 & 12.1286 & 0.049911 & 0.7177 \\
Exact angular & 0.0 & Shallow & 0.01585 & 1.3372 & 0.000772 & 3.6732 \\
Exact angular & 0.0 & Deep & 0.10000 & 12.2967 & 0.030438 & 0.3853 \\
Exact angular & 0.2 & Deep & 0.10000 & 12.2599 & 0.028420 & 0.4238 \\
Exact angular & 0.2 & Shallow & 0.15849 & 1.3750 & 0.000080 & 0.8753 \\
\end{tabular}
\end{ruledtabular}
\end{table}

\section{Trajectory selection and observed parameter dependence}

Each local family uses 16 logarithmically spaced $\eta$ values from $0.01$ to $10$.
Consecutive complex energies are assigned one-to-one with explicit unmatched
choices and a $0.2\eV$ gate. A selected trajectory has at least five points,
a positive width, an interior local minimum of
$|dE/d\ln\eta|<0.002\eV$, and projected action residual below $10^{-8}$.
Residuals are also checked at the two neighbouring points.
The finite-difference derivative is compared with Eq.~(\ref{eq:s-condition}) at
the representative points. These are numerical selection criteria, not confidence
limits or a complete definition of a physical resonance.

Different minima on an identical trajectory are retained in the underlying record;
the lowest-speed one represents that trajectory in tables. Distinct trajectories
are not merged solely because their energies are close. Historical member labels
are assigned after this extraction. Cross-family matching uses a stricter
$0.02\eV$ gate, with separately labelled exploratory $0.2\eV$ matches when a cut
causes larger shifts. All unmatched candidates and assignment ambiguities are
preserved. Proximity matching does not prove eigenvector identity across bases or
identify the outgoing channel sheet of an S-matrix singularity.
\begin{table}[tbp]
\caption{\label{tab:response-check}Completed $L=0$ sensitivity diagnostics at $R_{\mathrm{CAP}}=60\amu$. 
$B=E_{2s}-E_r$ and $\GC$ are in meV. At $\Delta=0$ the historical seventh candidate is matched after blind extraction; 
at $\Delta=0.2\eV$ the shallowest selected stationary candidate below the actual $2s$ threshold is shown. 
This descriptor does not establish a physical state count or cross-threshold continuation. Each row uses a complete 16-point $\eta$ scan.}
\begin{ruledtabular}
\begin{tabular}{lcrrr}
Basis & $\epsilon_S$ & $\Delta$ (eV) & $B$ & $\GC$ \\
\hline
Original & $10^{-12}$ & 0.0 & 51.421 & 15.696 \\
Original & $10^{-12}$ & 0.2 & 1.853 & 5.441 \\
Response 24 & $10^{-12}$ & 0.0 & 52.002 & 15.859 \\
Response 24 & $10^{-12}$ & 0.2 & 3.682 & 5.384 \\
Response 24 & $10^{-11}$ & 0.0 & 52.227 & 17.059 \\
Response 24 & $10^{-11}$ & 0.2 & 3.694 & 5.847 \\
Response 24 & $10^{-10}$ & 0.0 & 52.732 & 18.606 \\
Response 24 & $10^{-10}$ & 0.2 & 3.782 & 6.508 \\
Exact angular & $10^{-12}$ & 0.0 & 51.992 & 15.915 \\
Exact angular & $10^{-12}$ & 0.2 & 3.675 & 5.399 \\
Exact angular & $10^{-11}$ & 0.0 & 52.191 & 16.950 \\
Exact angular & $10^{-11}$ & 0.2 & 3.687 & 5.800 \\
Exact angular & $10^{-10}$ & 0.0 & 52.738 & 18.579 \\
Exact angular & $10^{-10}$ & 0.2 & 3.784 & 6.501 \\
Response 24, $R_{\max}=800\amu$ & $10^{-12}$ & 0.0 & 52.000 & 15.855 \\
Response 24, $R_{\max}=800\amu$ & $10^{-12}$ & 0.2 & 3.681 & 5.383 \\
\end{tabular}
\end{ruledtabular}
\end{table}
\begin{figure}[tbp]
\includegraphics[width=0.94\textwidth]{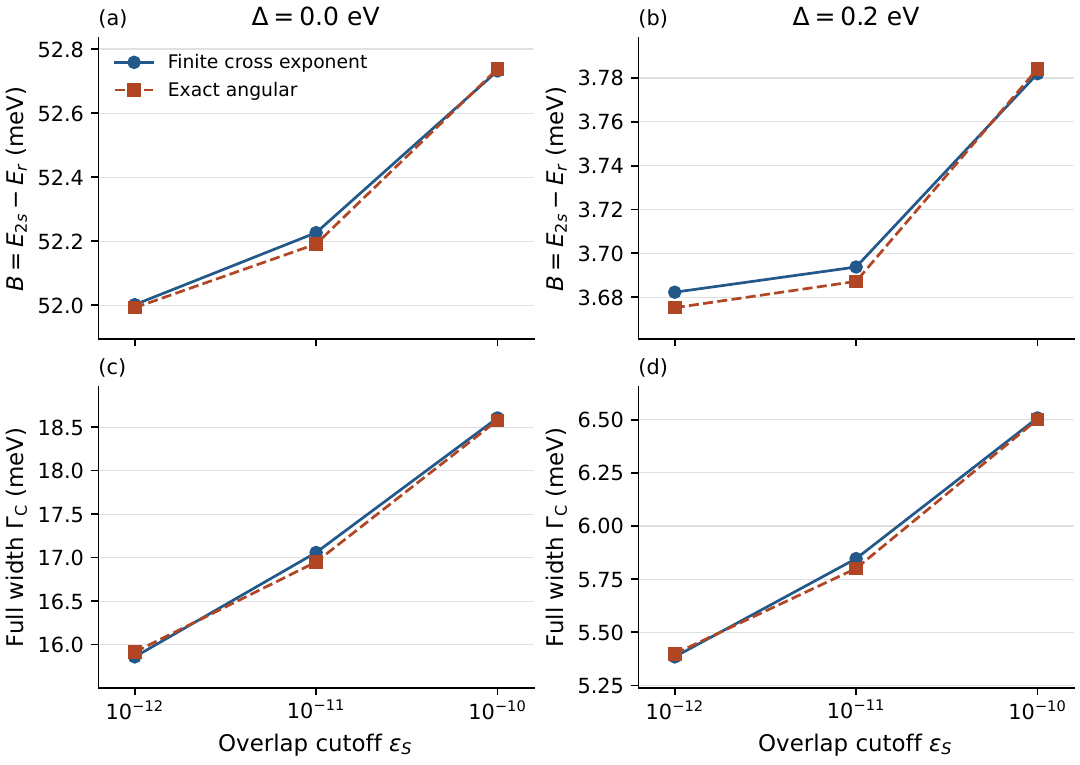}
\caption{\label{fig:s-cutoff}Observed cutoff sensitivity in the two enhanced $L=0$
angular representations. At degeneracy the seventh historical candidate is matched
after blind trajectory extraction; at $\Delta=0.2\eV$ each point denotes the
shallowest selected stationary candidate below its actual atomic $2s$ threshold.
Panels (a) and (b) show $B$, panels (c) and (d) the full width $\GC$.
Every point uses a complete 16-point CAP scan. Lines guide the eye between three
discrete cutoffs and are not confidence bands or a proof of state identity.
\AIplotnote}
\end{figure}

\subsection{Width sensitivity at finite splitting in $L=0$}
At $\Delta=0.2\eV$, the seventh $0^+$ position varies by about
$0.1\meV$ over the listed diagnostics. Its width decreases from
$6.51$ to $5.38\meV$ over the completed stationary scans as the
overlap cut is loosened from $10^{-10}$ to $10^{-12}$
(Fig.~\ref{fig:s-cutoff}). At $3\times10^{-13}$ and $10^{-13}$,
fixed-$\eta$ snapshots give $5.26$ and $5.15\meV$, respectively;
these are not completed stationary scans. Together, the comparisons
support the reduction from about $16$ to $5$--$6\meV$ reported in
the main text. The working-cut value $5.38\meV$ is retained for
comparison without assigning a converged width or a final
uncertainty interval.

\subsection{Fixed-strength diagnostics at finer overlap cuts}
The finite-cross-exponent enhanced $L=0$ basis also has individual
$\eta=0.1$ results at cuts $3\times10^{-13}$ and $10^{-13}$.
Table~\ref{tab:s-finer} reports those points without promoting them
to complete CAP scans or appending them to an interior-stationary
sequence. At degeneracy the point nearest the seventh-member
position is shown; at $0.2\eV$ the shallow neighbour is shown.

\begin{table}[tbp]
\caption{\label{tab:s-finer}Existing fixed-$\eta$ $L=0$ diagnostics,
with $B$ and full width in meV. These four points are not new
stationary representatives.}
\begin{ruledtabular}
\begin{tabular}{crrrrr}
$\epsilon_S$ & $N$ & $\Delta$ (eV) & $\eta$ & $B$ & $\GC$\\\hline
$3\times10^{-13}$ & 4362 & 0.0 & 0.1 & 51.974 & 15.511 \\
$3\times10^{-13}$ & 4362 & 0.2 & 0.1 & 3.691 & 5.260 \\
$10^{-13}$ & 4489 & 0.0 & 0.1 & 51.955 & 15.200 \\
$10^{-13}$ & 4489 & 0.2 & 0.1 & 3.700 & 5.149 \\
\end{tabular}
\end{ruledtabular}
\end{table}

The shallow threshold distance stays near $3.7\meV$ while widths
decrease as directions are admitted. The looser cut has less separation
from the overlap's arithmetic noise scale: the retained cutoff divided
by the magnitude of the most negative computed overlap eigenvalue is
about 6.8 and 2.3 for these two settings. This ratio is a diagnostic,
not a full matrix-error bound. The points help describe the observed
trend but do not supply a final uncertainty interval or a reason to
replace all working-cut widths by their endpoint values.

\subsection{Angular and absorber controls in $L=0$}
\label{sec:s-l0-n1}
The exact scalar angular factor is constructed as
\begin{equation}
 \bm r\cdot\bm R=
 \frac{|\bm r+q\bm R|^2-|\bm r-q\bm R|^2}{4q},
 \qquad q=\sqrt{a/b},
\end{equation}
with a common factorized Gaussian $\exp(-br^2-aR^2)$.
This removes higher odd partial waves from the finite-cross-exponent
approximation. At $10^{-12}$, the seventh-member position and width
at degeneracy change by $0.0095$ and $0.0557\meV$; the shallow
$0.2\eV$ values change from $3.682/5.384$ to $3.675/5.399\meV$.
This angular comparison is much smaller than the overlap-cut width
dependence at the tested settings.

An all-contraction atomic-$1s$ projector absorber was tested at
$\Delta=0,0.2\eV$ with 28 points and four doubled-order controls.
The seventh-member position/width changes at degeneracy are at most
$0.0013/0.011\meV$; the shallow $0.2\eV$ changes are at most
$0.00018/0.0040\meV$. The deepest candidate is more sensitive,
up to $0.082/1.14\meV$ in the same comparison.
Replacing the existing $t\mu(n=2)$ coefficient-block absorber by
an all-contraction $2s+2p$ projector changes the shallow position/width
by at most $0.00040/0.00024\meV$ at degeneracy and
$0.000125/0.000154\meV$ at $0.2\eV$. Quadrature and metric bounds
were checked independently. These are $L=0$ operator-definition
controls; the corresponding $L=1$ projector test is reported in
Sec.~\ref{sec:s-l1-n1}.
In particular, they do not add the newly open $d\mu(2s)$ outgoing
boundary above the split lower threshold.

\subsection{Radial range of the channel grid in $L=0$}
The channel-grid range of the production basis is $400\amu$. As a range
control, the geometric grid (ratio $1.06$) was continued to $800\amu$ by
appending 12 grid points, i.e.\ 360 contracted functions (both atoms,
$1s$/$2s$/$2p$, five complex-range partners each), while every other
function, the radial-response replacements, and the absorber definition
were kept unchanged. The extended basis has 6880 contractions, of which
4511 are retained at $\epsilon_S=10^{-12}$ (4221 of 6400 for the
production basis). Table~\ref{tab:s-rmax} pairs the below-threshold
representatives of the two complete 16-point scans by nearest complex
energy.

\begin{table}[tbp]
\caption{\label{tab:s-rmax}Radial-range control of the enhanced $L=0$
basis: below-threshold representatives with the channel grid ending at
$400$ and $800\amu$, $\epsilon_S=10^{-12}$, $R_{\mathrm{CAP}}=60\amu$,
complete 16-point $\eta$ scans. $B=E_{2s}-E_r$ and $\GC$ in meV;
$\delta B=B_{800}-B_{400}$ and $\delta\GC=\GC^{800}-\GC^{400}$.
Labels follow the energy order below the threshold of each family and are
descriptive only; parenthesised labels mark candidates present only in
the extended basis.}
\begin{ruledtabular}
\begin{tabular}{crrrrrr}
$j$ & $B_{400}$ & $\GC^{400}$ & $B_{800}$ & $\GC^{800}$ & $\delta B$ & $\delta\GC$\\\hline
\multicolumn{7}{l}{$\Delta=0$}\\
1 & 5022.762 & 1343.522 & 5022.520 & 1343.446 & -0.242 & -0.076 \\
2 & 2334.545 & 671.788 & 2334.459 & 671.724 & -0.087 & -0.065 \\
3 & 1086.942 & 322.778 & 1086.896 & 322.727 & -0.047 & -0.051 \\
4 & 508.038 & 153.214 & 508.015 & 153.190 & -0.023 & -0.024 \\
5 & 237.612 & 72.134 & 237.602 & 72.122 & -0.011 & -0.012 \\
6 & 111.156 & 33.845 & 111.152 & 33.842 & -0.004 & -0.004 \\
7 & 52.002 & 15.859 & 52.000 & 15.855 & -0.002 & -0.004 \\
8 & 24.299 & 7.449 & 24.325 & 7.423 & +0.026 & -0.025 \\
9 & 11.280 & 3.496 & 11.375 & 3.475 & +0.095 & -0.020 \\
10 & 4.433 & 2.631 & 5.288 & 1.658 & +0.855 & -0.973 \\
(11) & --- & --- & 2.387 & 0.779 & --- & --- \\
(12) & --- & --- & 0.623 & 0.779 & --- & --- \\
\multicolumn{7}{l}{$\Delta=0.2\eV$}\\
1 & 4920.680 & 1342.218 & 4920.435 & 1342.134 & -0.244 & -0.084 \\
2 & 2234.459 & 670.576 & 2234.365 & 670.514 & -0.094 & -0.063 \\
3 & 989.258 & 321.391 & 989.212 & 321.340 & -0.046 & -0.051 \\
4 & 414.330 & 150.996 & 414.307 & 150.972 & -0.023 & -0.024 \\
5 & 151.393 & 67.976 & 151.383 & 67.965 & -0.010 & -0.012 \\
6 & 39.311 & 26.224 & 39.308 & 26.221 & -0.003 & -0.003 \\
7 & 3.682 & 5.384 & 3.681 & 5.383 & -0.001 & -0.002 \\
\end{tabular}
\end{ruledtabular}
\end{table}

Members one through seven move by at most $0.24\meV$ in position and
$0.08\meV$ in width, the shifts decreasing monotonically along the
sequence to $0.002/0.004\meV$ for member seven at degeneracy and
$0.001/0.002\meV$ at $\Delta=0.2\eV$. The extended grid resolves members
eight and nine at degeneracy ($B=24.3$ and $11.4\meV$, $\GC=7.4$ and
$3.5\meV$), which the $400\amu$ grid already places within $0.03$ and
$0.10\meV$; the successive ratios of the $800\amu$ sequence are $2.138$
in binding and $2.131$--$2.136$ in width for members five through nine,
against the dipole value $2.137$. The tenth candidate ($B\simeq5\meV$)
moves by $0.9\meV$, and two shallower candidates appear only in the
extended basis with a width ratio of one between them; none of these
three is resolved by the present range. Above the split $2s$ threshold
at $0.2\eV$, the stationary candidates of both grids form a discretized
continuum of the open $d\mu(2s)+t$ channel: their kinetic energies scale
as $\varepsilon_k\simeq c\,k^2$ with $c\simeq1.5\meV$ for the $400\amu$
grid and $0.29\meV$ for the $800\amu$ grid, so they are box levels rather
than resonances. Members eight and nine are not added to the Coulomb
baseline table of the main text, whose $0^+$ entries are the untruncated
QZ set.

\section{Calibration of the reduced models}

Each fitted input set keeps the position and width of a candidate together from
one basis and one overlap cutoff. For $L=0$, members 2--7 at $\Delta=0$ determine
the complex boundary. The model is continued in $\Delta$ on the specified channel
sheet, and the endpoint is checked with a larger contour radius and tighter
integration tolerance. Input sets that fail the stated matching requirements are
reported as excluded; neither positions nor widths are silently taken from another run.
The reference boundary for the original $L=1$ radial basis uses members
3--7 of the \texttt{l1k1chonly} run. Each boundary is linear in complex
binding energy $\varepsilon=E^{(d)}_{2p}(0)-E$, in eV, with the
Coulomb $2p$ threshold kept as the reference at finite splitting.
Specifically, $L_a=c_0+c_1\varepsilon$ denotes
$(1/v_a)dv_a/dx$ at the matching radius, where
$x=\ln(R/\amu)$ and $u_a=\sqrt{R}\,v_a$.
Its fitted coefficients are held fixed as $\Delta$ varies.
For the Coulomb dipole matrix, the atomic radial integral has magnitude
$3\sqrt{3}\amu$. Its lowest eigenvalues are $-68.698$ and $-66.756$
for $L=0$ and $1$, giving consecutive-binding ratios $2.1371$ and
$2.1608$ when the complex quantum defect is constant.
\begin{table}[tbp]
\caption{\label{tab:s-model}$L=0$ seventh-member model predictions at $\Delta=0.2$ eV. Each boundary is fitted to positions and widths together from one completed $\Delta=0$ run, members 2--7. $B$ is measured from that model\textquotesingle s lower threshold. These are run-dependent predictions, not an accepted uncertainty interval or a proof of physical pole identity.}
\begin{ruledtabular}
\begin{tabular}{lcrr}
Input basis & $\epsilon_S$ & $B$ (meV) & $\Gamma$ (meV) \\
\hline
Original & $10^{-12}$ & 3.692921 & 5.331122 \\
Response 24 & $10^{-12}$ & 3.776500 & 5.444400 \\
Response 24 & $10^{-11}$ & 3.787121 & 5.908994 \\
Response 24 & $10^{-10}$ & 3.877357 & 6.582238 \\
Exact angular & $10^{-12}$ & 3.769298 & 5.462472 \\
Exact angular & $10^{-11}$ & 3.779455 & 5.864304 \\
Exact angular & $10^{-10}$ & 3.877767 & 6.574253 \\
\end{tabular}
\end{ruledtabular}
\end{table}
The largest endpoint changes over these 7 fits are $6.04\times10^{-8}\meV$ in position and $2.10\times10^{-7}\meV$ in full width. They compare radii $8000$ and $16000a_{d\mu}$, contour angles $-30^\circ$ and $-40^\circ$, and relative integration tolerances $2\times10^{-10}$ and $2\times10^{-12}$. These endpoint controls test integration settings, not the accuracy of the fitted inputs.

For $L=1$ the boundary is fitted to the stationary members 5--8 of each enhanced
zero-splitting family; members 1--4 of the 16-point enhanced scans have no interior
speed minimum and do not enter the fit. Table~\ref{tab:s-model-l1} lists the fitted
boundaries, the seventh-member prediction at $\Delta=0.2\eV$ and the tracking
limits, together with the reference boundary fitted to the original radial basis.
The two enhanced input sets agree to $0.04\meV$ in $B_7$, $0.04\meV$ in
$\Gamma_7$ and $0.001\eV$ in $\Delta_c$; the original-input boundary differs by
$0.5\meV$, $1.0\meV$ and $0.02\eV$, respectively.
\begin{table}[tbp]
\caption{\label{tab:s-model-l1}$L=1$ model boundaries and predictions.
$L_a(\varepsilon)=c_0+c_1\varepsilon$ is the attractive-channel logarithmic
derivative at $R_m=20\amu$ ($\varepsilon$ in eV), fitted to the $\Delta=0$
poles of the input named in the first column. $B_7$ and $\Gamma_7$ are the
seventh-member native binding and full width at $\Delta=0.2\eV$;
$\Delta_c^{(7)}$ and $\Delta_c^{(8)}$ are the real-binding tracking limits
($0.05\meV$ rule) of members seven and eight.}
\begin{ruledtabular}
\begin{tabular}{lccrrrr}
Input & $c_0$ & $c_1$ (eV$^{-1}$) & $B_7$ (meV) & $\Gamma_7$ (meV) & $\Delta_c^{(7)}$ (eV) & $\Delta_c^{(8)}$ (eV)\\\hline
Enhanced, 24 terms & -3.487-11.648i & -7.574+0.507i & 1.788 & 3.827 & 0.255 & 0.115 \\
Enhanced, 32 terms & -3.869-11.361i & -6.231-1.704i & 1.827 & 3.792 & 0.256 & 0.115 \\
Original (\texttt{l1k1chonly}) & -3.046-10.302i & -0.375-0.068i & 1.297 & 4.783 & 0.236 & --- \\
\end{tabular}
\end{ruledtabular}
\end{table}

The spread among these fits is an observed dependence on numerical input choices.
It describes input sensitivity, not a statistical interval or a total error
bound. The integration tolerances and fit residuals characterize the reduced
calculation and are reported separately from the three-body basis and
operator comparisons.

\subsection{Complete model comparison and dependence on the fitted input}
\begin{table}[t]
\caption{\label{tab:s-v2-model-full}Enhanced $L=1$ three-body representatives
(24-term basis) and the two-threshold model refitted to the zero-splitting
members $5$--$8$ of the 24-term and 32-term bases; the last two columns use
the boundary fitted to the original radial basis. Bindings and full widths
are in meV; each binding uses its own threshold, which differ by
$0.065\meV$ at $0.2\eV$. The $0.24\eV$ three-body entries are
single-family candidates that the extended-grid basis does not
reproduce (Sec.~\ref{sec:s-l1-rmax}); they are shown for
completeness and are not used as a test of the model.}
\begin{ruledtabular}
\begin{tabular}{lrrrrrrrrr}
 & & \multicolumn{2}{c}{Three-body} & \multicolumn{2}{c}{Refit, 24-term} &
 \multicolumn{2}{c}{Refit, 32-term} & \multicolumn{2}{c}{Original input}\\
Member & $\Delta$ (eV) & $B$ & $\GC$ & $B$ & $\Gamma$ & $B$ & $\Gamma$ & $B$ & $\Gamma$\\\hline
6 & 0.00 & 103.42 & 31.36 & 103.47 & 33.13 & 103.21 & 32.51 & 101.15 & 40.32 \\
6 & 0.20 & 32.55 & 23.24 & 32.73 & 22.53 & 32.98 & 22.36 & 31.06 & 29.94 \\
6 & 0.24 & 24.41 & 20.50 & 24.64 & 19.42 & 24.94 & 19.37 & --- & --- \\
7 & 0.00 & 47.96 & 15.62 & 47.95 & 16.05 & 47.69 & 15.66 & 46.77 & 18.70 \\
7 & 0.20 & 1.73 & 3.89 & 1.79 & 3.83 & 1.83 & 3.79 & 1.30 & 4.78 \\
7 & 0.24 & 0.31 & 2.06 & 0.33 & 1.92 & 0.37 & 1.93 & --- & --- \\
\end{tabular}
\end{ruledtabular}
\end{table}

For the enhanced 24-term calibration at $\Delta=0.2\eV$, the
predicted seventh-member $L=1$ position lies within the admitted
three-body range, while its width is about $0.01\meV$ below the
lower edge. The sixth-member predicted width is $0.71\meV$ below
the working three-body value and outside the $23.1$--$26.4\meV$
control range. The 24- and 32-term model predictions agree within
$0.05\meV$ for the seventh member; their sixth-member positions
and widths differ by $0.25$ and $0.17\meV$, respectively.
Across these finite-splitting comparisons, the largest differences
from the principal three-body values are about $0.4\meV$ in
position and $0.9\meV$ in width. The zero-splitting rows instead
measure calibration residuals, reaching $1.8\meV$ in the sixth-member
width. Neither the model--calculation differences nor the spread
over retained spaces constitutes an accuracy estimate.

The original-input columns in Table~\ref{tab:s-v2-model-full}
show the dependence on the zero-splitting calibration. That boundary
is calibrated to input widths of $40.0$ and $19.0\meV$ for members six
and seven and reproduces them at $\Delta=0$ as $40.3$ and $18.7\meV$,
whereas the enhanced input gives $31.4$ and $15.6\meV$.
The resulting predictions at $\Delta=0.2\eV$ are also larger:
$29.94$ and $4.78\meV$, compared with the enhanced three-body
values $23.24$ and $3.89\meV$. This comparison illustrates why
the input spectrum must be specified when assessing the
finite-splitting model response. The calibration is, however,
insensitive to inclusion of the eighth-member input: refitting
the 24-term boundary to members 5--7 only changes the
seventh-member prediction at $\Delta=0.2\eV$ from $1.788/3.827$
to $1.799/3.874\meV$ and its tracking limit from $0.255$ to
$0.254\eV$.
For $L=0$, the enhanced 24-term input at $\epsilon_S=10^{-12}$
(members 2--7 fitted) predicts the seventh member at
$\Delta=0.2\eV$ at $B=3.777\meV$, $\Gamma=5.444\meV$, against the
three-body $3.682$ and $5.384\meV$; the binding difference of
$0.094\meV$ reduces to a position difference of about $0.030\meV$
when both are referred to the common $d+t+\mu$ breakup zero, the
remainder being the atomic threshold offset of
Sec.~\ref{sec:s-spectrum}. The exact-angular input gives
$3.675/5.399\meV$ for the three-body values and a comparison of the
same size.
For context, the original first-order-calibrated $L=0$ sweep has maximum
model differences of $3.882\meV$ in position and $3.208\meV$ in width
across all 71 matched member-$2$--$8$ points including degeneracy; the
supplemental tables retain this earlier comparison separately.

Short-range absorption also changes the apparent threshold scale. For
member seven, multiplying the zero-splitting input widths by $0.5,1,2$
at fixed real energies changes $\Delta_c/B_7(0)$ from
$7.00$ to $6.37$ to $4.95$ for $L=0$, and from $5.59$ to $5.04$ to
$3.94$ for $L=1$. Here $\Delta_c$ is the original algorithm's
tracking limit, defined by a $0.05\meV$ real-binding stopping rule or
iteration failure. These controlled changes show that a scaling
description must retain short-range loss as well as angular momentum.
They are sensitivity experiments, not physical factor-of-two error bars.
With the enhanced $L=1$ inputs the seventh-member tracking limit
moves from $0.236$ to $0.255$--$0.256\eV$, and the eighth-member
limit is $0.115\eV$. For $L=0$ the original-input boundary gives
$0.327\eV$ for the seventh and $0.149\eV$ for the eighth member.
These are operational endpoints of the tracking algorithm, where the
real part of the pole reaches the $2s$ threshold; they are not points
at which a pole disappears.

\section{Precision checks and fixed-model diagnostics}
Independent high-precision matrix-element samples identify cancellation
in the overlap tail. The measured projected metric and eigenvector
condition checks distinguish numerical arithmetic from changes in the
retained basis. No complete higher-precision molecular spectrum is used
in the present results. The numerical conclusions rest on the completed
radial, cutoff, angular and absorber controls given above.

The original fixed-boundary model was also varied over matching radii
$10$--$30\amu$, polynomial orders zero through two, fitted member sets,
and repulsive-channel boundaries. At $\Delta=0.2\eV$, the seventh-member
native bindings span $3.621$--$3.722\meV$ for $L=0$ and
$1.270$--$1.317\meV$ for $L=1$. These integration and fit-setting
ranges do not include the enhanced-input width changes in the main text.
Halving and doubling input widths instead gives binding ranges
$2.436$--$4.002\meV$ and $-0.815$--$1.807\meV$, respectively.
This changes a model input and illustrates loss dependence.

\subsection{Continued modes and elastic scattering}
\label{sec:pole-continuation}
To determine the character of a continued solution, we match the
inner boundary to asymptotic channel solutions along complex radial
contours. The sign of each channel momentum is retained during
continuation. Let $F_-$ and $F_+$ have incoming and outgoing first
columns in $d\mu(2s)+t$, respectively, with the remaining columns
decaying in $2p$. The matching matrices are
$J_\pm=F_\pm'(R_m)-Y_0F_\pm(R_m)$, where $Y_0$ is the inner
logarithmic derivative in the radial coordinate $R$. They give
\begin{equation}
 S_{ss}(E)=-\frac{\det J_-}{\det J_+},
 \qquad u_s\sim\tfrac12(h^-+S_{ss}h^+),
 \label{eq:jost-ratio}
\end{equation}
where $h^-$ and $h^+$ are incoming and outgoing channel functions.
This comparison distinguishes a pole on a continued channel sheet
from an elastic zero on the outgoing sheet~\cite{newton1961,eden1964}.
At real energies, asymptotic matching gives the optical elastic
element $S=(1+iK)/(1-iK)$, with complex phase
$\delta=(2i)^{-1}\log S$ and loss probability $1-|S|^2$.
The determinant normalization and numerical controls are given below.

\begin{figure*}[t]
\includegraphics[width=0.96\textwidth]{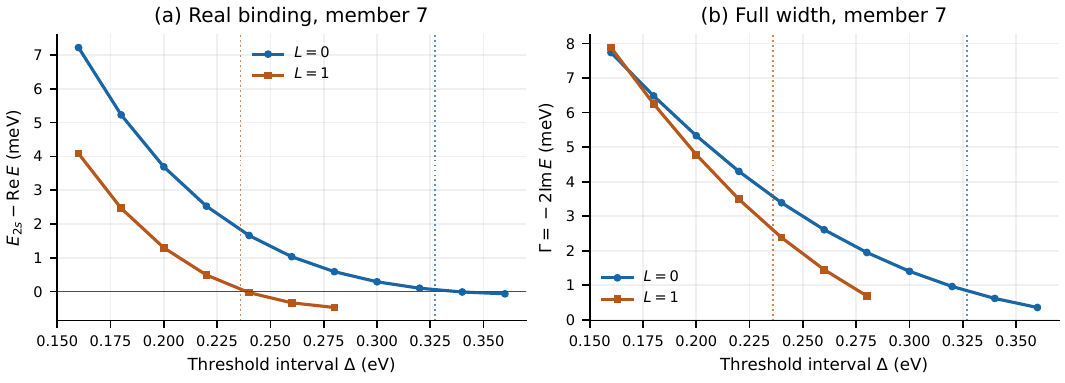}
\caption{\label{fig:pole-continuation}Seventh-member continuation in the
fixed original model, showing native binding and full width. Dotted lines
mark the earlier operational stopping values. The sheet is continued from
closed $d\mu(2s)$; four tested above-threshold modes correspond to elastic
zeros on the outgoing sheet. These curves use the original-input boundary;
with the enhanced $L=1$ inputs the tracking limit moves to $0.255\eV$.
\AIplotnote}
\end{figure*}

Complex-contour continuation extends the seventh-member model solution
beyond the original tracking limits near $0.327\eV$ and $0.236\eV$
for $L=0$ and $1$. Its real part crosses the lower threshold with a
nonzero imaginary part, as shown in Fig.~\ref{fig:pole-continuation}.
A real-part crossing is therefore distinct from collision with a real
branch point.

The continued $2s$ momentum acquires the incoming sign relative to the
open-channel outgoing branch. At $L=0$, $\Delta=0.34,0.36\eV$, and
$L=1$, $\Delta=0.24,0.28\eV$, Eq.~(\ref{eq:jost-ratio}) gives
$|S_{ss}(E)|<4\times10^{-10}$ in 24 numerical controls. The outgoing
matching matrix remains nonsingular, with singular-value ratio above
$0.17$. These four continued modes are elastic zeros on the outgoing
sheet, rather than outgoing poles at those same energies. This
classification applies to the fixed reduced model; it is distinct from
the enhanced three-body near-threshold candidate at $0.24\eV$.

At real energies the opened $d\mu(2s)+t$ channel exhibits substantial
finite-energy loss. At $0.1\eV$ above its threshold and
$\Delta=0.2\eV$, the model gives $|S|=0.2886$ and $0.2012$ for
$L=0$ and $1$, or partial-wave loss probabilities about $92\%$ and
$96\%$. The finite-energy plateau should not be extrapolated to the
zero-energy limit. Quantitative ultralow-energy amplitudes require
additional matching-radius convergence checks and are not reported
here. Apart from the first sampled interval at $\Delta=0.3\eV$, where
it rises by about $0.4\pi$ within $1.5\meV$ of the $2s$ threshold, the
sampled elastic phase decreases; neither its sign nor the absence of a
$\pi$ rise alone excludes poles in a multichannel absorptive problem. The continuation values and numerical controls are given below.

\subsection{Normalization and numerical controls}
The continued seventh-member values are listed in Table~\ref{tab:s-poles}.
In the complex-contour matching, the fundamental-solution normalization
is retained in both incoming and outgoing determinants. It is propagated
through $(\log\det F)'=\mathrm{tr}(F'F^{-1})$, using the same radial
coordinate for both derivatives. This normalization is needed for the
elastic amplitude obtained from the determinant ratio.
Thirty radius, angle and tolerance controls at six parameter points
change position and width by less than $10^{-5}$ and
$4\times10^{-6}\meV$ in this fixed model. At the four classified
above-threshold points, 24 additional controls give elastic
$|S_{ss}|<4\times10^{-10}$ with a nonsingular outgoing matching
matrix. These tolerances quantify the fixed reduced equations, not
three-body uncertainty.
\begin{table}[tbp]
\caption{\label{tab:s-poles}Seventh-member solutions on the sheet continued
from closed $d\mu(2s)$. Negative $B$ places the real part above its lower
threshold. The four tested above-threshold modes are also zeros of the
outgoing-sheet elastic element.}
\begin{ruledtabular}
\begin{tabular}{rrrr}
$L$ & $\Delta$ (eV) & $B$ (meV) & $\Gamma$ (meV)\\\hline
$0$ & $0.20$ & $+3.692947$ & $5.331183$ \\
$0$ & $0.32$ & $+0.098763$ & $0.964180$ \\
$0$ & $0.34$ & $-0.012170$ & $0.618590$ \\
$0$ & $0.36$ & $-0.065978$ & $0.355928$ \\
$1$ & $0.20$ & $+1.296202$ & $4.782277$ \\
$1$ & $0.22$ & $+0.490301$ & $3.489164$ \\
$1$ & $0.24$ & $-0.030700$ & $2.380029$ \\
$1$ & $0.26$ & $-0.332633$ & $1.450592$ \\
$1$ & $0.28$ & $-0.470823$ & $0.689924$ \\
\end{tabular}
\end{ruledtabular}
\end{table}

\section{Earlier Coulomb and effective-splitting comparisons}
The tables and figures in this section retain the original inputs for
comparison with the completed enhanced calculation. The untruncated
$L=0$ QZ estimates, the $10^{-12}$ sweep, and the enhanced working
configuration are identified separately. The original $L=1$ range
centres were descriptive inputs, not a complete envelope of all runs.
They are used only to show the input dependence of the earlier
quantum-defect fits; the main text reports the enhanced sequence.

\begin{table}[tbp]
\caption{\label{tab:setup}Energy scale and validation. Energies in eV from the
$d+t+\mu$ breakup. The ground-state entries use the $400$-function inner basis;
reference comparisons retain their original basis settings.}
\begin{ruledtabular}
\begin{tabular}{lrr}
Quantity & This work & Reference \\
\hline
$\tmu(1s)+d$ threshold        & $-2711.2417$ & $-2711.242$ \\
$\dmu(1s)+t$ threshold        & $-2663.2005$ & $-2663.200$ \\
$\tmu(2)+d$ threshold         & $-677.8104$  & $-677.810$  \\
$\dmu(2)+t$ threshold         & $-665.8001$  & $-665.800$  \\[2pt]
$\dtmu(J{=}0,v{=}0)$, total   & $-3030.3785$ & $-3030.38$\footnotemark[1] \\
\quad binding from $\tmu(1s)+d$ & $319.1368$ & $319.14$\footnotemark[1] \\
$\dtmu(J{=}0,v{=}1)$, binding & $34.78$      & $34.83$\footnotemark[1] \\[2pt]
$L=0$ levels below $\tmu(2)+d$ (11) &
   \multicolumn{2}{r}{max.\ deviation $0.40$\footnotemark[2]} \\
$L=1$ levels below $\tmu(2)+d$ (3) &
   \multicolumn{2}{r}{max.\ deviation $0.004$\footnotemark[3]} \\
\end{tabular}
\end{ruledtabular}
\footnotetext[1]{Standard $\mu$CF values; see also Ref.~\cite{korobov2000}.}
\footnotetext[2]{Against Ref.~\cite{wenzhu2025}.}
\footnotetext[3]{Against Ref.~\cite{kilic2005}, with the largest ($10\,000$-function) $1^{-}$
basis; $0.09$ with the production basis.}
\end{table}

\begin{table}[tbp]
\caption{\label{tab:L0}$L^{\pi}=0^{+}$ poles in the $n=2$ window, $E=E_{r}-i\GC/2$, in eV.  This
work: $5150$ contracted functions, untruncated QZ endpoint of the original basis; the
comparison scales for these values are defined in Sec.~\ref{sec:s-spectrum}.  The widths of Ref.~\cite{wenzhu2025} carry
$\pm0.2\eV$.}
\begin{ruledtabular}
\begin{tabular}{cccccc}
 & \multicolumn{2}{c}{This work} & \multicolumn{2}{c}{Wen--Zhu~\cite{wenzhu2025}} &
   K\i l\i \c{c}~\cite{kilic2004} \\
$j$ & $E_{r}$ & $\GC$ & $E_{r}$ & $\Gamma$ & $E_{r}$ ($\Gamma$) \\
\hline
1 & $-670.810$ & $1.273$ & $-670.73$ & $1.2$ & $-670.81$ $(1.2)$ \\
2 & $-668.130$ & $0.635$ & $-668.18$ & $0.8$ & $-668.13$ $(0.6)$ \\
3 & $-666.885$ & $0.304$ & $-666.87$ & $0.4$ & \\
4 & $-666.307$ & $0.144$ & $-666.30$ & $0.2$ & \\
5 & $-666.037$ & $0.068$ & $-666.03$ & \footnotemark[1] & \\
6 & $-665.911$ & $0.032$ & \multicolumn{2}{c}{---} & \\
7 & $-665.851$ & $0.015$ & \multicolumn{2}{c}{---} & \\
\end{tabular}
\end{ruledtabular}
\footnotetext[1]{Reported without a resolved width.}
\end{table}

The complex-defect fits below use either the original $L=0$ QZ
inputs, the original $L=1$ range centres, or one coherent $L=1$ run.
They are descriptive fits to earlier inputs and have not been updated
by refitting the enhanced $L=1$ data. The two $L=1$ eighth-member
predictions are approximately $7.78$ and $8.71\meV$ in full width.
\begin{table*}[tbp]
\caption{\label{tab:qdt}Descriptive dipole quantum-defect fits to separately identified original inputs.
The reference energy is $1\eV$. Residuals are obtained by self-consistently inverting the fitted
defect law at each member; they are not molecular error estimates. The range-centre data and
single-run data must not be combined into one fit.}
\begin{ruledtabular}
\begin{tabular}{lccc}
 & $0^+$ original QZ & $1^-$ range centres & $1^-$ single run \\
\hline
$\lambda_{\min}$ & $-68.698$ & $-66.756$ & $-66.756$ \\
$\alpha_L$ & $8.2734$ & $8.1551$ & $8.1551$ \\
$\exp(2\pi/\alpha_L)$ & $2.1371$ & $2.1608$ & $2.1608$ \\
Fitted members & $1\text{--}7$ & $1\text{--}7$ & $2\text{--}7$ \\
$\mu_0$ & $3.1113-0.1896i$ & $3.0428-0.2312i$ & $3.0515-0.2579i$ \\
$\mu_1$ (eV$^{-1}$) & $0.0048+0.0047i$ & $0.0026+0.0029i$ & $0.0135+0.0175i$ \\
Maximum $|\delta E|$ / $|\delta\Gamma|$, 2--7 (meV) & $6.06 / 1.10$ & $10.18 / 53.60$ & $3.56 / 7.56$ \\
Extrapolated member 8: $E_r$ (eV) & $-665.824282$ & $-665.821718$ & $-665.821782$ \\
Extrapolated member 8: $\Gamma$ (meV) & $7.0015$ & $7.7754$ & $8.7115$ \\
\end{tabular}
\end{ruledtabular}
\end{table*}

\begin{figure*}[tbp]
\includegraphics[width=\textwidth]{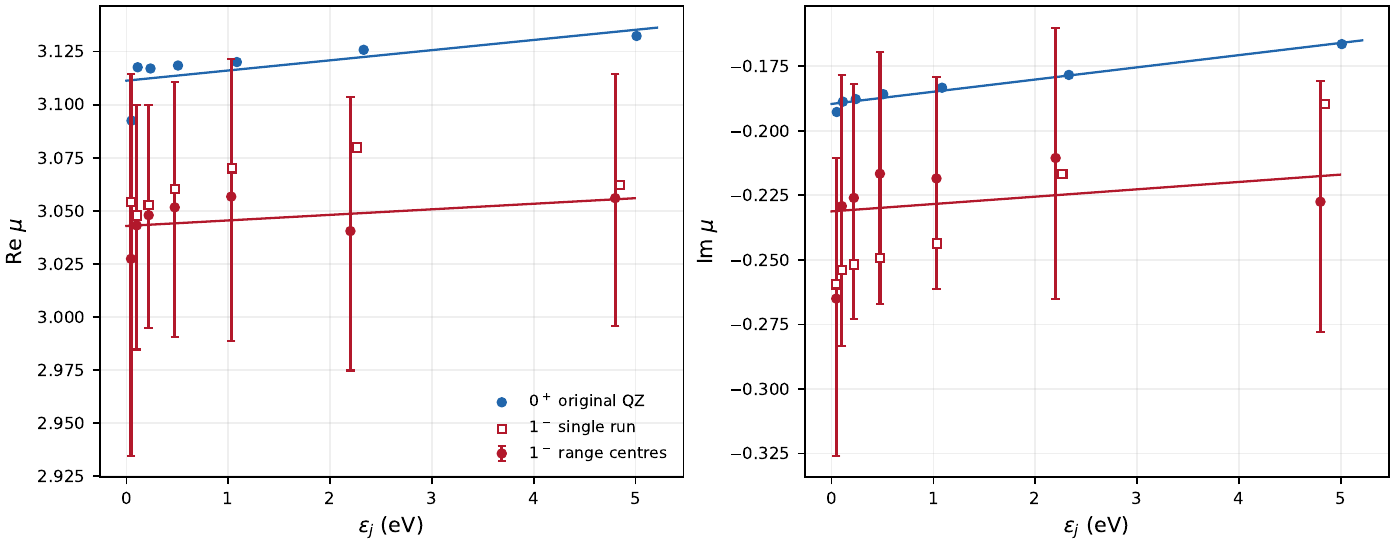}
\caption{\label{fig:qd}Complex dipole quantum defect
$\mu_j=j+(\alpha_L/2\pi)\ln[(\varepsilon_j-i\Gamma_j/2)/(1\eV)]$.
Solid lines fit the seven circles in each angular sector. Blue circles are the original
$0^+$ QZ values (comparison scales in Table~I of the main text); red circles are the original
$1^-$ range centres. Bars show exact rectangular propagation of the original ranges,
not confidence intervals. Open squares are one coherent $1^-$ run; its separate
members-$2$--$7$ fit is listed in Table~\ref{tab:qdt}. \AIplotnote}
\end{figure*}

The original $L=0$ effective-splitting sweep used a first-order
atomic calibration. A nominal $0.2\eV$ displacement produces an actual
$2s$ shift of about $-0.200683\eV$; the nonlinear calibration used in
the enhanced calculation instead imposes the actual interval.
Across all 71 original member-$2$--$8$ matched points, including zero
splitting, model position and full-width differences reach
$3.882$ and $3.208\meV$. Restricting to members $3$--$8$ at positive
splitting gives 53 pairs and maxima $1.315$ and $1.314\meV$.
These are different subsets. Neither statistic describes the complete
enhanced $L=1$ comparison.

\begin{figure*}[tbp]
\includegraphics[width=\textwidth]{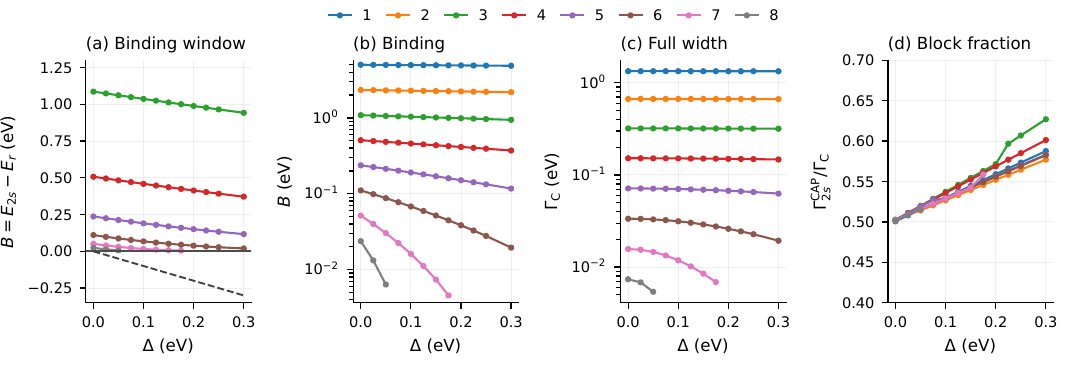}
\caption{\label{fig:sweep}Original $0^{+}$ sweep with first-order atomic calibration,
redrawn from the saved candidate rows. The coordinate is $B=E_{2s}-E_r$.
In panel (a), the solid gray line marks $B=0$ and the dashed line marks the higher
$2p$ threshold, $B=E_{2s}-E_{2p}\simeq-\Delta$, including its stored small shift.
Panel (a) magnifies the shallow region; panel (b) includes all retained bindings.
Panel (d) shows the CAP-block fractions of that calculation, not physical branching ratios.
Missing later points retain the historical unresolved status. \AIplotnote}
\end{figure*}

\begin{figure*}[tbp]
\includegraphics[width=\textwidth]{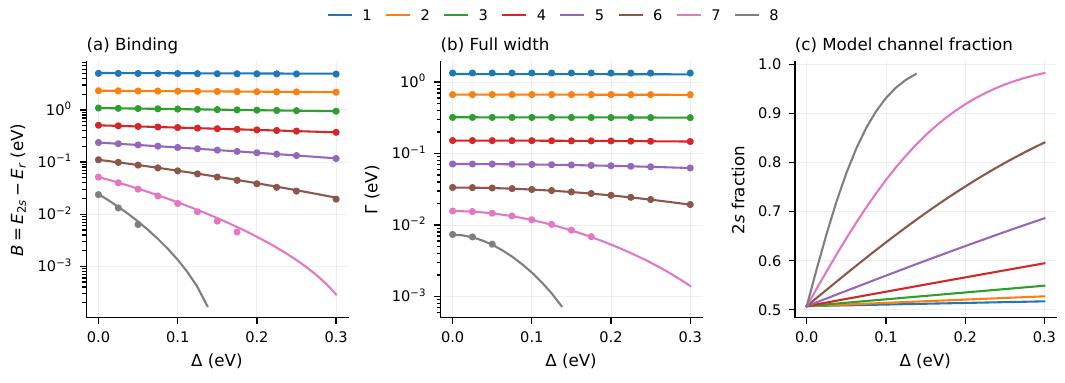}
\caption{\label{fig:mqdt0}Two-threshold MQDT (lines) against the ab initio sweep (symbols) for
$0^{+}$: binding below the $\dmu(2s)+t$ threshold, width, and $2s$ fraction.  The short-range
logarithmic derivative is fitted at $\Delta=0$; the $\Delta$ dependence is a parameter-free
prediction within the first-order calibration convention of that sweep. \AIplotnote}
\end{figure*}

\begin{table}[tbp]
\caption{\label{tab:eig}
Elastic optical-model phase and amplitude over the sampled intervals. The scans exclude the last $20\meV$ below the upper threshold and start $0.1\meV$ above the lower threshold. The five-interval row gives the largest phase change over five consecutive grid intervals. Monotonicity is not a pole-exclusion criterion. Matched at $\kappa_{2p}R_{\max}=40$; ultralow-energy amplitudes are not reported, since they require larger matching radii.}

\begin{ruledtabular}
\begin{tabular}{lcc}
 & $0^{+}$ & $1^{-}$ \\
\hline
$\Delta\delta/\pi$, $\Delta = 0.05\eV$ & $-2.56$ & $-2.42$ \\
\phantom{$\Delta\delta/\pi$,} $\Delta = 0.10\eV$ & $-2.84$ & $-2.63$ \\
\phantom{$\Delta\delta/\pi$,} $\Delta = 0.20\eV$ & $-2.94$ & $-2.67$ \\
\phantom{$\Delta\delta/\pi$,} $\Delta = 0.30\eV$ & $-2.41$ & $-2.23$ \\
Max.\ change over five intervals & $-(0.020\text{--}0.022)\pi$ & $-(0.020\text{--}0.021)\pi$ \\[3pt]
$\mathrm{Im}\,\delta/\pi$ (plateau) & $0.198$ & $0.255$ \\
$|S|$ (plateau) & $0.288$ & $0.200$ \\
\end{tabular}
\end{ruledtabular}
\end{table}

\begin{figure*}[tbp]
\includegraphics[width=\textwidth]{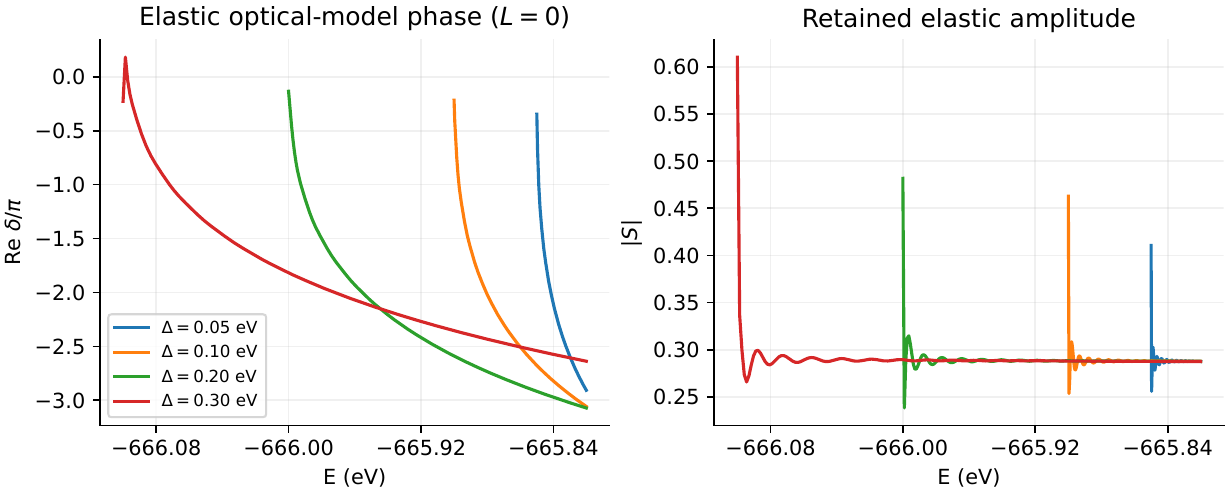}
\caption{\label{fig:eig}
Elastic phase (left) and retained amplitude $|S|$ (right) of the opened $\dmu(2s)+t$ channel. Curves use the finite sampling interval of the original scans. The phase alone does not determine pole survival, and the finite-energy plateau does not describe the zero-energy limit. \AIplotnote}

\end{figure*}

\noindent\begin{minipage}{\textwidth}
\section{Data and reproducibility}
The accompanying code and data package~\cite{zenodo2026tdmu} provides frozen
basis inputs, saved CAP spectra, atomic and reduced-model records, and
analysis scripts. Its files are accessible to the editors and referees during
the review and will become public upon publication. The file \path{REPRODUCE.md} gives a step-by-step
path to every table and figure, and \path{docs/SUPPLEMENT_EVIDENCE_MAP.md}
maps the supplemental results to their data and programs. Complete scans,
fixed-strength diagnostics, and earlier-input comparisons retain their
separate labels. The default replay reconstructs selected analyses from
saved spectra; it does not solve new three-body eigenproblems. Calculation
paths are described in \path{REPRODUCIBILITY.md}.

The ECG optimizer used for the compact $L=0$ block and the historical
Fortran comparison backend are not distributed; their frozen parameters and
comparison reports are provided as archived evidence. New three-body calculations require
rebuilding the excluded matrix caches. The original optimization and
backend checks therefore cannot be fully rerun from this package.
\end{minipage}
\bibliography{refs}